\documentclass[longauth]{aa} 

\usepackage{graphicx}
\usepackage{txfonts}
\usepackage{hyperref}
\usepackage{multirow}
\usepackage{marvosym}
\usepackage[euler]{textgreek}
\usepackage{xcolor}
\usepackage{chemformula}
\usepackage{setspace}

\newcommand{\host}{WASP-18}
\newcommand{\planet}{WASP-18\,b}
\newcommand{\cheops}{CHEOPS}
\newcommand{\tess}{TESS}

\newcommand{\spitzer}{Spitzer}
\newcommand{\gaia}{Gaia}
\newcommand{\jwst}{JWST}
\newcommand{\kepler}{Kepler}
\newcommand{\batman}{\texttt{batman}}

\newcommand{\pyratbay}{\textsc{Pyrat Bay}}
\definecolor{royalblue}{rgb}{0.255, 0.412, 0.882}
\definecolor{salmon}{rgb}{0.980, 0.502, 0.447}
\definecolor{darkorchid}{rgb}{0.596, 0.196, 0.800}
\definecolor{xkcdgreen}{rgb}{0.082, 0.690, 0.102}

\usepackage{pifont}
\newcommand{\cmark}{\ding{51}}
\newcommand{\xmark}{\ding{55}}

\newcommand{\tbc}[1]{{\bf \color{orange} [TBC\expandafter\ifx\expandafter\relax\detokenize{#1}\relax\else\textnormal{\bf: #1}\fi]}}
\newcommand{\tbd}[1]{{\bf \color{orange} [TBD\expandafter\ifx\expandafter\relax\detokenize{#1}\relax\else\textnormal{\bf: #1}\fi]}}

\newcommand{\au}{au}
\newcommand{\um}{\textmu m}

\begin{document} 

\title{
{Dark skies of the slightly eccentric \planet{}\\from its optical-to-infrared dayside emission}
\thanks{This work makes use of CHEOPS data from the Guaranteed Time Observation (GTO) programmes \texttt{CH\_PR100012} and \texttt{CH\_PR100016}.}\textsuperscript{,}\thanks{Raw and detrended light curves are available at the CDS via anonymous ftp to \href{ftp://cdsarc.u-strasbg.fr}{cdsarc.u-strasbg.fr} (\href{ftp://130.79.128.5}{130.79.128.5}) or via \url{http://cdsweb.u-strasbg.fr/cgi-bin/qcat?J/A+A/}.}}
\titlerunning{Dark skies of the slightly eccentric \planet{}}

\author{
        A.~Deline\inst{\ref{inst:Geneva},}\thanks{\email{adrien.deline@unige.ch}} \and
        P.~E.~Cubillos\inst{\ref{inst:Torino_INAF},\ref{inst:Graz_OAW}} \and
        L.~Carone\inst{\ref{inst:Graz_OAW}} \and
        B.-O.~Demory\inst{\ref{inst:Bern_CSH},\ref{inst:Bern}}\,$^{\href{https://orcid.org/0000-0002-9355-5165}{\protect\includegraphics[height=0.19cm]{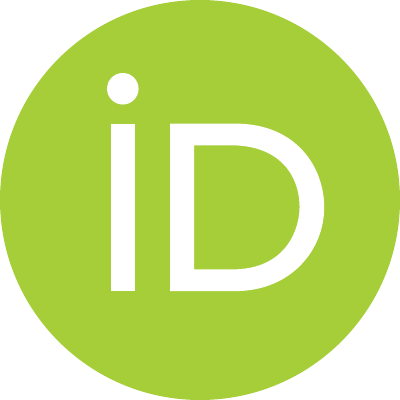}}}$ \and
        M.~Lendl\inst{\ref{inst:Geneva}}\,$^{\href{https://orcid.org/0000-0001-9699-1459}{\protect\includegraphics[height=0.19cm]{images/orcid.pdf}}}$ \and
        W.~Benz\inst{\ref{inst:Bern},\ref{inst:Bern_CSH}}\,$^{\href{https://orcid.org/0000-0001-7896-6479}{\protect\includegraphics[height=0.19cm]{images/orcid.pdf}}}$ \and
        A.~Brandeker\inst{\ref{inst:Stockholm}}\,$^{\href{https://orcid.org/0000-0002-7201-7536}{\protect\includegraphics[height=0.19cm]{images/orcid.pdf}}}$ \and
        M.~N.~G\"unther\inst{\ref{inst:ESTEC}}\,$^{\href{https://orcid.org/0000-0002-3164-9086}{\protect\includegraphics[height=0.19cm]{images/orcid.pdf}}}$ \and
        A.~Heitzmann\inst{\ref{inst:Geneva}}\,$^{\href{https://orcid.org/0000-0002-8091-7526}{\protect\includegraphics[height=0.19cm]{images/orcid.pdf}}}$ \and
        S.~C.~C.~Barros\inst{\ref{inst:Porto_CAUP},\ref{inst:Porto}}\,$^{\href{https://orcid.org/0000-0003-2434-3625}{\protect\includegraphics[height=0.19cm]{images/orcid.pdf}}}$ \and
        L.~Kreidberg\inst{\ref{inst:MPIA}} \and
        G.~Bruno\inst{\ref{inst:Catania_INAF}}\,$^{\href{https://orcid.org/0000-0002-3288-0802}{\protect\includegraphics[height=0.19cm]{images/orcid.pdf}}}$ \and
        {D.~Kitzmann\inst{\ref{inst:Bern},\ref{inst:Bern_CSH}} \and}
        A.~Bonfanti\inst{\ref{inst:Graz_OAW}}\,$^{\href{https://orcid.org/0000-0002-1916-5935}{\protect\includegraphics[height=0.19cm]{images/orcid.pdf}}}$ \and
        M.~Farnir\inst{\ref{inst:Liege}} \and
        C.~M.~Persson\inst{\ref{inst:Chalmers}} \and
        S.~G.~Sousa\inst{\ref{inst:Porto_CAUP}}\,$^{\href{https://orcid.org/0000-0001-9047-2965}{\protect\includegraphics[height=0.19cm]{images/orcid.pdf}}}$ \and
        T.~G.~Wilson\inst{\ref{inst:Warwick}}\,$^{\href{https://orcid.org/0000-0001-8749-1962}{\protect\includegraphics[height=0.19cm]{images/orcid.pdf}}}$ \and
        D.~Ehrenreich\inst{\ref{inst:Geneva},\ref{inst:Geneva_CVU}}\,$^{\href{https://orcid.org/0000-0001-9704-5405}{\protect\includegraphics[height=0.19cm]{images/orcid.pdf}}}$ \and
        {V.~Singh\inst{\ref{inst:Catania_INAF}}\,$^{\href{https://orcid.org/0000-0002-7485-6309}{\protect\includegraphics[height=0.19cm]{images/orcid.pdf}}}$ \and}
        {N.~Iro\inst{\ref{inst:DLR}} \and}
        Y.~Alibert\inst{\ref{inst:Bern_CSH},\ref{inst:Bern}}\,$^{\href{https://orcid.org/0000-0002-4644-8818}{\protect\includegraphics[height=0.19cm]{images/orcid.pdf}}}$ \and
        R.~Alonso\inst{\ref{inst:IAC},\ref{inst:LaLaguna}}\,$^{\href{https://orcid.org/0000-0001-8462-8126}{\protect\includegraphics[height=0.19cm]{images/orcid.pdf}}}$ \and
        T.~B\'arczy\inst{\ref{inst:Admatis}}\,$^{\href{https://orcid.org/0000-0002-7822-4413}{\protect\includegraphics[height=0.19cm]{images/orcid.pdf}}}$ \and
        D.~Barrado~Navascues\inst{\ref{inst:INTA}}\,$^{\href{https://orcid.org/0000-0002-5971-9242}{\protect\includegraphics[height=0.19cm]{images/orcid.pdf}}}$ \and
        W.~Baumjohann\inst{\ref{inst:Graz_OAW}}\,$^{\href{https://orcid.org/0000-0001-6271-0110}{\protect\includegraphics[height=0.19cm]{images/orcid.pdf}}}$ \and
        M.~Bergomi\inst{\ref{inst:Padova_INAF}}\,$^{\href{https://orcid.org/0000-0001-7564-2233}{\protect\includegraphics[height=0.19cm]{images/orcid.pdf}}}$ \and
        N.~Billot\inst{\ref{inst:Geneva}}\,$^{\href{https://orcid.org/0000-0003-3429-3836}{\protect\includegraphics[height=0.19cm]{images/orcid.pdf}}}$ \and
        L.~Borsato\inst{\ref{inst:Padova_INAF}}\,$^{\href{https://orcid.org/0000-0003-0066-9268}{\protect\includegraphics[height=0.19cm]{images/orcid.pdf}}}$ \and
        C.~Broeg\inst{\ref{inst:Bern},\ref{inst:Bern_CSH}}\,$^{\href{https://orcid.org/0000-0001-5132-2614}{\protect\includegraphics[height=0.19cm]{images/orcid.pdf}}}$ \and
        M.-D.~Busch\inst{\ref{inst:Bern}} \and
        A.~Collier~Cameron\inst{\ref{inst:StAndrews}}\,$^{\href{https://orcid.org/0000-0002-8863-7828}{\protect\includegraphics[height=0.19cm]{images/orcid.pdf}}}$ \and
        A.~C.~M.~Correia\inst{\ref{inst:Coimbra}} \and
        Sz.~Csizmadia\inst{\ref{inst:DLR}}\,$^{\href{https://orcid.org/0000-0001-6803-9698}{\protect\includegraphics[height=0.19cm]{images/orcid.pdf}}}$ \and
        M.~B.~Davies\inst{\ref{inst:Lund}}\,$^{\href{https://orcid.org/0000-0001-6080-1190}{\protect\includegraphics[height=0.19cm]{images/orcid.pdf}}}$ \and
        M.~Deleuil\inst{\ref{inst:LAM}}\,$^{\href{https://orcid.org/0000-0001-6036-0225}{\protect\includegraphics[height=0.19cm]{images/orcid.pdf}}}$ \and
        L.~Delrez\inst{\ref{inst:Liege_ARU},\ref{inst:Liege},\ref{inst:Leuven}}\,$^{\href{https://orcid.org/0000-0001-6108-4808}{\protect\includegraphics[height=0.19cm]{images/orcid.pdf}}}$ \and
        O.~D.~S.~Demangeon\inst{\ref{inst:Porto_CAUP},\ref{inst:Porto}}\,$^{\href{https://orcid.org/0000-0001-7918-0355}{\protect\includegraphics[height=0.19cm]{images/orcid.pdf}}}$ \and
        A.~Derekas\inst{\ref{inst:ELTE_Gothard}} \and
        B.~Edwards\inst{\ref{inst:SRON}} \and
        A.~Erikson\inst{\ref{inst:DLR}} \and
        A.~Fortier\inst{\ref{inst:Bern},\ref{inst:Bern_CSH}}\,$^{\href{https://orcid.org/0000-0001-8450-3374}{\protect\includegraphics[height=0.19cm]{images/orcid.pdf}}}$ \and
        L.~Fossati\inst{\ref{inst:Graz_OAW}}\,$^{\href{https://orcid.org/0000-0003-4426-9530}{\protect\includegraphics[height=0.19cm]{images/orcid.pdf}}}$ \and
        M.~Fridlund\inst{\ref{inst:Leiden},\ref{inst:Chalmers}}\,$^{\href{https://orcid.org/0000-0002-0855-8426}{\protect\includegraphics[height=0.19cm]{images/orcid.pdf}}}$ \and
        D.~Gandolfi\inst{\ref{inst:Torino}}\,$^{\href{https://orcid.org/0000-0001-8627-9628}{\protect\includegraphics[height=0.19cm]{images/orcid.pdf}}}$ \and
        K.~Gazeas\inst{\ref{inst:Athens}}\,$^{\href{https://orcid.org/0000-0002-8855-3923}{\protect\includegraphics[height=0.19cm]{images/orcid.pdf}}}$ \and
        M.~Gillon\inst{\ref{inst:Liege_ARU}}\,$^{\href{https://orcid.org/0000-0003-1462-7739}{\protect\includegraphics[height=0.19cm]{images/orcid.pdf}}}$ \and
        M.~G\"udel\inst{\ref{inst:Vienna}} \and
        J.~Hasiba\inst{\ref{inst:Graz_OAW}} \and
        Ch.~Helling\inst{\ref{inst:Graz_OAW},\ref{inst:Graz}} \and
        K.~G.~Isaak\inst{\ref{inst:ESTEC}}\,$^{\href{https://orcid.org/0000-0001-8585-1717}{\protect\includegraphics[height=0.19cm]{images/orcid.pdf}}}$ \and
        L.~L.~Kiss\inst{\ref{inst:Konkoly},\ref{inst:ELTE_Eotvos_Lorand}} \and
        J.~Korth\inst{\ref{inst:Lund_Observatory}} \and
        K.~W.~F.~Lam\inst{\ref{inst:DLR}}\,$^{\href{https://orcid.org/0000-0002-9910-6088}{\protect\includegraphics[height=0.19cm]{images/orcid.pdf}}}$ \and
        J.~Laskar\inst{\ref{inst:IMCCE}}\,$^{\href{https://orcid.org/0000-0003-2634-789X}{\protect\includegraphics[height=0.19cm]{images/orcid.pdf}}}$ \and
        A.~Lecavelier~des~\'Etangs\inst{\ref{inst:IAP}}\,$^{\href{https://orcid.org/0000-0002-5637-5253}{\protect\includegraphics[height=0.19cm]{images/orcid.pdf}}}$ \and
        D.~Magrin\inst{\ref{inst:Padova_INAF}}\,$^{\href{https://orcid.org/0000-0003-0312-313X}{\protect\includegraphics[height=0.19cm]{images/orcid.pdf}}}$ \and
        P.~F.~L.~Maxted\inst{\ref{inst:Keele}}\,$^{\href{https://orcid.org/0000-0003-3794-1317}{\protect\includegraphics[height=0.19cm]{images/orcid.pdf}}}$ \and
        B.~Mer\'in\inst{\ref{inst:ESAC}}\,$^{\href{https://orcid.org/0000-0002-8555-3012}{\protect\includegraphics[height=0.19cm]{images/orcid.pdf}}}$ \and
        C.~Mordasini\inst{\ref{inst:Bern},\ref{inst:Bern_CSH}} \and
        V.~Nascimbeni\inst{\ref{inst:Padova_INAF}}\,$^{\href{https://orcid.org/0000-0001-9770-1214}{\protect\includegraphics[height=0.19cm]{images/orcid.pdf}}}$ \and
        G.~Olofsson\inst{\ref{inst:Stockholm}}\,$^{\href{https://orcid.org/0000-0003-3747-7120}{\protect\includegraphics[height=0.19cm]{images/orcid.pdf}}}$ \and
        R.~Ottensamer\inst{\ref{inst:Vienna}} \and
        I.~Pagano\inst{\ref{inst:Catania_INAF}}\,$^{\href{https://orcid.org/0000-0001-9573-4928}{\protect\includegraphics[height=0.19cm]{images/orcid.pdf}}}$ \and
        E.~Pall\'e\inst{\ref{inst:IAC},\ref{inst:LaLaguna}}\,$^{\href{https://orcid.org/0000-0003-0987-1593}{\protect\includegraphics[height=0.19cm]{images/orcid.pdf}}}$ \and
        G.~Peter\inst{\ref{inst:DLR_IOSS}}\,$^{\href{https://orcid.org/0000-0001-6101-2513}{\protect\includegraphics[height=0.19cm]{images/orcid.pdf}}}$ \and
        D.~Piazza\inst{\ref{inst:Bern}} \and
        G.~Piotto\inst{\ref{inst:Padova_INAF},\ref{inst:Padova}}\,$^{\href{https://orcid.org/0000-0002-9937-6387}{\protect\includegraphics[height=0.19cm]{images/orcid.pdf}}}$ \and
        D.~Pollacco\inst{\ref{inst:Warwick}} \and
        D.~Queloz\inst{\ref{inst:ETHZ},\ref{inst:Cambridge}}\,$^{\href{https://orcid.org/0000-0002-3012-0316}{\protect\includegraphics[height=0.19cm]{images/orcid.pdf}}}$ \and
        R.~Ragazzoni\inst{\ref{inst:Padova_INAF},\ref{inst:Padova}}\,$^{\href{https://orcid.org/0000-0002-7697-5555}{\protect\includegraphics[height=0.19cm]{images/orcid.pdf}}}$ \and
        N.~Rando\inst{\ref{inst:ESTEC}} \and
        F.~Ratti\inst{\ref{inst:ESTEC}} \and
        H.~Rauer\inst{\ref{inst:DLR},\ref{inst:Berlin}}\,$^{\href{https://orcid.org/0000-0002-6510-1828}{\protect\includegraphics[height=0.19cm]{images/orcid.pdf}}}$ \and
        I.~Ribas\inst{\ref{inst:ICE},\ref{inst:IEEC}}\,$^{\href{https://orcid.org/0000-0002-6689-0312}{\protect\includegraphics[height=0.19cm]{images/orcid.pdf}}}$ \and
        N.~C.~Santos\inst{\ref{inst:Porto_CAUP},\ref{inst:Porto}}\,$^{\href{https://orcid.org/0000-0003-4422-2919}{\protect\includegraphics[height=0.19cm]{images/orcid.pdf}}}$ \and
        G.~Scandariato\inst{\ref{inst:Catania_INAF}}\,$^{\href{https://orcid.org/0000-0003-2029-0626}{\protect\includegraphics[height=0.19cm]{images/orcid.pdf}}}$ \and
        D.~S\'egransan\inst{\ref{inst:Geneva}}\,$^{\href{https://orcid.org/0000-0003-2355-8034}{\protect\includegraphics[height=0.19cm]{images/orcid.pdf}}}$ \and
        A.~E.~Simon\inst{\ref{inst:Bern},\ref{inst:Bern_CSH}}\,$^{\href{https://orcid.org/0000-0001-9773-2600}{\protect\includegraphics[height=0.19cm]{images/orcid.pdf}}}$ \and
        A.~M.~S.~Smith\inst{\ref{inst:DLR}}\,$^{\href{https://orcid.org/0000-0002-2386-4341}{\protect\includegraphics[height=0.19cm]{images/orcid.pdf}}}$ \and
        M.~Stalport\inst{\ref{inst:Liege},\ref{inst:Liege_ARU}} \and
        S.~Sulis\inst{\ref{inst:LAM}}\,$^{\href{https://orcid.org/0000-0001-8783-526X}{\protect\includegraphics[height=0.19cm]{images/orcid.pdf}}}$ \and
        Gy.~M.~Szab\'o\inst{\ref{inst:ELTE_Gothard},\ref{inst:HUN-REN-ELTE}}\,$^{\href{https://orcid.org/0000-0002-0606-7930}{\protect\includegraphics[height=0.19cm]{images/orcid.pdf}}}$ \and
        S.~Udry\inst{\ref{inst:Geneva}}\,$^{\href{https://orcid.org/0000-0001-7576-6236}{\protect\includegraphics[height=0.19cm]{images/orcid.pdf}}}$ \and
        V.~Van~Grootel\inst{\ref{inst:Liege}}\,$^{\href{https://orcid.org/0000-0003-2144-4316}{\protect\includegraphics[height=0.19cm]{images/orcid.pdf}}}$ \and
        J.~Venturini\inst{\ref{inst:Geneva}}\,$^{\href{https://orcid.org/0000-0001-9527-2903}{\protect\includegraphics[height=0.19cm]{images/orcid.pdf}}}$ \and
        E.~Villaver\inst{\ref{inst:IAC},\ref{inst:LaLaguna}} \and
        N.~A.~Walton\inst{\ref{inst:Cambridge_IoA}}\,$^{\href{https://orcid.org/0000-0003-3983-8778}{\protect\includegraphics[height=0.19cm]{images/orcid.pdf}}}$ \and
        K.~Westerdorff\inst{\ref{inst:DLR_IOSS}}
        }
\authorrunning{A.~Deline et al.}

\institute{
        Department of Astronomy, University of Geneva, Chemin Pegasi 51, 1290 Versoix, Switzerland\label{inst:Geneva}
        \and
        INAF, Osservatorio Astrofisico di Torino, Via Osservatorio, 20, I-10025 Pino Torinese To, Italy\label{inst:Torino_INAF}
        \and
        Space Research Institute, Austrian Academy of Sciences, Schmiedlstra\ss e 6, A-8042 Graz, Austria\label{inst:Graz_OAW}
        \and
        Center for Space and Habitability, University of Bern, Gesellschaftsstrasse 6, 3012 Bern, Switzerland\label{inst:Bern_CSH}
        \and
        Weltraumforschung und Planetologie, Physikalisches Institut, University of Bern, Gesellschaftsstrasse 6, 3012 Bern, Switzerland\label{inst:Bern}
        \and
        Department of Astronomy, Stockholm University, AlbaNova University Center, 10691 Stockholm, Sweden\label{inst:Stockholm}
        \and
        European Space Agency (ESA), European Space Research and Technology Centre (ESTEC), Keplerlaan 1, 2201 AZ Noordwijk, The Netherlands\label{inst:ESTEC}
        \and
        Instituto de Astrofisica e Ciencias do Espaco, Universidade do Porto, CAUP, Rua das Estrelas, 4150-762 Porto, Portugal\label{inst:Porto_CAUP}
        \and
        Departamento de Fisica e Astronomia, Faculdade de Ciencias, Universidade do Porto, Rua do Campo Alegre, 4169-007 Porto, Portugal\label{inst:Porto}
        \and
        Max Planck Institute for Astronomy, K\"onigstuhl 17, 69117 Heidelberg, Germany\label{inst:MPIA}
        \and
        INAF, Osservatorio Astrofisico di Catania, Via S. Sofia 78, 95123 Catania, Italy\label{inst:Catania_INAF}
        \and
        Space sciences, Technologies and Astrophysics Research (STAR) Institute, Universit\'e de Li\`ege, All\'ee du 6 Ao\^ut 19C, 4000 Li\`ege, Belgium\label{inst:Liege}
        \and
        Department of Space, Earth and Environment, Chalmers University of Technology, Onsala Space Observatory, 439 92 Onsala, Sweden\label{inst:Chalmers}
        \and
        Department of Physics, University of Warwick, Gibbet Hill Road, Coventry CV4 7AL, United Kingdom\label{inst:Warwick}
        \and
        Centre Vie dans l’Univers, Facult\'e des sciences, Universit\'e de Gen\`eve, Quai Ernest-Ansermet 30, 1211 Gen\`eve 4, Switzerland\label{inst:Geneva_CVU}
        \and
        Institute of Planetary Research, German Aerospace Center (DLR), Rutherfordstrasse 2, 12489 Berlin, Germany\label{inst:DLR}
        \and
        Instituto de Astrof\'isica de Canarias, V\'ia L\'actea s/n, 38200 La Laguna, Tenerife, Spain\label{inst:IAC}
        \and
        Departamento de Astrof\'isica, Universidad de La Laguna, Astrof\'isico Francisco Sanchez s/n, 38206 La Laguna, Tenerife, Spain\label{inst:LaLaguna}
        \and
        Admatis, 5. Kandó K\'alm\'an Street, 3534 Miskolc, Hungary\label{inst:Admatis}
        \and
        Depto. de Astrof\'isica, Centro de Astrobiolog\'ia (CSIC-INTA), ESAC campus, 28692 Villanueva de la Ca\~nada (Madrid), Spain\label{inst:INTA}
        \and
        INAF, Osservatorio Astronomico di Padova, Vicolo dell'Osservatorio 5, 35122 Padova, Italy\label{inst:Padova_INAF}
        \and
        Centre for Exoplanet Science, SUPA School of Physics and Astronomy, University of St Andrews, North Haugh, St Andrews KY16 9SS, UK\label{inst:StAndrews}
        \and
        CFisUC, Department of Physics, University of Coimbra, 3004-516 Coimbra, Portugal\label{inst:Coimbra}
        \and
        Centre for Mathematical Sciences, Lund University, Box 118, 221 00 Lund, Sweden\label{inst:Lund}
        \and
        Aix Marseille Univ, CNRS, CNES, LAM, 38 rue Fr\'ed\'eric Joliot-Curie, 13388 Marseille, France\label{inst:LAM}
        \and
        Astrobiology Research Unit, Universit\'e de Li\`ege, All\'ee du 6 Ao\^ut 19C, B-4000 Li\`ege, Belgium\label{inst:Liege_ARU}
        \and
        Institute of Astronomy, KU Leuven, Celestijnenlaan 200D, 3001 Leuven, Belgium\label{inst:Leuven}
        \and
        ELTE Gothard Astrophysical Observatory, 9700 Szombathely, Szent Imre h. u. 112, Hungary\label{inst:ELTE_Gothard}
        \and
        SRON Netherlands Institute for Space Research, Niels Bohrweg 4, 2333 CA Leiden, Netherlands\label{inst:SRON}
        \and
        Leiden Observatory, University of Leiden, PO Box 9513, 2300 RA Leiden, The Netherlands\label{inst:Leiden}
        \and
        Dipartimento di Fisica, Universit\`a degli Studi di Torino, via Pietro Giuria 1, I-10125, Torino, Italy\label{inst:Torino}
        \and
        National and Kapodistrian University of Athens, Department of Physics, University Campus, Zografos GR-157 84, Athens, Greece\label{inst:Athens}
        \and
        Department of Astrophysics, University of Vienna, T\"urkenschanzstrasse 17, 1180 Vienna, Austria\label{inst:Vienna}
        \newpage
        \and
        Institute for Theoretical Physics and Computational Physics, Graz University of Technology, Petersgasse 16, 8010 Graz, Austria\label{inst:Graz}
        \and
        Konkoly Observatory, Research Centre for Astronomy and Earth Sciences, 1121 Budapest, Konkoly Thege Miklós út 15-17, Hungary\label{inst:Konkoly}
        \and
        ELTE E\"otv\"os Lor\'and University, Institute of Physics, P\'azm\'any P\'eter s\'et\'any 1/A, 1117 Budapest, Hungary\label{inst:ELTE_Eotvos_Lorand}
        \and
        Lund Observatory, Division of Astrophysics, Department of Physics, Lund University, Box 43, 22100 Lund, Sweden\label{inst:Lund_Observatory}
        \and
        IMCCE, UMR8028 CNRS, Observatoire de Paris, PSL Univ., Sorbonne Univ., 77 av. Denfert-Rochereau, 75014 Paris, France\label{inst:IMCCE}
        \and
        Institut d'astrophysique de Paris, UMR7095 CNRS, Universit\'e Pierre \& Marie Curie, 98bis blvd. Arago, 75014 Paris, France\label{inst:IAP}
        \and
        Astrophysics Group, Lennard Jones Building, Keele University, Staffordshire, ST5 5BG, United Kingdom\label{inst:Keele}
        \and
        European Space Agency, ESA - European Space Astronomy Centre, Camino Bajo del Castillo s/n, 28692 Villanueva de la Ca\~nada, Madrid, Spain\label{inst:ESAC}
        \and
        Institute of Optical Sensor Systems, German Aerospace Center (DLR), Rutherfordstra{\ss}e 2, 12489 Berlin, Germany\label{inst:DLR_IOSS}
        \and
        Dipartimento di Fisica e Astronomia "Galileo Galilei", Universit\`a degli Studi di Padova, Vicolo dell'Osservatorio 3, 35122 Padova, Italy\label{inst:Padova}
        \and
        ETH Zurich, Department of Physics, Wolfgang-Pauli-Strasse 2, CH-8093 Zurich, Switzerland\label{inst:ETHZ}
        \and
        Cavendish Laboratory, JJ Thomson Avenue, Cambridge CB3 0HE, UK\label{inst:Cambridge}
        \and
        Institut fuer Geologische Wissenschaften, Freie Universit\"at Berlin, Maltheserstra{\ss}e 74-100,12249 Berlin, Germany\label{inst:Berlin}
        \and
        Institut de Ciencies de l'Espai (ICE, CSIC), Campus UAB, Can Magrans s/n, 08193 Bellaterra, Spain\label{inst:ICE}
        \and
        Institut d'Estudis Espacials de Catalunya (IEEC), 08860 Castelldefels (Barcelona), Spain\label{inst:IEEC}
        \and
        HUN-REN-ELTE Exoplanet Research Group, Szent Imre h. u. 112., Szombathely, H-9700, Hungary\label{inst:HUN-REN-ELTE}
        \and
        Institute of Astronomy, University of Cambridge, Madingley Road, Cambridge, CB3 0HA, United Kingdom\label{inst:Cambridge_IoA}
        }

\date{Received May 31, 2024; accepted May 1, 2025}

\abstract
    {Ultra-hot Jupiters (UHJs) are gas giant exoplanets that are strongly irradiated by their star, setting intense molecular dissociation that leads to atmospheric chemistry dominated by ions and atoms. These conditions inhibit day-to-night heat redistribution, which results in high temperature contrasts. Phase-curve observations over several passbands offer insights on the thermal structure and properties of these extreme atmospheres.}
    {We aim to perform a joint analysis of {multiple} observations of \planet{} from the visible to the mid-infrared, using data from \cheops{}, \tess{}, and \spitzer{}. Our purpose is to characterise the {planetary atmosphere} with a consistent view over the large wavelength range covered, including \jwst{} {data}.}
    {We implemented a model for the planetary signal including transits, occultations, phase signal, ellipsoidal variations, Doppler boosting, and light travel time. We performed a joint fit of more than 250 eclipse events and derived the atmospheric properties using general circulation models (GCMs) and retrieval analyses.}
    {We obtained new ephemerides with unprecedented precisions of 1\,second and 1.4\,millisecond on the time of inferior conjunction and orbital period, respectively. We computed a planetary radius of $R_p=1.1926\pm0.0077\,R_J$ with a precision of 0.65\% (or 550\,km). Based on a timing inconsistency with \jwst{}, we discuss and confirm the orbital eccentricity ($e=0.00852\pm0.00091$). We also constrain the argument of periastron to {$\omega=261.9^{+1.3}_{-1.4}\,\deg$}. We show that the large dayside emission implies the presence of magnetic drag and super-solar metallicity. We find a steep thermally inverted gradient in the planetary atmosphere, which is common for UHJs. We detected the presence of strong CO emission lines at 4.5\,\textmu m from an excess of dayside brightness in the \spitzer{}/IRAC/Channel\,2 passband. Using these models to constrain the reflected contribution in the \cheops{} passband, we derived an extremely low geometric albedo of $A_g^\text{\cheops{}}={0.027}\pm{0.011}$.}
    {The orbital eccentricity remains a potential challenge for planetary dynamics that might require further study given the short-period massive planet and despite the young age of the system. The characterisation of the atmosphere of \planet{} reveals the necessity to account for magnetic friction and super-solar metallicity to explain the full picture of the dayside emission. We find the planetary dayside to be extremely unreflective; however, when juxtaposing \tess{} and \cheops{} data, we get hints of increased scattering efficiency in the visible, likely due to Rayleigh scattering.}


\maketitle

\section{Introduction} \label{sec:intro}
    
    Most exoplanetary properties, such as temperatures, radii, compositions, and orbital architectures, span a broad range of values that go well beyond those of our Solar System. One of the categories of extra-solar planets that best illustrates this variety are hot Jupiters. These gas giants orbiting close to their stars exhibit extreme conditions in their massive atmospheres with temperatures above $\sim$\,1000~K. At such small planet-to-star separations, hot Jupiters tend to synchronise their rotation and revolution periods due to immense tidal forces, which results in the stellar insolation always heating the same planetary hemisphere. Tidally locked objects receive large amounts of energy on their permanent dayside, which may lead to high day-to-night contrast and/or strong winds redistributing heat on a planetary scale \citep{Showman2002_atmos_circulation, Showman2010_atmos_circulation, Komacek2016_atmos_circulation, Zhang2020_atmos_circulation}.
    The hottest planets in this category are dubbed ultra-hot Jupiters (UHJs), whose atmospheres enter a specific regime where the dayside is dominated by atomic hydrogen \citep{Bell2018}, the major source of spectral continuum opacity is induced by hydrogen anions (H$^-$) \citep{Arcangeli2018, Lothringer2018, Parmentier2018}, and the atmospheric composition of atoms and ions resembles that of a star \citep{Kitzmann2018}.
    The thermal emission coming from UHJs is so strong that it can be observed from the infrared up to the optical with large occultation signals when the planetary dayside is hidden by the host star. Observations over a large range of wavelengths are thus decisive to get key insights on UHJs.
    
    The UHJ known as \planet{} \citep{Hellier2009_W18b} is orbiting an early F6-type star with a very short orbital period of 0.94~day and a separation of 0.02~\au{} (semi-major axis). The planet is a peculiar object of ten~Jupiter masses, which is only slightly larger than Jupiter ($1.2\,R_J$), which makes it denser than the Earth (more than five~times denser than Jupiter). With such properties, \planet{} resides at the intersection between brown-dwarf and planet populations. It was recently observed by the Characterising Exoplanet Satellite (\cheops{}; \citealt{Benz2021_CHEOPS}) and the Transiting Exoplanet Survey Satellite (\tess{}; \citealt{Ricker2015_TESS}). These observations continue to contribute to the extensive characterisation of the planet with several instruments, including the \spitzer{} \citep{Werner2004_Spitzer} and James~Webb space telescopes (\jwst{}; \citealt{Gardner2006_JWST}). Recent works have reported the detection of several species in the atmosphere of \planet{} including $\text{H}^-$, $\text{H}_2\text{O}$, $\text{OH}$, and $\text{CO}$ \citep{Changeat2022_HJ, Brogi2023_W18b, Yan2023_W18b_W76b, Coulombe2023}. Upper limits on the geometric albedo have been computed in the \tess{} passband \citep{Shporer2019_W18b, Blazek2022_albedos}.
    
    In this work, we report the new \cheops{} and \tess{} phase-curve observations of \planet{}. We jointly analysed these data sets with already published \tess{} observations \citep{Shporer2019_W18b}. We also included \spitzer{} observations: 4~occultations previously analysed in \cite{Nymeyer2011_WASP-18b} and 10~occultations observed in 2015 and firstly reported in \cite{2023AJ....165..104D}. We first start by presenting refined stellar properties of \host{} in Section~\ref{sec:star}. Sections~\ref{sec:obs} and~\ref{sec:lc} detail the data sets and the modelling framework used in this analysis. We present our results and how we improved the constraints on the planetary parameters (Section~\ref{sec:results}). Finally, we discuss the implications on the atmospheric properties of the planet in Section~\ref{sec:atmos}.

\section{Stellar properties of \host} \label{sec:star}
    
    In the framework of this study, we derived the properties of the host star \host{} (HD~10069; TOI-185) listed in Table~\ref{tab:star}, following the methods described below.
    
    \begin{table}
        \caption{Properties of the star \host.}
        \label{tab:star}
        \centering
        \resizebox{\columnwidth}{!}{
            \begin{tabular}{lcr}
                \hline\hline
                Parameter & Value & Source \\
                \hline
                \multirow{5}{*}{Names \& aliases} & \host{} & \multirow{5}{*}{Simbad \tablefootmark{1}} \\
                & HD\,10069 & \\
                & TOI-185 & \\
                & TIC\,100100827 & \\
                & Gaia~DR2~4955371367334610048 & \\
                \hline
                V-band magnitude & $\sim 9.30$ & Simbad \tablefootmark{1} \\
                Gaia G-band magnitude & $9.16617\pm0.00024$ & Gaia archive \tablefootmark{2} \\
                \hline
                $T_\text{eff}\ \left[K\right]$ & $6332\pm60$ & spectroscopy \\
                $M_\star\ \left[M_\odot\right]$ & $1.245^{+0.057}_{-0.058}$ & evolution model \\
                $R_\star\ \left[R_\odot\right]$ & $1.256\pm0.008$ & IRFM \\
                $\log g\ \left[\log_{10}\left(\text{cm}.\text{s}^{-2}\right)\right]$ & $4.34\pm0.10$ & spectroscopy \\
                $v_\text{mic}\ \left[\text{km}.\text{s}^{-1}\right]$ & $1.50$ & \cite{Bruntt2008_spectroscopy}   \\
                $v_\text{mac}\ \left[\text{km}.\text{s}^{-1}\right]$ & $5.60$ &  \cite{Doyle2014_vmac}  \\
                $\left[\text{Fe}/\text{H}\right]$ & $0.08\pm0.04$ & spectroscopy \\
                $t_\star\ \left[\text{Gyr}\right]$ & $1.4\pm0.7$ & evolution model \\
                $L_\star\ \left[L_\odot\right]$ & $2.285\pm0.091$ & $L_\star=4\pi R_\star^2 \sigma_\text{SB} T_\text{eff}^4$ \\
                $v_\star\sin{i_\star}\ \left[\text{km}.\text{s}^{-1}\right]$ & $11.2\pm0.6$ & spectroscopy \\
                \hline\hline
            \end{tabular}
        }
        \tablefoot{
            The methods used to obtain the stellar parameters are described in the text. The stellar luminosity $L_\star$ is computed following the Stefan–Boltzmann law and using the Stefan–Boltzmann constant $\sigma_\text{SB}$.
            \tablefoottext{1}{SIMBAD astronomical database from the Centre de Donn\'ees astronomiques de Strasbourg\footnote{\url{http://simbad.u-strasbg.fr/simbad/}}.}
            \tablefoottext{2}{Archive of the Gaia mission of the European Space Agency\footnote{\url{https://gea.esac.esa.int/archive/}}}
        }
    \end{table}
    
    We used co-added high-resolution spectra obtained with the HARPS spectrograph \citep{Mayor2003_HARPS, Pepe2004_HARPS} and analysed them with the spectral synthesis tool \texttt{SME}\footnote{\url{http://www.stsci.edu/~valenti/sme.html}} \citep[Spectroscopy Made Easy;][]{Valenti1996_SME, Piskunov2017_SME}. This software fits the observations to a set of computed synthetic spectra for a chosen set of parameters based on atomic and molecular line data  from the Vienna Atomic Line Database\footnote{\url{http://vald.astro.uu.se}} \citep[VALD;][]{Piskunov1995_VALD, Ryabchikova2015_VALD}. We chose the ATLAS12 \citep{Kurucz2013_ATLAS12} stellar atmosphere grid and modelled the stellar effective temperature, $T_\text{eff}$, the surface gravity, $\log g$, abundances, and the projected rotational velocity, $v\,\sin i_\star$. 
    We used narrow and unblended lines between $6200\,\text{\AA}$ and~$6600\,\text{\AA}$ to derive abundances and $v\,\sin i_\star$. 
    The micro- \citep{Bruntt2008_spectroscopy} and macro-turbulent \citep{Doyle2014_vmac} velocities were kept fixed to $1.50\,\text{km}.\text{s}^{-1}$ and $5.60\,\text{km}.\text{s}^{-1}$, respectively. We refer to \cite{Persson2018_K2-216} for further details on the modelling. Our results suggest that \host{} is an early F6\textsc{v} star with $T_\text{eff} = 6332 \pm 60\,K$. All results are  listed in Table \ref{tab:star} and are within $1\,\sigma$ agreement with the \gaia{}~DR3 results and the values listed on the NASA archive\footnote{\url{https://exoplanetarchive.ipac.caltech.edu/}}.
    
    We used our stellar spectroscopic parameters to determine the stellar radius of \host{} using a MCMC modified infrared flux method \citep[IRFM;][]{Blackwell1977,Schanche2020}. From the stellar properties $T_\text{eff}$, $\log g$, and $\left[\text{Fe}/\text{H}\right]$, we constrained stellar atmospheric models from two catalogues \citep{Kurucz1993,Castelli2003}. We then constructed spectral energy distributions (SEDs) from which we computed synthetic photometry. We compared the photometry to the broadband observations in the following passbands: \gaia{} $G$, $G_\mathrm{BP}$, and $G_\mathrm{RP}$, 2MASS $J$, $H$, and $K$, and WISE $W1$ and $W2$ \citep{Skrutskie2006,Wright2010,2023A&A...674A...1G}. We derived the stellar bolometric flux that we had converted into the stellar angular diameter using the measured effective temperature. The angular diameter and offset-corrected \gaia{} parallax \citep{Lindegren2021} were combined to produce the stellar radius. To account and correct for atmospheric model uncertainties, we conducted a Bayesian modelling, averaging of the posterior distributions of the radius produced via this process with both atmospheric catalogues.

    The effective temperature, $T_\text{eff}$, metallicity, $\left[\text{Fe}/\text{H}\right]$, and radius, $R_\star$ along with their uncertainties, were used as basic inputs to infer the isochronal mass, $M_\star$, and age, $t_\star$, from stellar evolutionary models. We derived two pairs $\left\{M_\star, t_\star\right\}$ from two different approaches. The first pair of mass and age were estimated by applying the isochrone placement algorithm \citep{Bonfanti2015_stellar_age, Bonfanti2016_stellar_age} that interpolates the input stellar parameters within pre-computed grids of isochrones and tracks of \texttt{PARSEC}\footnote{\textsl{PA}dova and T\textsl{R}ieste \textsl{S}tellar \textsl{E}volutionary \textsl{C}ode: \url{http://stev.oapd.inaf.it/cgi-bin/cmd}} v1.2S \citep{Marigo2017_PARSEC}. As outlined in \citet{Bonfanti2016_stellar_age}, the isochrone placement code also implements the gyrochronological relation by \cite{Barnes2010_stellar_rotation} and, thus, we further provided the algorithm with the projected equatorial velocity of the star ($v_\star\sin{i_\star}$) to benefit from the synergy between isochrone interpolation and gyrochronology \citep[see e.g.][]{Angus2019_stellar_age}, thereby improving the convergence. The second pair of mass and age estimates were derived from the Code Li\`egeois d'\'Evolution Stellaire \citep[\texttt{CL\'ES};][]{Scuflaire2008_CLES}, which computes the best-fit evolutionary track accounting for the basic input set of stellar parameters following a Levenberg-Marquardt minimisation scheme \citep{Salmon2021_AlphaCen}. As detailed in \citet{Bonfanti2021_HD108236}, we checked the mutual consistency between the two respective pairs of outcomes via a $\chi^2$-based criterion and then merged the results obtaining $M_\star=1.245_{-0.058}^{+0.057}~M_\odot$ and $t_\star=1.4\pm0.7$~Gyr.

    We also note that the star \host{} has a very low magnetic activity as indicated by several indices \citep[e.g. Fig.~2 of][]{Fossati2013_WASP-12}. This unusually quiet behaviour and its possible causes have been discussed in several works \citep[e.g.][]{Lanza2014_chromospheric_flux, Fossati2015_chromospheric_flux, Lanza2024_WASP-18}.
                
\section{Observations and data reduction} \label{sec:obs}
        
        \subsection{\cheops{} observations} \label{ssec:cheops_obs}
            
            We obtained 38 observations of \host{} with \cheops{} spanning three observability seasons and a total time range of 2~years and 2~months. These visits were part of the Guaranteed Time Observation programmes of the \cheops{} consortium (\texttt{CH\_PR100012} and \texttt{CH\_PR100016}) and covered 25~occultations, 11~transits, and a complete phase-curve. One of the visits, executed on 2022-09-29, was interrupted to perform a collision avoidance manoeuvre and the very short data set (of 1~hour) was not included in the analyses of this work. The visits from programme \texttt{CH\_PR100012} (transits) were observed with an exposure time of 53\,sec, and the visits from programme \texttt{CH\_PR100016} (occultations and phase curve) were observed with an exposure time of 50\,sec. Table~\ref{tab:cheops_log} provides a detailed overview of all the \cheops{} observations {with their corresponding raw light curves displayed in Fig.~\ref{fig:cheops_raw_data}.}.

                The photometric light curve was extracted from the raw data using the Data Reduction Pipeline (DRP; \citealt{Hoyer2020_CHEOPS_DRP}) that performs circular aperture photometry on images corrected for several effects of instrumental (bias offset, gain, dark current, hot pixels, and flat field) and astrophysical (cosmic rays, stray light, and smearing trails) origins. The DRP provides photometry for a range of aperture radii to make it possible to select the best option. To determine the aperture size with the best photometric precision, we first estimated the white noise level from the raw light curves. For each aperture, we computed the point-to-point difference of the light curves to remove any signal (planetary or red noise) and calculated the standard deviations of the resulting flat data sets using a method robust to outliers. These values were then divided by $\sqrt{2}$ to correct for the point-to-point difference step and obtain a reliable estimate of the white noise level. We then selected the aperture of 24\,pixels that consistently gives the smallest noise level over all the 37 visits.
                
                Across all the visits, we identified 400 out of 13760 points (2.91 \%) that were flagged by the DRP (e.g. out-of-range temperatures, Earth occultation, and high cosmic ray hits). We also detected 148 outliers (1.11 \%) by performing 3.5-$\sigma$ clipping after subtracting the best-fit results of the modelling described in Section~\ref{sec:lc}. Finally, we flagged 407 (3.05 \%) data points that had background values above $7.8\,10^5$\,electrons. This threshold was fixed as the point above which the flux does not correlate monotonically with the background level (a similar approach as that shown in Fig.~2 of \citealt{Deline2022_WASP-189b}). The total of discarded points was 509 out of 13360 (3.81 \%) since some of the outliers also had high background values.
                
                Due to the orbital configuration of the spacecraft and the requirements on its thermal stability, \cheops{} photometry is generally affected by short interruptions every $\sim$100\,min (the orbital period of \cheops{}) mostly due to Earth occultations and also systematic trends induced by background level variations or close-by stars. These systematics usually strongly correlate with the roll angle of the spacecraft, the background level, and the photometric centroid of the target (e.g.~\citealt{2020A&A...643A..94L, Delrez2021_nu2Lupi, Morris2021_55Cnc, Barros2022_WASP-103b_deformation, Deline2022_WASP-189b, Krenn2023_HD189733b, Bonfanti2024_TOI-732, Demangeon2024}). We included the photometric variability caused by systematics in our global model using Gaussian processes (GPs) and polynomial correlation, as detailed in Section~\ref{ssec:systematics}.

        \subsection{\tess{} observations} \label{ssec:tess_obs}
                
                The Transiting Exoplanet Survey Satellite (\tess{}; \citealt{Ricker2015_TESS}) observed the star \host{} in sectors 2, 3, 29, 30, and 69, spanning 5~years from August~2018 to September~2023, and covering up to 115~phase curves of planet\,b. We downloaded the data sets from the Mikulski Archive for Space Telescopes (MAST\footnote{\url{https://archive.stsci.edu/}}). The photometry of sectors 2 and 3 was available in a 2 minute cadence, while sectors 29, 30, and 69 had both 2 minute and 20 second cadences. To optimise the computational cost our our analysis, we only used the 2 minute cadence data for all sectors.
                \tess{} photometry was provided by the Science Processing Operations Center~(SPOC) in two formats that have undergone different levels of processing. The simple aperture photometry (SAP) is a sum of pixel values within predefined non-circular apertures computed on calibrated images. The pre-search data conditioning SAP (PDCSAP) goes further by correcting the SAP photometry using information from the most common systematic features stored in the so-called cotrending basis vectors (CBVs; \citealt{Kinemuchi2012_Kepler_systematics}). The resulting PDCSAP photometry is usually much cleaner than SAP, with fewer long-term trends. However, the PDCSAP flux can sometimes feature systematic trends that are not present in the SAP or, in the worst cases, have some low-amplitude astrophysical signals removed. After a careful inspection of the \tess{} light curves from all sectors, we found that the PDCSAP provides a better photometry with significant corrections of red noise for all sectors, except for sector~3 for which PDCSAP flux has strong systematics that are not present in the SAP flux. We thus selected the cleaner SAP for sector 3 and PDCSAP for the other sectors (2, 29, 30, and 69) as the base \tess{} flux for this work.
                We used the \textsc{quality} flag provided for the SPOC pipeline to discard 16565 out of 93313 data points (17.75 \%).
                Given the focus of our analysis is to measure the phase-curve signal of \planet{}, we have not included any complex time-dependent systematic correction (e.g. spline functions or Gaussian processes) as we might lose our ability to accurately retrieve the planetary signal. We thus discarded parts of the light curves with remaining trends or high noise to keep only the cleanest photometry of each sector, as shown in Fig.~\ref{fig:tess_raw_data} and listed in Table~\ref{tab:tess_cut}.
                From the best-fit residuals obtained with the model described in Section~\ref{sec:lc}, we performed a 4-$\sigma$ clipping to remove 28 outliers out of the remaining 67190 photometric points.
                The final \tess{} light curves that we obtained after discarding trends and outliers cover about 101 full orbital period of \planet.

        \subsection{\spitzer{} observations} \label{ssec:spitzer_obs}

                We recovered the observational data of \host{} with the Infrared Array Camera (IRAC) of the \spitzer{} space telescope \citep{Werner2004_Spitzer} from the \spitzer{} Heritage Archive~(SHA\footnote{\url{https://irsa.ipac.caltech.edu/applications/Spitzer/SHA/}}). We downloaded all available data sets from three General Observing (GO) programmes: 50517 (PI:~Harrington), 60185 (PI:~Maxted), and 11099 (PI:~Kreidberg). Programme~50517 observed two~planetary occultations in December 2008, both observed simultaneously with the IRAC channels~1 and~3, and~2 and~4, respectively \citep{Nymeyer2011_WASP-18b}. Programme~60185 included two full-orbit observations of \planet{} in January and August 2010 with the IRAC channels 1 and 2, respectively \citep{Maxted2013_WASP-18b}. 
                Programme~11099 covered ten~occultations of planet~b with the IRAC channel~2 in September 2015 that were published in \cite{2023AJ....165..104D}. A detailed log of the \spitzer{} observations is listed in Table~\ref{tab:spitzer_log}.
                
                We re-extracted the photometry and applied a pre-processing correction that models the IRAC intra-pixel sensitivity \citep{Ingalls2016_Spitzer} using the bilinearly interpolated subpixel sensitivity~(BLISS) mapping \citep{Stevenson2012_BLISS}.  We also included the possibility for a linear decorrelation as a function of the full-width at half maximum (FWHM) of the pixel response function (PRF).
                The modelling uncertainties of this correction were added quadratically to the errorbars of the data to ensure consistent error propagation. The final corrected light curves were sampled at cadences of 10.8 and 13.6~seconds (channels 1 and 2 and channels 3 and 4, respectively) for programme~50517, 27.4~seconds for programme~60185, and 129.4~seconds for programme~11099. \cite{Demory2016_55Cnce} and \cite{Bourrier2022_HD3167} provide a detailed description of this pre-processing step.
                
                When analysing the full-orbit light curves of programme~60185 (PI:~Maxted), we noticed abnormal levels of red noise. We made several reduction attempts varying the noise parametrisation in order to mitigate the systematics without success. We also reduced the data with another independent pipeline \citep{Stevenson2012_BLISS, StevensonEtal2012apjGJ346nonPlanets, CampoEtal2011apjWASP12b, CubillosEtal2013apjWASP8b, CubillosEtal2014apjTrES1, CubillosEtal2017apjRednoise, BellEtal2019mnrasWASP12bPhaseCurve}, but we ended up not being able to detrend the systematics from the astrophysical signal. The residual red noise was significant and the astrophysical signal depended heavily on the modelling choices without a clear optimal procedure (e.g. which data points or systematics models to include). This resulted in significant inconsistencies for some parameters when comparing these two phase curves with the other light curves from \spitzer{}, but also \cheops{} and \tess{}, in particular, for the noise levels, the occultation depths, and the mid-transit times. These inconsistencies were biasing our results when jointly analysing all the light curves. Without being able to explain the source of this red noise and without being able to retrieve reliable results out of these data sets, we chose to discard these two \spitzer{} phase curves.
                
                We computed a joint fit to the remaining \spitzer{} data using the modelling of astrophysical signal and systematic noise described in Section~\ref{sec:lc} and performed a 4.5-sigma clipping to the best-fit residuals. This resulted in flagging and discarding a total of four~outliers out of 8519 data points (0.05\%).

\section{Light-curve analysis} \label{sec:lc}

        \subsection{Normalisation and noise modelling} \label{ssec:systematics}
                
                We adopted a similar normalisation and noise modelling for each of the observations analysed in this work; namely, each \cheops{} visit, each \tess{} orbit (2 orbits per sector), and each \spitzer{} visit. At least two~parameters were fitted for each observation: the reference flux value used to normalise the photometric flux (see Section~\ref{ssec:astro_model} for details about what a normalised flux of~1 corresponds to) and the noise jitter, $\sigma_\text{w}$, that is added (or subtracted if $\sigma_\text{w}<0$) quadratically to the photometric error bars to account for the under- or over-estimation of the uncertainties of the data points.
                For each of the \cheops{} and \spitzer{} visits, we evaluated whether an additional linear slope was necessary to properly model the data. For this, we fit a full model including slopes for all visits to the data and identified every visit with a significant slope (value inconsistent with~0 by $>3\sigma$). All other slopes were fixed to~0 in our final model. This resulted in fitting a slope for \cheops{} visits 3, 9, 31, 32, 33, and 35, and for \spitzer{} visits 3, 7, 8, 9, and 10.
                
                These corrections are enough for the \tess{} and \spitzer{} data sets, as the photometric light curves have been pre-processed with the PDCSAP and the BLISS mapping, respectively.
                
                \cheops{} photometry, however, is not corrected and is strongly affected by systematics, mainly due to background level variations and flux modulation due to the spacecraft rolling around its line of sight, which results in having the field of view rotating around the target star. We included a correction for the induced systematic noise as a function of the background flux and the roll angle of the satellite.
                We modelled the flux variations as a function of the background with the following formula:
                \begin{equation} \label{eq:bkg}
                        F=a_\text{bkg}\left[\log_{10}\!\left(B/B_0\right)\right]^2+b_\text{bkg}\log_{10}\!\left(B/B_0\right),
                \end{equation}
                where $F$ is the photometric flux, $B$ is the background level, $B_0$ is a fixed reference background level, and $a_\text{bkg}$ and $b_\text{bkg}$ are fitted parameters. We compared the Bayesian information criterion (BIC) of the model with only $a_\text{bkg}$ free ($b_\text{bkg}=0$), only $b_\text{bkg}$ free ($a_\text{bkg}=0$), or both parameters free. The latter option was shown to be significantly favoured ($\Delta\text{BIC}>70$) and we selected it for our flux-background model. We determined the fixed value of $B_0=188\,425.83\,\text{e}^-$ from a Gaussian fit to the histogram of all background values, which provides a mean (or median) estimate that is robust to outlier.
                
                The remaining systematic noise in CHEOPS light curves is strongly modulated by the rotation of the field of view, either due to background stars or any diffuse light source (e.g. Earth straylight). These effects on the photometry repeat themselves with each one of CHEOPS' orbits and can be accurately modelled as a function of the spacecraft roll angle. We adopted a Gaussian process (GP) modelling of the noise as a function of the roll angle, using a Mat\'ern-3/2 kernel from the \texttt{celerite2} package \citep{celerite1, celerite2}. This means that our GP was fitted to the residual flux phase-folded on the spacecraft roll-angle values, which are ranging roughly from $15\deg$ to $322\deg$. Given the versatility of GPs, we chose to use the same set of hyper-parameters (the standard deviation of the process $\sigma_\text{GP}$ and the correlation scale in roll-angle unit $\rho_\text{GP}$) for all CHEOPS visits, with each visit being modelled independently from the others. This means that every visit has its roll-angle dependency modelled with the same hyper-parameters, but only using the data points of that specific visit. This choice was found to be a good compromise between modelling flexibility and the number of free parameters.

                All the steps normalising the flux and modelling the noise were conducted simultaneously with the astrophysical model fit.

        \subsection{Astrophysical model} \label{ssec:astro_model}
                
                \subsubsection{Light travel time} \label{sssec:ltt}
                        
                        We modelled the astrophysical signal of the \host{} system by dividing it into two additive contributions: the planetary flux and the stellar flux. Both contributions are accounting for the light travel time (LTT) through the planetary system and synchronised at the time of inferior conjunction, $T_0$ (mid-transit time). The corresponding correction is done by converting the observation times, $t_\text{obs}$, into LTT-corrected times, $t_\text{corr}$, using the following equation:
                        \begin{equation}
                                \label{eq:ltt}
                                t_\text{corr} = t_\text{obs} - \frac{a}{c}\,\left[1-\cos\!\left(2\pi\frac{t_\text{obs}-T_0}{P}\right)\right]\,\sin\!\left(i\right),
                        \end{equation}
                        where $a$ is the semi-major axis of the planetary orbit, $c$ is the speed of light, $T_0$ is the time of inferior conjunction, and $P$ and $i$ are the orbital period and inclination, respectively. We note that Eq.~\ref{eq:ltt} assumes a circular orbit. We implemented a LTT correction for eccentric orbits but the computational time is much longer and we estimated the model difference to be smaller than $\pm0.15\,\text{ppm}$ (assuming the eccentricity value of 0.0091 from \citealt{Nymeyer2011_WASP-18b}). Therefore, we opted for the circular approximation of the LTT correction. Every time we estimated the model values (i.e. at each step of our parameter space exploration), we first computed the parameter-dependent LTT correction and then calculated our model at times $t_\text{corr}$. The amplitude of the correction for \planet{} is of the order of 20\,seconds. 
                        
                        Our model fits directly for the orbital period, $P$, the orbital inclination, $i$, and the normalised semi-major axis, $a/R_\star$, where $R_\star$ is the stellar radius. To compute the LTT correction, we included in our model a free parameter $R_\star$ with a normal prior based on the value derived from the IRFM method (see Section~\ref{sec:star} and Table~\ref{tab:star}). This new parameter allows us to compute the LTT, while accounting for uncertainties on the stellar radius.

                \subsubsection{Planetary flux} \label{sssec:planet_model}
                        
                        We modelled the flux coming from the planet as a combination of the phase curve of the planet and an occultation model when the planet is hidden by the star. The phase curve is computed by a sine function with a possible phase offset with respect to the orbital phase of the planet. This was implemented using the following equation {(details provided in Appendix~\ref{sec:pc_eq})}:
                        \begin{equation} \label{eq:pc}
                                F_\text{p} = \frac{1-\sin\!\left(\omega+\nu-\Delta\phi\right)\sin\!\left(i\right)}{2}\left(F_\text{max}-F_\text{min}\right)+F_\text{min},
                        \end{equation}
                        where $\omega$ is the argument of periastron\footnote{The argument of periastron $\omega$ is defined in such a way that the inferior conjunction (planetary transit) occurs at a true anomaly value of $\nu_\text{inf}=\pi/2-\omega$ (see Appendix~\ref{sec:pc_eq}).}, $\nu$ is the true anomaly, $\Delta\phi$ is the phase offset (e.g. hotspot offset), $i$ is the orbital inclination, and $F_\text{min}$ and $F_\text{max}$ are the minimum and maximum values of the phase curve as seen from $i=90\deg$, respectively. We note that when $\Delta\phi=0$, we have $F_\text{min}$ and $F_\text{max}$ being the nightside and dayside fluxes, respectively. Also, if $i\neq90\,\deg$, then the value of $F_\text{max}$ is greater than the observed occultation depth (even if $\Delta\phi=0$).
                        
                        We computed the planetary flux model by multiplying the phase-curve, $F_p$, by an occultation model, $F_\text{occ}$, from the \texttt{batman} package \citep{batman} that has been normalised in such a way that the out-of-occultation value is~1 and the in-occultation value is~0. The occultation model has up to seven free parameters: the mid-transit time, $T_0$, the orbital period, $P$, the planet-to-star radii ratio, $R_p/R_\star$, the normalised semi-major axis, $a/R_\star$, the orbital inclination, $i$, and the eccentricity, $e$, and the argument of periastron, $\omega$, implemented in the form: $\left\{e\cos\omega, e\sin\omega\right\}$.

                \subsubsection{Stellar flux} \label{sssec:star_model}
                        
                        Our model of the stellar flux only includes variability induced by the planet: the ellipsoidal variations, the Doppler boosting and the planetary transit. Other sources of variability, such as granulation or star spots, have not been detected in the data sets and were not modelled. Their small contribution (if any) is thus accounted for by the noise jitter, $\sigma_\text{w}$, described in Section~\ref{ssec:systematics}. The stellar flux can be expressed with the following equation:
                        \begin{equation}\label{eq:star_flux}
                                F_\star = \left(F_\text{EV}+F_\text{DB}\right) \times F_\text{tra},
                        \end{equation}
                        where $F_\text{EV}$ is the ellipsoidal variation, $F_\text{DB}$ is the Doppler boosting, and $F_\text{tra}$ is the flux dip induced by the planetary transit.
                        
\begin{figure}
                                \centering
                                \includegraphics[width=.95\hsize]{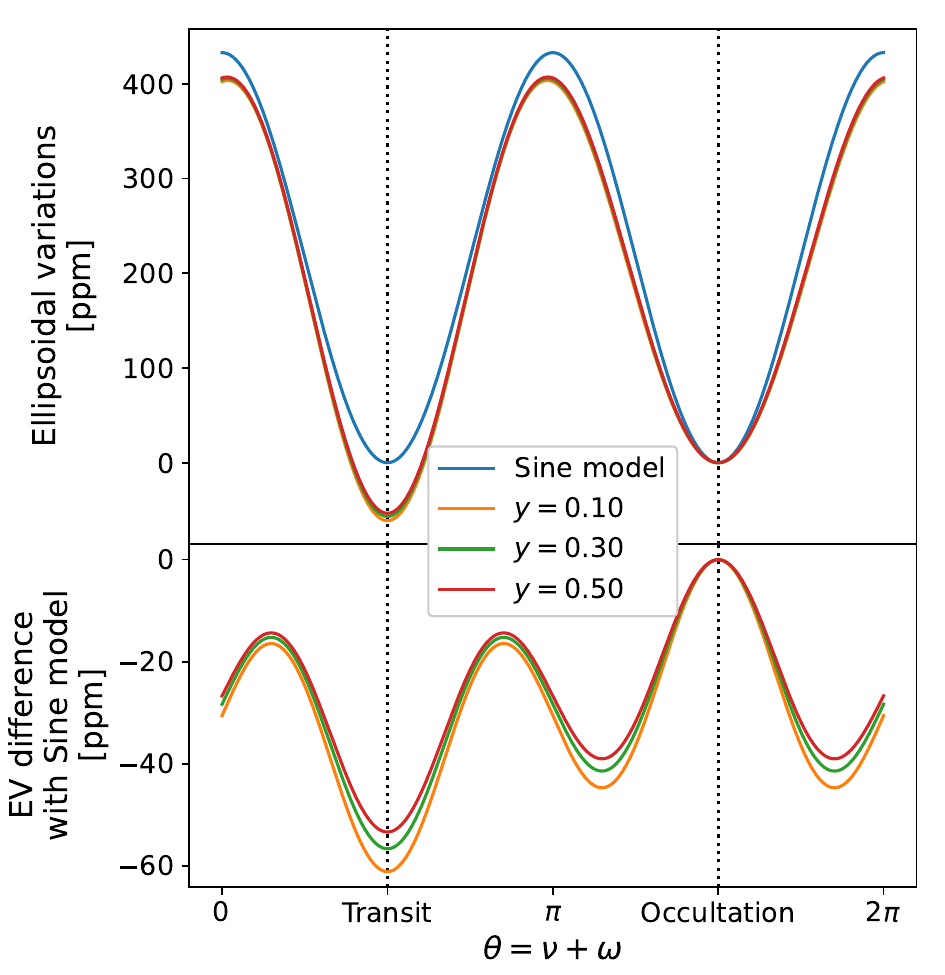}
                                \caption{
                                {Top.} Ellipsoidal variations modelled with and without secondary terms for the best-fit parameters in the \cheops{} passband. The x-axis represents the orbital phase of the planet $\theta=\nu+\omega$, where $\nu$ and $\omega$ are the true anomaly and the argument of periastron, respectively.
                                The blue curve ({Sine model}) is computed as a simple sine wave, i.e. with $A_1=A_3=0$. The other models include the secondary terms that are computed for several values of the gravity-darkening coefficient $y$.
                                {Bottom.} Differences between the different EV models and the {Sine model}. The peak-to-peak difference in amplitude is of the order of 50~ppm for the realistic range of GDC values ($0.1 \leq y \leq 0.5$).
                                The effect of accounting for $A_1$ and $A_3$ in the modelling of EV impacts mainly the difference in baseline level between transit and occultation. The baseline flux during transit will be lower than that during occultation (effect similar to a {negative} nightside flux). If not taken into account, this may lead to overestimating the occultation depth or underestimating the nightside flux from a planet.
                                }
                                \label{fig:cheops_ev}
                        \end{figure}

                        The ellipsoidal variations (EVs) are caused by the deformation of the stellar sphere due to the gravitational pull (tidal forces) of the planet. The bulge created on the stellar surface rotates together with the planet, in and out of view of the observer, which in turn induces photometric variability. This effect is well described in \cite{Esteves2013_PC} where they present a model derived from the work on binary stars of \cite{Morris1985_EV} (Eqs.~1-3), which itself is based on the proposed cosine series expansion of \cite{Kopal1959_Binaries} (Eq.~IV-2-37). {For consistency, Appendix~\ref{sec:ev_eq} details the steps to derive the EV model used in this work resulting in the following equations (Eqs.~\ref{eq:ev}~to~\ref{eq:ev_bev}).} Similarly to Eq.~8 of \cite{Esteves2013_PC}, we express the ellipsoidal variations of circular orbits using the first three dominating terms of the expansion:
                        \begin{equation}\label{eq:ev}
                                F_\text{EV} = A_\text{EV}\,\left[\cos\!\left(2\theta\right) - A_1\sin\!\left(\theta\right) + A_3\sin\!\left(3\theta\right) + A_0\right],
                        \end{equation}
                        where $\theta=\omega+\nu$ with $\omega$ the argument of periastron and $\nu$ the true anomaly. We chose to define the zero-flux level $F_\text{EV}=0$ at mid-occultation time (i.e. $\theta=3\pi/2$) by adding a constant offset $A_0=1-A_1-A_3$ to the EV model. Similarly to \cite{Esteves2013_PC}, we can write the coefficients $A_\text{EV}$, $A_1$, and $A_2$ for circular orbits as follows:
                        \begin{equation}\label{eq:ev_amp}
                                A_\text{EV} = \alpha_\text{EV}\frac{M_p}{M_\star}\left(\frac{R_\star}{a}\right)^3\sin^2\!\left(i\right),
                        \end{equation}
                        \begin{equation}\label{eq:ev_amp1}
                                A_1 = 3\,\beta_\text{EV}\,\frac{R_\star}{a}\,\frac{5\sin^2\!\left(i\right)-4}{\sin\!\left(i\right)},
                        \end{equation}
                        \begin{equation}\label{eq:ev_amp3}
                                A_3 = 5\,\beta_\text{EV}\,\frac{R_\star}{a}\,\sin\!\left(i\right),
                        \end{equation}
                        where $M_p$ and $M_\star$ are the masses of the planet and the star, respectively, $R_\star$ is the stellar radius, $a$ is the semi-major axis, and $i$ is the orbital inclination. The coefficients $\alpha_\text{EV}$ and $\beta_\text{EV}$ can be expressed as a function of limb-darkening and gravity-darkening coefficients (LDC and GDC, respectively). \cite{Esteves2013_PC} only uses the linear LDC, and we recalculated the expression to account for the quadratic term based on \cite{Kopal1959_Binaries} and using the limb-darkening law from \cite{Manduca1977_LD}\footnote{The quadratic limb-darkening law proposed by \cite{Manduca1977_LD} is $\mathcal{I}\!\left(\mu\right)=1-u_1\left(1-\mu\right)-u_2\left(1-\mu\right)^2$, where $\mu=\sqrt{1-x^2}$ with $x$ being the normalised radial coordinate on the stellar disc, $\mathcal{I}$ is the local intensity attenuation with respect the center of the stellar disc, and $u_1$ and $u_2$ are the limb-darkening coefficients.} implemented in \texttt{batman} for instance. We obtain the following expressions:
                        \begin{equation}\label{eq:ev_aev}
                                \alpha_\text{EV}=\frac{3}{10}\frac{15+u_1+2\,u_2}{6-2\,u_1-u_2}\left(1+y\right),
                        \end{equation}
                        \begin{equation}\label{eq:ev_bev}
                                \beta_\text{EV}=\frac{5}{168}\frac{35\,u_1+22\,u_2}{15+u_1+2\,u_2}\frac{2+y}{1+y},
                        \end{equation}
                        where $u_1$ and $u_2$ are the LDC, and $y$ is the GDC. Note that when $u_2=0$, we retrieve the equations~12 and~13 of \cite{Esteves2013_PC}. These expressions allow us to include the quadratic LDC values fitted to the transit light curve to compute the ellipsoidal variations. Given that $a/R_\star$ and $i$ can be derived from the transit/occultation models, the EV model only introduces two additional free parameters: the amplitude $A_\text{EV}$ and the GDC $y$ (via $\beta_\text{EV}$). We note that not accounting for the secondary terms in the EV model (i.e. $A_1=A_3=0$) results in a difference in EV amplitude of the order of 50\,ppm in the \cheops{} passband, which is detectable with the photometric precision (see Fig.~\ref{fig:cheops_ev}).
                        
                        We computed the gravity-darkening coefficient for each passband of our data sets. For \spitzer{} and \tess{}, we used directly the values of $y$ tabulated in \cite{Claret2011_LGDC_SPITZER} and \cite{Claret2017_LGDC_TESS}. Based on the stellar properties $T_\text{eff}$, $\log g$, $\left[\text{Fe}/\text{H}\right]$ listed in Table~\ref{tab:star}, we selected all $y$ values within $\pm4\sigma$ of the stellar parameters, and computed the mean and standard deviation. For \cheops{}, \cite{Claret2021_LGDC_CHEOPS} provides two values, $y_1$ and $y_2$, that relates to the GDC as $y=\beta_1\,y_1+ y_2$, where $\beta_1$ is the gravity-darkening exponent (GDE). From the work of \cite{Claret2004_beta1}, we extracted the values of the GDE using the stellar parameters and their uncertainties and computed $\beta_1=0.305\pm0.004$ for \host{}. Table~\ref{tab:gdc} lists the values of the gravity-darkening coefficients obtained for the different passbands.
                        
                        \begin{table}
                                \caption{List of the gravity-darkening coefficients (GDC) for \host{}.}
                                \label{tab:gdc}
                                \centering
                                \begin{tabular}{cc}
                                        \hline\hline
                                        Instrument & GDC \\
                                        passband & $y$ \\
                                        \hline
                                        \cheops{} & $0.296\pm0.009$ \\
                                        \tess{} & $0.230\pm0.013$ \\
                                        \spitzer{}/IRAC/Channel\,1 & $0.081\pm0.004$ \\
                                        \spitzer{}/IRAC/Channel\,2 & $0.093\pm0.009$ \\
                                        \spitzer{}/IRAC/Channel\,3 & $0.086\pm0.006$ \\
                                        \spitzer{}/IRAC/Channel\,4 & $0.078\pm0.004$ \\
                                        \hline\hline
                                \end{tabular}
                                \tablefoot{The GDC value of \cheops{} was validated against the value for the passband of \gaia{}, which is very similar to that of CHEOPS \citep{Deline2020_CHEOPS}. Using tabulated values of \cite{Claret2019_LGDC_Gaia}, we computed $y=0.282\pm0.016$, which is fully consistent with our \cheops{} value.
                                }
                        \end{table}
                        
                        We estimated that variations of the GDC of about 0.01 (order of magnitude of our uncertainties) will induce changes of about 0.2\,ppm in our EV model. Therefore, for our analysis, we neglected these error bars and fixed the values of the GDC to their mean values for all passbands.
                        
                        The Doppler boosting (DB) is a combination of two effects related to the motion of the star induced by the gravitational pull of the planet: the first contribution is the Doppler beaming that concentrates the stellar flux in the direction of motion; and the second contribution is the Doppler effect creating a passband-dependent variability due to the blue- and red-shift of the stellar light. The sources and modelling of DB are discussed in more details in the literature \citep[e.g.][]{Hills1974_EB, Rybicki1979_Radiative_processes, Loeb2003_Doppler, Zucker2007_Beaming, Bloemen2011_Beaming, Barclay2012_TrES-2, Esteves2013_PC}. Overall, the photometric effect of DB is proportional to the radial velocity of the star and can thus be modelled with the following expression:
                        \begin{equation}
                                F_\text{DB} = -A_\text{DB}\left[\cos\!\left(\omega+\nu\right)+e\cos\!\left(\omega\right)\right],
                        \end{equation}
                        where $A_\text{DB}$ is the amplitude of the Doppler boosting, $\omega$ is the argument of periastron, $\nu$ is the true anomaly, and $e$ is the eccentricity.
                        
                        Finally, the stellar flux variability due to ellipsoidal variations and Doppler boosting is further modulated by the partial flux loss when the planet transits its host. We thus multiply the sum of the two effects by a transit model (see Eq.~\ref{eq:star_flux}) also from the \texttt{batman} package \citep{batman} sharing the same parameters as the occultation model (Section~\ref{sssec:planet_model}), but also including two quadratic limb-darkening parameters, $u_1$ and $u_2$.
                        
                        The final model including all astrophysical signals from the planetary system is therefore:
                        \begin{equation} \label{eq:final_model}
                                F_\text{tot} = F_p \times F_\text{occ} + \left(F_\text{EV} + F_\text{DB}\right) \times F_\text{tra}.
                        \end{equation}

        \subsection{Parameter space exploration and joint fit} \label{ssec:fit}
                
                The total number of parameters of our model for all passbands can be up to 187, among which 137 are nuisance parameters for noise modelling (84 for \cheops{}, 20 for \tess{}, and 33 for \spitzer{}). The 50 remaining parameters define the planetary model described in Section~\ref{ssec:astro_model}, with eight related to the planetary orbit ($T_0$, $P$, $R_p/R_\star$, $a/R_\star$, $i$, $e\cos\omega$, $e\sin\omega$, $R_\star$) and seven for each of the six passbands ($F_\text{max}$, $F_\text{min}$, $\Delta\phi$, $A_\text{EV}$, $A_\text{DB}$, $u_1$, $u_2$). We did not fit for the passband-dependent variations of $R_p/R_\star$ as the only strong constraints on this parameter came from the transits of \cheops{} and \tess{}, which have similar passbands hence similar expected planetary radii (negligible difference due to Rayleigh scattering). Moreover, \spitzer{} occultations would only poorly constrain $R_p/R_\star$ from the durations of the occultations.
                
                Given \spitzer{} data only covers planetary occultations, we fixed the phase-curve and transit parameters ($F_\text{min}$, $A_\text{EV}$, $A_\text{DB}$, $u_1$, $u_2$) to~0 as they could not be constrained by the light curves. In addition, we fitted the data with all parameters free and we obtained values for the minimum PC level $F_\text{min}$ (\cheops{} and \tess{}) and the phase offset $\Delta\phi$ consistent with zero (less than $3\,\sigma$ significance). These parameters were thus fixed to~0 for the final analysis. To further reduce the size of the parameter space and considerably improve the computational time, we assumed the orbit of \planet{} to be circular ($e\cos\omega=e\sin\omega=0$). This choice is {motivated by the small value of the} eccentricity ($e\sim0.009$) reported in previous works \citep{Hellier2009_W18b, Triaud2010_HJs, Nymeyer2011_WASP-18b, Csizmadia2019_WASP-18b}. {Furthermore, a small eccentricity with the peculiar orbital orientation of \planet{} ($\omega\sim270\deg$) could} be explained by radial-velocity signals induced by the bulge of the tidally deformed star \citep{Arras2012_tidal_RV}\footnote{{\cite{Arras2012_tidal_RV} estimate that tides in the \host{} system have a radial-velocity effect of $\sim$\,32\,m/s whereas the eccentric effect is $\sim$\,15\,m/s (see $K_\text{tide}$ and $eK_\text{orb}$ in their Table~1). However, \cite{Bernabo2024_WASP-19} show that neglecting stellar rotation, as done in \cite{Arras2012_tidal_RV}, leads to overestimating the tidal effect by a factor $P_\text{rot}/P_\text{orb}$ (see their Eq.~10). \cite{Csizmadia2019_WASP-18b} estimates for \planet{} that $P_\text{rot}/P_\text{orb}\sim5.8$, which means the tides effect on RV is of the order of 5.5\,m/s. As this value is still comparable to the eccentric effect of 15\,m/s, the tidal origin of the eccentric RV~signal remains possible.}}. In total, we fixed 30~parameters, reducing the total number of free parameters to~157.
                
                We explored the large parameter space with the Nested sampling method \citep{Skilling2004_NS, Skilling2006_NS} implemented in the package \texttt{dynesty} \citep{Speagle2020_dynesty, Koposov2023_dynesty}. We used a Dynamic Nested Sampler \citep{Higson2019_DNS} with bounding option using multiple ellipsoid bounds \citep{Feroz2009_multinest} and a random slice sampling \citep{Neal2003_slice_sampling, Handley2015_polychord, Handley2015_polychord2}.
                We started the run with 2000 live points, which corresponds to more than 10~times the number of dimensions and is good compromise between computational time and good convergence. The weight and stopping functions of the dynamic sampling were set to the default values, i.e. a relative fractional importance of 80\% on the posterior for the weight, and a stopping criterion of $d\log\mathcal{Z}\leq0.01$ (with $\mathcal{Z}$ being the estimated Bayesian evidence). The run ended with a total number of 333\,506 points sampling the parameter space.

\section{Refined system properties} \label{sec:results}
        
        \subsection{Planetary parameters} \label{ssec:param}
                
                The final values of the planetary parameters obtained from our analysis are listed in Table~\ref{tab:parameters}. The values of the nuisance parameters fitted for each instrument are listed in appendix (Tables~\ref{tab:cheops_syst}, \ref{tab:tess_syst} and \ref{tab:spitzer_syst} for \cheops{}, \tess{} and \spitzer{}, respectively).
                
\begin{figure*}
                        \centering
                        \includegraphics[width=.95\hsize,trim={1.2cm 1.2cm 10cm 6cm},clip]{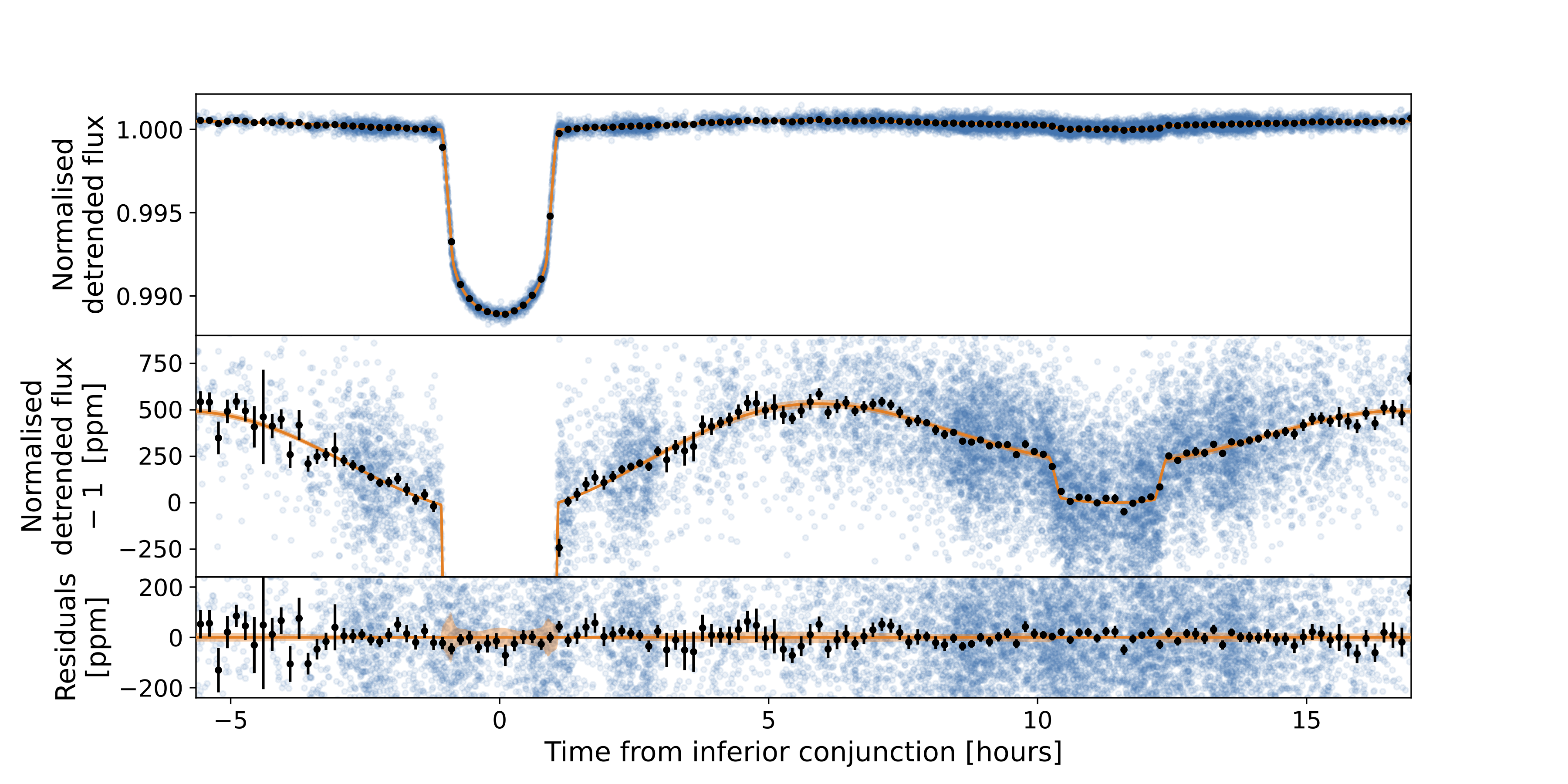}
                        \caption{\cheops{} phase-folded phase curve. Data points are shown in blue, and their 10-min binned counterparts are shown in black with error bars. The mean and $3\,\sigma$ uncertainty of the model are shown in orange (line and shaded area, respectively) and have been computed from a set of 2000~models randomly drawn from the posterior distribution. From top to bottom are show the full phase curve, a zoomed-in version to highlight the phase-curve signal, and the residuals in ppm.}
                        \label{fig:cheops_lc}
                \end{figure*}

\begin{figure*}
                        \centering
                        \includegraphics[width=.95\hsize,trim={1.2cm 1.2cm 10cm 6cm},clip]{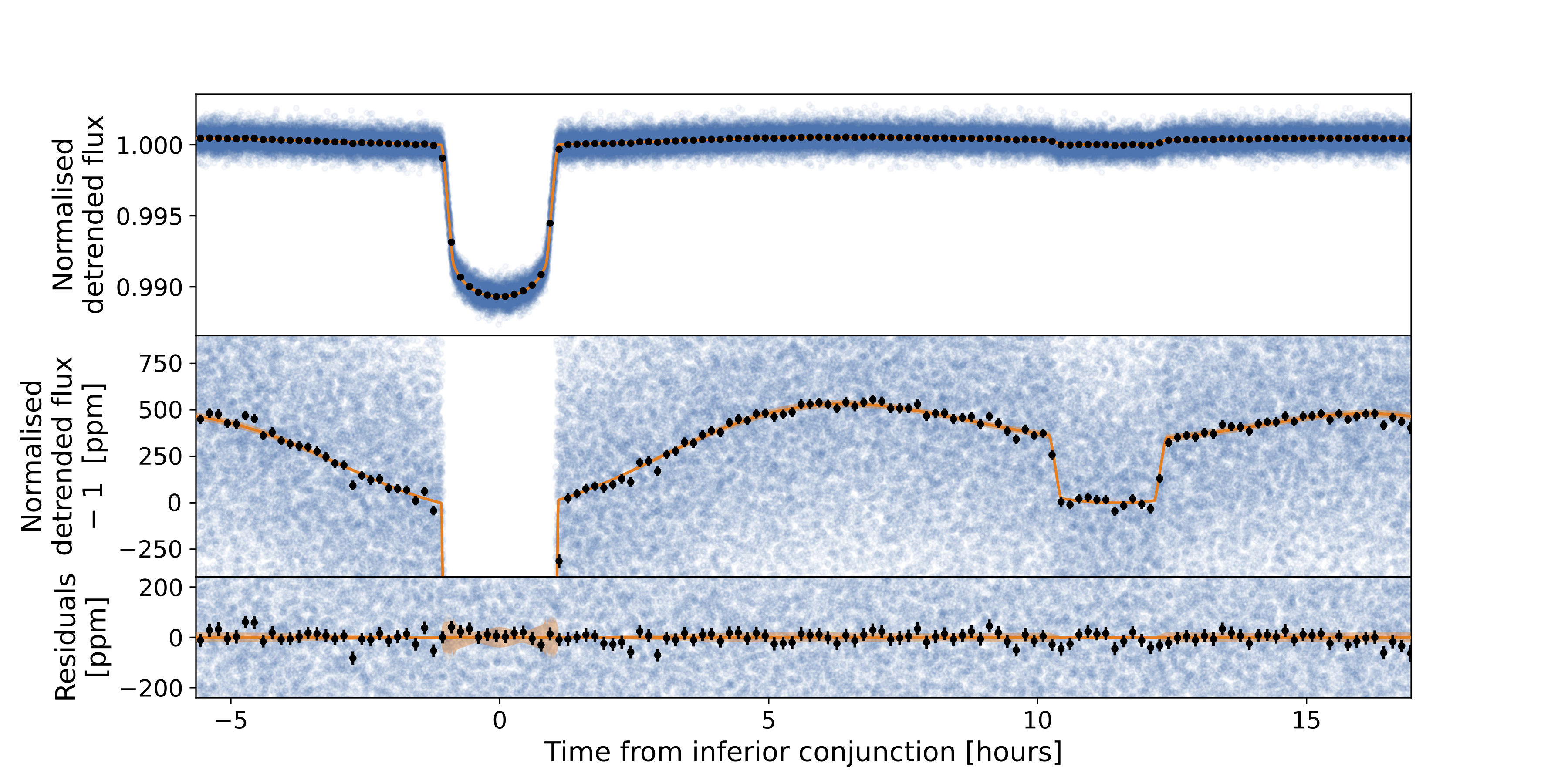}
                        \caption{\tess{} phase-folded phase curve. Data points are shown in blue, and their 10-min binned counterparts are shown in black with error bars. The mean and $3\,\sigma$ uncertainty of the model are shown in orange (line and shaded area, respectively) and have been computed from a set of 2000~models randomly drawn from the posterior distribution. From top to bottom are show the full phase curve, a zoomed-in version to highlight the phase-curve signal, and the residuals in ppm.}
                        \label{fig:tess_lc}
                \end{figure*}

\begin{figure*}
                        \centering
                        \includegraphics[width=.95\hsize,trim={0cm 0cm 0cm 0cm},clip]{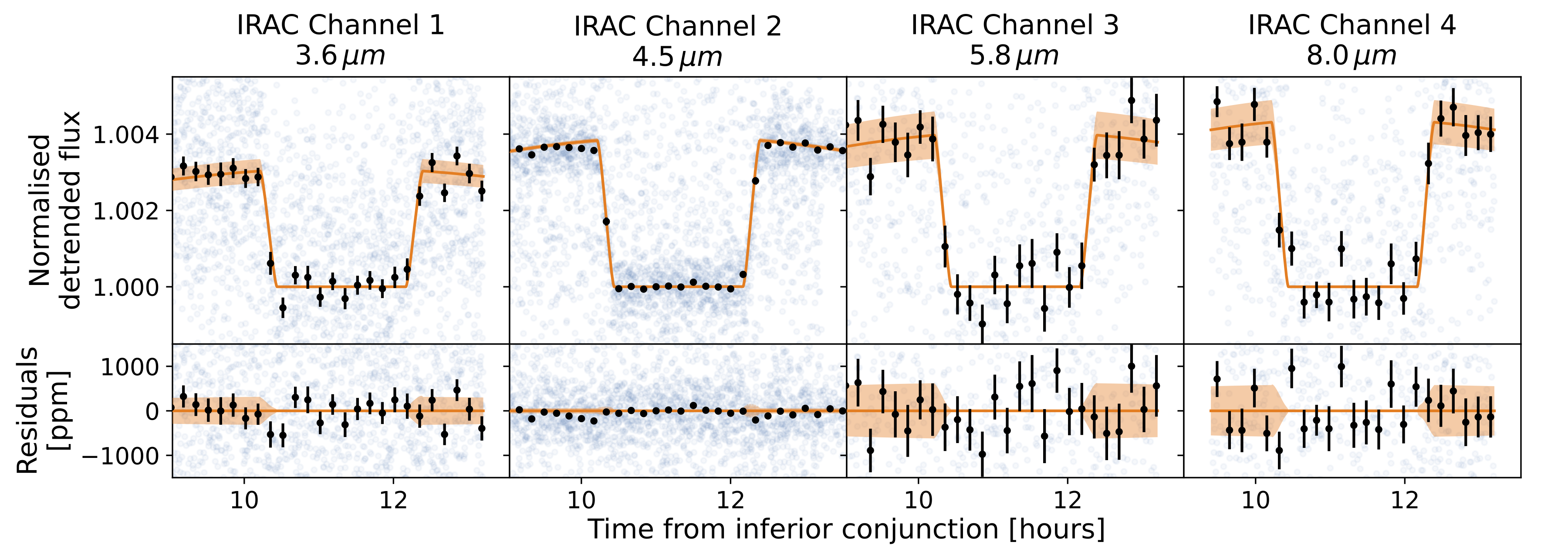}
                        \caption{\spitzer{} phase-folded occultations. Data points are shown in blue, and their 10-min binned counterparts are shown in black with error bars. The mean and $3\,\sigma$ uncertainty of the model are shown in orange (line and shaded area, respectively) and have been computed from a set of 2000~models randomly drawn from the posterior distribution. The 4 IRAC channels 1, 2, 3, 4, centred at 3.6\,\textmu m, 4.5\,\textmu m, 5.8\,\textmu m, 8.0\,\textmu m, respectively, are shown at the top, with their corresponding residuals in ppm at the bottom.}
                        \label{fig:spitzer_lc}
                \end{figure*}

                The final detrended light curves are presented in Figs.~\ref{fig:cheops_lc}, \ref{fig:tess_lc}, and~\ref{fig:spitzer_lc}, for the 36~\cheops{} visits, the 5~\tess{} sectors, and the 14~\spitzer{} visits, respectively.
                
                \begin{table*}
                        \caption{Parameter values for the \planet{} system.}
                        \label{tab:parameters}
                        \centering
                        {
                                \begin{tabular}{llccr}
                                        \hline\hline
                                        Fitted parameter & Symbol & Value & Prior & Unit \\
                                        \hline
                                        Time of inferior conjunction & $T_0$ & ${375.169847}\pm{0.000019}$ & $\mathcal{U}\!\left(375.1, 375.2\right)$ & $\text{BJD}_\text{TDB} - 2\,458\,000$ \\
                                        Orbital period & P & ${0.941452379}\pm{0.000000016}$ & -- & days \\
                                        Planet-to-star radii ratio & $R_p/R_\star$ & ${0.09757}_{-{0.00013}}^{+{0.00012}}$ & -- & -- \\
                                        Normalised semi-major axis & $a/R_\star$ & ${3.493}\pm{0.011}$ & -- & -- \\
                                        Orbital inclination & $i$ & ${84.08}\pm{0.17}$ & $\mathcal{U}\!\left(0, 90\right)$ & deg \\
                                        \multirow{2}{*}{Eccentricity / argument of periastron} & $e\,\cos\omega$ & 0\tablefootmark{$\ast$} & fixed & -- \\
                                        & $e\,\sin\omega$ & 0\tablefootmark{$\ast$} & fixed & -- \\
                                        Stellar radius & $R_\star$ & ${1.2561}\pm{0.0079}$ & $\mathcal{N}\!\left(1.256, 0.008\right)$ & $R_\odot$ \\
                                        \hline
                                        \cheops{}\\
                                        \qquad Dayside flux\tablefootmark{$\dagger$} & $F_\text{max}^\text{\cheops{}}$ & ${211.9}_{-{7.5}}^{+{7.6}}$ & $\mathcal{U}\!\left(0, +\infty\right)$ & ppm \\
                                        \qquad EV amplitude & $A_\text{EV}^\text{\cheops{}}$ & ${217.0}\pm{5.0}$ & $\mathcal{U}\!\left(0, +\infty\right)$ & ppm \\
                                        \qquad DB amplitude & $A_\text{DB}^\text{\cheops{}}$ & ${18.8}\pm{4.3}$ & $\mathcal{U}\!\left(0, +\infty\right)$ & ppm \\
                                        \multirow{2}{*}{\qquad LDC} & $u_1^\text{\cheops{}}$ & ${0.357}\pm{0.018}$ & -- & -- \\
                                        & $u_2^\text{\cheops{}}$ & ${0.229}_{-{0.028}}^{+{0.029}}$ & -- & -- \\
                                        \hline
                                        \tess{}\\
                                        \qquad Dayside flux\tablefootmark{$\dagger$} & $F_\text{max}^\text{\tess{}}$ & ${340.4}\pm{7.0}$ & $\mathcal{U}\!\left(0, +\infty\right)$ & ppm \\
                                        \qquad EV amplitude & $A_\text{EV}^\text{\tess{}}$ & ${170.8}_{-{3.5}}^{+{3.4}}$ & $\mathcal{U}\!\left(0, +\infty\right)$ & ppm \\
                                        \qquad DB amplitude & $A_\text{DB}^\text{\tess{}}$ & ${26.2}_{-{3.0}}^{+{3.1}}$ & $\mathcal{U}\!\left(0, +\infty\right)$ & ppm \\
                                        \multirow{2}{*}{\qquad LDC} & $u_1^\text{\tess{}}$ & ${0.293}\pm{0.018}$ & -- & -- \\
                                        & $u_2^\text{\tess{}}$ & ${0.168}\pm{0.031}$ & -- & -- \\
                                        \hline
                                        \spitzer{}/IRAC \\
                                        \qquad Channel 1 dayside flux\tablefootmark{$\dagger$} & $F_\text{max}^\text{\spitzer{}\,1}$ & ${3106}_{-{111}}^{+{112}}$ & $\mathcal{U}\!\left(0, +\infty\right)$ & ppm \\
                                        \qquad Channel 2 dayside flux\tablefootmark{$\dagger$} & $F_\text{max}^\text{\spitzer{}\,2}$ & ${3935}\pm{25}$ & $\mathcal{U}\!\left(0, +\infty\right)$ & ppm \\
                                        \qquad Channel 3 dayside flux\tablefootmark{$\dagger$} & $F_\text{max}^\text{\spitzer{}\,3}$ & ${4090}\pm{230}$ & $\mathcal{U}\!\left(0, +\infty\right)$ & ppm \\
                                        \qquad Channel 4 dayside flux\tablefootmark{$\dagger$} & $F_\text{max}^\text{\spitzer{}\,4}$ & ${4360}_{-{210}}^{+{200}}$ & $\mathcal{U}\!\left(0, +\infty\right)$ & ppm \\
                                        \hline\hline
                                        Derived parameters \\
                                        \hline
                                        Optimal time of inferior conjunction & $T_{0,\,\text{opt}}$ & $1251.662013\pm{0.000011}$ & & $\text{BJD}_\text{TDB} - 2\,458\,000$ \\
                                        Planetary radius & $R_p$ & ${1.1926}\pm{0.0077}$ & & $R_J$ \\
                                        Semi-major axis & $a$ & ${0.02041}_{-{0.00014}}^{+{0.00015}}$ & & \au{} \\
                                        Impact parameter & $b$ & ${0.3605}_{-{0.0095}}^{+{0.0091}}$ & & $R_\star$ \\
                                        Eccentricity & $e$ & 0\tablefootmark{$\ast$} & & -- \\
                                        \cheops{} occultation depth & $\delta_\text{occ}^\text{\cheops{}}$ & ${211.3}\pm{7.5}$ & & ppm \\
                                        \tess{} occultation depth & $\delta_\text{occ}^\text{\tess{}}$ & ${339.5}_{-{7.0}}^{+{6.9}}$ & & ppm \\
                                        \spitzer{}/IRAC\,Ch.\,1 occ. depth & $\delta_\text{occ}^\text{\spitzer{}\,1}$ & ${3098}\pm{111}$ & & ppm \\
                                        \spitzer{}/IRAC\,Ch.\,2 occ. depth & $\delta_\text{occ}^\text{\spitzer{}\,2}$ & ${3925}\pm{25}$ & & ppm \\
                                        \spitzer{}/IRAC\,Ch.\,3 occ. depth & $\delta_\text{occ}^\text{\spitzer{}\,3}$ & ${4080}\pm{230}$ & & ppm \\
                                        \spitzer{}/IRAC\,Ch.\,4 occ. depth & $\delta_\text{occ}^\text{\spitzer{}\,4}$ & ${4350}_{-{210}}^{+{200}}$ & & ppm \\
                                        Transit duration & $T_{14}$ & ${2.1790}\pm{0.0017}$ & & hours \\
                                        Max. equilibrium dayside temperature & $T_\text{day,\,max}$ & ${3061}\pm{29}$\tablefootmark{$\ddagger$} & & K \\
                                        \hline\hline
                                \end{tabular}
                                }
                                \tablefoot{
                                The top part of the table lists the fitted parameters with their corresponding prior probabilities. Uniform prior probabilities are represented with $\mathcal{U}\!\left(x_\text{min}, x_\text{max}\right)$, where $x_\text{min}$ and $x_\text{max}$ are the minimum and maximum allowed values, respectively. Normal (Gaussian) prior probabilities are written as $\mathcal{N}\!\left(\mu, \sigma\right)$, where $\mu$ and $\sigma$ are the mean and standard deviation of the normal distribution, respectively. The lower part of the table shows the values of parameters derived from the sampled parameter space.
                                \tablefoottext{$\ast$}{We fixed the eccentricity to~0 despite the non-zero values reported in the literature \citep{Hellier2009_W18b, Triaud2010_HJs, Nymeyer2011_WASP-18b, Csizmadia2019_WASP-18b} because {of the small eccentricity value and the possible tidal origin of} the eccentric radial-velocity signal {(see discussion in Section~\ref{ssec:fit})}.}
                                \tablefoottext{$\dagger$}{The maximum values of the phase curve $F_\text{max}$ are equal to the dayside fluxes because the phase offset $\Delta\phi$ was fixed to 0.}
                                \tablefoottext{$\ddagger$}{Equilibrium dayside temperature assuming full absorption of the incoming radiation ($A_B=0$), no heat redistribution ($\varepsilon=0$, i.e. immediate re-radiation) and black-body spectral energy distributions for both the planet and the star, i.e. $T_\text{day,\,max}=\left(2/3\right)^{0.25}\!\sqrt{R_\star/a}\; T_\text{eff}$.}
                                }
                        \end{table*}
                
                We retrieve a time of inferior conjunction $T_0$ consistent at $1.15\,\sigma$ with the value of \cite{Shporer2019_W18b} with a precision improved from 2.2\,s to 1.6\,s. The uncertainty on $T_0$ reaches a minimum of less than 1\,s for the date $T_{0,\,\text{opt}}=2\,459\,251.66201\,\text{BJD}_\text{TDB}$ (i.e. on February 6, 2021).
                The precision of 1.4\,ms on the orbital period $P$ is also a significant improvement with respect to previously published results (138\,ms for \citealt{Shporer2019_W18b}, 21\,ms for \citealt{Cortes-Zuleta2020_TraMos}, or 6\,ms for \citealt{Coulombe2023}). The values of $P$ are all consistent within $1\,\sigma$.
                The normalised semi-major axis $a/R_\star$ and the orbital inclination $i$ are both consistent with the values from the literature, with the largest differences from \cite{Shporer2019_W18b} of $2.68\,\sigma$ and $2.15\,\sigma$, respectively. This might be explained by the fact that the analysis of \cite{Shporer2019_W18b} fitted the data with fixed limb-darkening coefficients and a fixed non-zero eccentricity.
                
                The planet-to-star radii ratio, $R_p/R_\star$, that we derived from our global fit is consistent with the values from \cite{Shporer2019_W18b} ($2.24\,\sigma$) and \cite{Coulombe2023} ($<1\,\sigma$). It is, however, significantly smaller than the value from \cite{Cortes-Zuleta2020_TraMos}. We refined the precision down to $0.13\,\%$ on $R_p/R_\star$, which converts into $0.65\,\%$, or $550\,\text{km}$, on the absolute planetary radius.
                
                In Section~\ref{ssec:fit}, we explained that the phase offset, $\Delta\phi$, and the minimum planetary flux, $F_\text{min}$, were fixed to zero as their values were always consistent with zero when let free. The difficulty to constrain $\Delta\phi$ mainly came from the strong degeneracy between a hotspot offset and the effect of Doppler boosting (strong DB can be counterbalanced by having large westward hotspot offset), leading to unrealistically large values for both parameters. We note that fixing $\Delta\phi$ to 0 makes the values of $F_\text{max}$ and $F_\text{min}$ equivalent to the dayside flux (or occultation depth) and nightside flux, respectively. We were able to fit for $F_\text{min}$ in both the \cheops{} and \tess{} passbands, as we analysed the full phase curves and we derived the following $3\,\sigma$ upper limits: $F_\text{min}^\text{\cheops{}}<12.3\,\text{ppm}$ and $F_\text{min}^\text{\tess{}}<18.4\,\text{ppm}$.

                When comparing the values of $F_\text{max}$, $A_\text{EV}$ and $A_\text{DB}$ for the TESS passband to previously published values, we obtain consistency within $\sim1\,\sigma$ for all parameters of \cite{Shporer2019_W18b} and \cite{Coulombe2023}, except for the EV of \cite{Shporer2019_W18b} that is marginally consistent at $2.89\,\sigma$. The precision on all three parameters is significantly improved. We note that the non-detection of the nightside flux is in agreement with the analysis of \cite{Coulombe2023}, and we lowered the $3\,\sigma$ upper limit by a factor of~2.4 (from 44\,ppm to 18.4\,ppm).
                
                Following Equations~\ref{eq:ev_amp} and~\ref{eq:ev_aev}, we could convert the detected EV amplitudes $A_\text{EV}$ in both \cheops{} and \tess{} passbands to planetary masses. We obtained $M_p^\text{CHEOPS}={10.03}_{-{0.52}}^{+{0.53}}\,M_J$ and $M_p^\text{TESS}={8.73}\pm{0.45}\,M_J$, which are consistent with each other ($1.89\,\sigma$) and with the literature. Despite their mutual consistency, the mass difference between the two passbands is probably due to inaccuracies of the GDC estimates (Table~\ref{tab:gdc}) between reality and theoretical modelling.

                Finally, the dayside flux values, $F_\text{max}$, that we derived for the four \spitzer{}/IRAC channels are all consistent ($<1.1\,\sigma$) with the literature \citep{Nymeyer2011_WASP-18b, Maxted2013_WASP-18b,2017ApJ...850L..32S,2020AJ....159..137G}.

        \subsection{Orbital eccentricity and argument of periastron} \label{ssec:ecc_jwst}
        
        We considered the planetary orbit to be circular for our fit given the small eccentricity reported in the literature and the fact that it may actually be explained by a spurious signal from the bulge of the tidally deformed star {(see discussion in Section~\ref{ssec:fit})}.
        However, \cite{Coulombe2023} recently observed the occultation of \planet{} with JWST and derived a mid-occultation time of $T_\text{occ}^\text{JWST}=2459802.881867\pm0.000092\,\text{BJD}_\text{TDB}$ that is inconsistent with our results. Indeed, the parameter set we compute gives an occultation time of $T_\text{occ}=2459802.8826151\pm0.000015\,\text{BJD}_\text{TDB}$ after including the LTT of $20.26\pm0.15\,\text{sec}$. This means that there is a significant timing inconsistency of $\Delta T_\text{occ}=-64.6\pm8.1\,\text{sec}$ ($7.98\,\sigma$).
        
        We first discarded the possibility of orbital period variation as the JWST observation occurred in-between the datasets considered in our analysis (nine \cheops{} visits within 2.5~months and \tess{} sector~69 fewer than 400~days after). We {then} considered the possibility of having an effect due to the thermal structure of the atmosphere and the fact that different passbands may probe different dayside brightness distribution. As shown in \cite{2012A&A...548A.128D}, this might result in mid-occultation time shifts with respect to expectations from the transit. However, even if we could not reach the required precision in the \tess{} data, we found a hint of a similar shift ($\Delta T_\text{occ}^\text{\tess{}}=-48\pm32\,\text{sec}$). In addition, the detection of orbital eccentricity reported from radial velocity \citep{Hellier2009_W18b, Triaud2010_HJs, Nymeyer2011_WASP-18b, Csizmadia2019_WASP-18b} {made us consider} the eccentric nature of the planetary orbit {as} the most probable source {of} the timing inconsistency.

\begin{figure}
                \centering
                \includegraphics[width=.95\hsize,trim={0.1cm 0.1cm 0.7cm 0.2cm},clip]{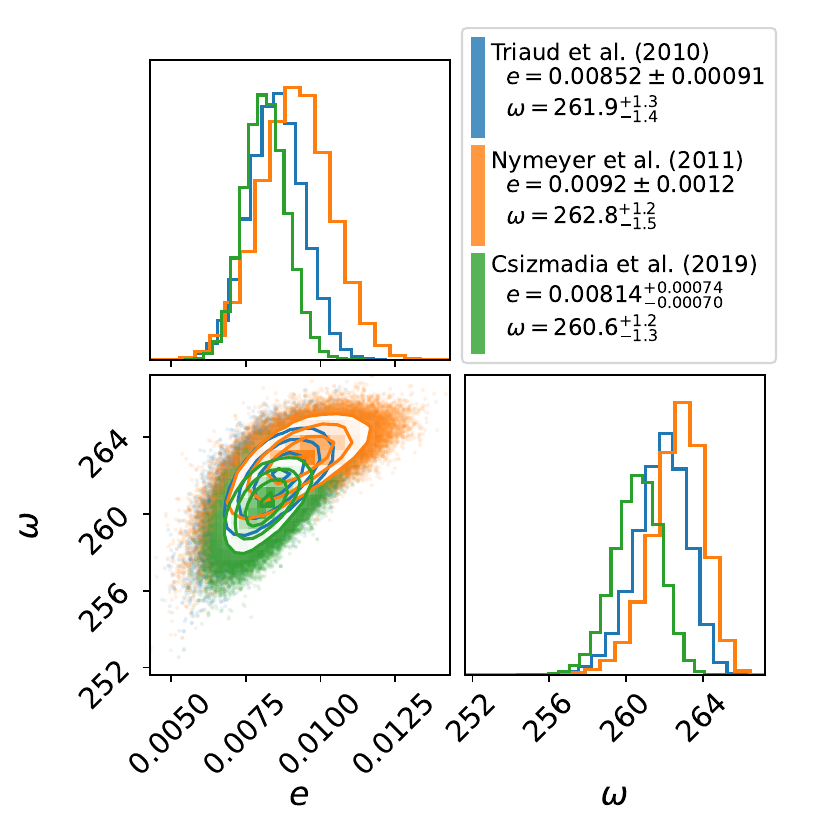}
                \caption{Correlation plot of the posterior distribution of the eccentricity $e$ and argument of periastron $\omega$ of the orbit of \planet{} that match the mid-occultation time $T_\text{occ}^\text{JWST}$ from \cite{Coulombe2023}. We used our posterior distribution on planetary parameters from Table~\ref{tab:parameters}, and Gaussian priors on $e\,\cos\omega$ and $e\,\sin\omega$ {from \cite{Triaud2010_HJs} in blue, \cite{Nymeyer2011_WASP-18b} in orange, and \cite{Csizmadia2019_WASP-18b} in green. Overall, the 3 posterior distributions are all consistent with each other within $<1.2\,\sigma$.}
                This plot made use of the package \texttt{corner.py} \citep{corner}.}
                \label{fig:e-w}
        \end{figure}

        We converted $\Delta T_\text{occ}$ to orbital phase and obtained an offset of $0.00079\pm0.00010$ corresponding to an occultation phase of $0.49921\pm0.00010$. Due to the high computational cost of running a global fit with an eccentric orbit, we derived constraints on the eccentricity $e$ and the argument of periastron $\omega$ indirectly. We {considered 3 different sets of prior values on $e\,\cos\omega$ and $e\,\sin\omega$ from the works of \cite{Triaud2010_HJs}, \cite{Nymeyer2011_WASP-18b} and \cite{Csizmadia2019_WASP-18b} that we combined with the likelihood probability of these 2 parameters to match the mid-occultation time $T_\text{occ}^\text{JWST}$ reported in \cite{Coulombe2023}}. We explored the 2D-parameter space $\left\{e\,\cos\omega, e\,\sin\omega\right\}$ with the Markov chain Monte Carlo (MCMC) algorithm \texttt{emcee} \citep{emcee, emcee_v3} implemented in Python. The results from our 3 different approaches (3 different priors) are fully consistent ($<1.2\,\sigma$) with each other (see posterior distributions in Fig.~\ref{fig:e-w}). Table~\ref{tab:e-w} shows the published values of $e$ and $\omega$ (used as priors in the form $e\,\cos\omega$ and $e\,\sin\omega$) and the results we obtained with the mid-occultation time measured with \jwst{}. Eccentricity values are largely consistent ($<0.2\,\sigma$) with the published values with a significant improvement in precision for the value of \cite{Csizmadia2019_WASP-18b}. The same is true for the argument of periastron but with a weaker consistency ($\leq1.8\,\sigma$). We also note that all posterior distributions of $\omega$ values converge towards a similar angle of $\sim$\,262\,$\deg$ despite the $3.2\,\sigma$ inconsistency between the published values of \cite{Nymeyer2011_WASP-18b} and \cite{Csizmadia2019_WASP-18b}. This highlights the strong constraint provided by the timing inconsistency with $T_\text{occ}^\text{JWST}$, which translates into unprecedented precision on the value of the argument of periastron (less than $\pm1.5\,\deg$). As the posterior distribution derived from the published values of \cite{Triaud2010_HJs} is the most conservative among our 3 approaches (see blue data in Fig.~\ref{fig:e-w}), we selected it for our reference orbital parameters.
        
        \begin{table}
        {
        \caption{Values of the eccentricity~$e$ and argument of periastron~$\omega$.}
        \label{tab:e-w}
        \centering
        \resizebox{\columnwidth}{!}{
            \begin{tabular}{lccc}
                \hline\hline
                \multirow{2}{*}{Publication} & & Published values & Posterior values \\
                & & (used as priors) & matching $T_\text{occ}^\text{JWST}$ \\
                \hline
                \multirow{2}{*}{\cite{Triaud2010_HJs}} & $e$ & $0.00848^{+0.00085}_{-0.00095}$ & $0.00852 \pm 0.00091$ \\
                & $\omega$ & $267.9^{+4.9}_{-4.3}$ & $261.9^{+1.3}_{-1.4}$ \\
                \multirow{2}{*}{\cite{Nymeyer2011_WASP-18b}} & $e$ & $0.0091 \pm 0.0012$ & $0.0092 \pm 0.0012$ \\
                & $\omega$ & $269 \pm 3$ & $262.8^{+1.2}_{-1.5}$ \\
                \multirow{2}{*}{\cite{Csizmadia2019_WASP-18b}} & $e$ & $0.0085 \pm 0.0020$ & $0.00814^{+0.00074}_{-0.00070}$ \\
                & $\omega$ & $257.3 \pm 2.1$ & $260.6^{+1.2}_{-1.3}$ \\
                \hline\hline
            \end{tabular}
        }
        \tablefoot{
            Values of the argument of periastron $\omega$ are reported in degrees.
            Published $e$ and $\omega$ values were used as priors in the form of $e\,\cos\omega$ and $e\,\sin\omega$.
            The argument of periastron is defined in such a way that the inferior conjunction occurs at $\omega+\nu=90\,\deg$, where $\nu$ is the true anomaly (see Eq.\ref{eq:pc} and Appendix~\ref{sec:pc_eq}).
        }
        }
    \end{table}

        Following the confirmation of the eccentric nature of the orbit of \planet{}, we computed the circularisation timescale $\tau_\text{circ}$ of the system using several formulae from the literature (Eq.~1 fo \citealt{2011A&A...525A..54B}; Eqs.~2 or~3 of \citealt{2006ApJ...649.1004A}). We found a 3\,$\sigma$ upper limit of $\tau_\text{circ} \lesssim 20\,\text{Myr}$, which is much shorter than the age of the system $t_\star=1.4\pm0.7\,\text{Gyr}$ (see Table~\ref{tab:star}). The planet should thus have had its orbit circularised and explaining what this is not the case is challenging. A possible explanation could be that an outer companion is slowing down the damping of the eccentricity of \planet{} by injecting energy into the planetary orbit inducing \citep[eccentricity excitation;][]{2007MNRAS.382.1768M,2012A&A...538A.105L}. \cite{2019AJ....158..243P} reported a candidate outer planet with an orbital period of 2.16\,days and a mass of $\sim$\,0.2\,M$_J$, but there is no firm detection of this object. Another perturbing body that could cause eccentricity excitation is the stellar companion \host{}\,B that is a late-M dwarf at a distance of about 3500\,\au{} \citep{Csizmadia2019_WASP-18b}. Further observations and analyses of the radial velocity and transit-time variations of the system might help understand the dynamical mechanisms at play and constrain the properties of the potential perturbing body.

        \subsection{Occultation depth variability} \label{ssec:odv}
                
                We checked for variability of the occultation depth in each instrument passband by fixing all parameters to their best-fit values from our global fit (see Table~\ref{tab:parameters}), letting free only the occultation depth (via the dayside flux $F_\text{max}$), the noise jitter, $\sigma_w$, and the normalisation flux, $f_0$. The 3D~parameter space was explored with the MCMC algorithm \texttt{emcee} \citep{emcee, emcee_v3}.
                
\begin{figure}
                \centering
                \includegraphics[width=\hsize,trim={0cm 0cm 0cm 0cm},clip]{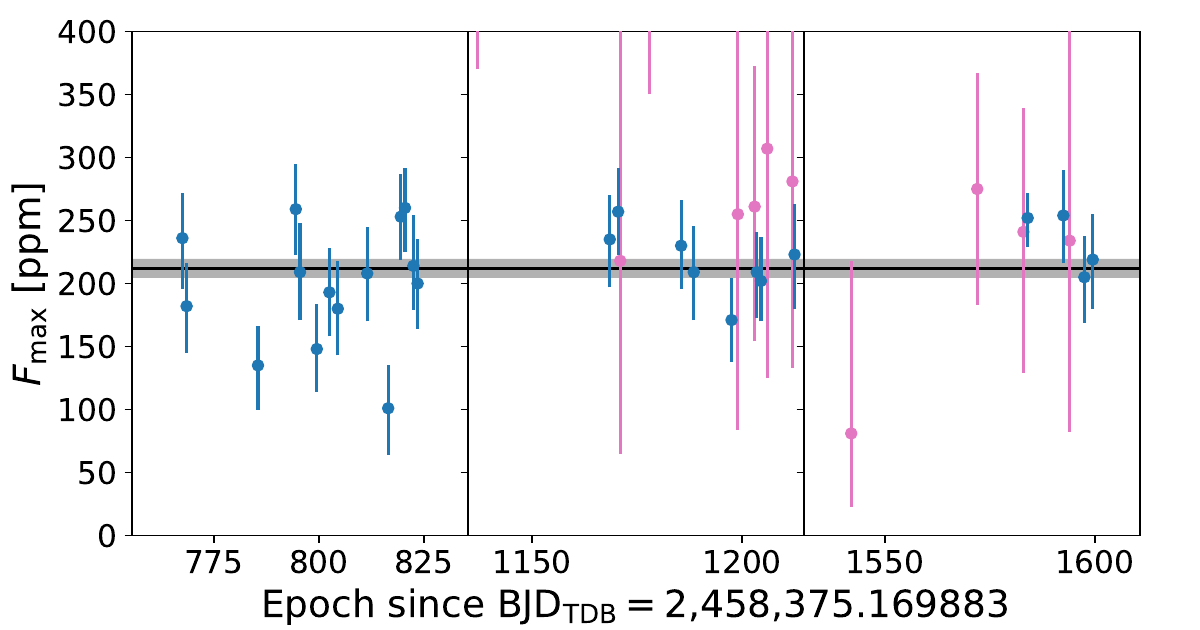}
                \caption{Dayside flux observed in the \cheops{} passband over the three seasons of observations (all three panels). The observations covering an occultation are represented in blue (including a phase curve), and the ones only covering transits in pink. The horizontal black line and shaded areas mark the value $F_\text{max}^\text{\cheops{}}={211.9}_{-{7.5}}^{+{7.6}}\,\text{ppm}$ from Table~\ref{tab:parameters}.}
                \label{fig:cheops_odv}
                \end{figure}

                We fitted each \cheops{} visit individually and extracted the posterior values of $F_\text{max}^\text{CHEOPS}$ for each epoch. The dayside flux values are represented in Fig.~\ref{fig:cheops_odv}. Naturally, the visits not covering any occultation event are poorly constrained, but we obtain a precision of the order of $30\,\text{ppm}$ for each of the other ones. We do not detect any strong sign of variability with discrepancies being below 3\,$\sigma$. To quantify the dayside flux scattering, we calculated the multiplicative factor to be applied to the errorbar to find a distribution consistent with a Normal distribution. For the \cheops{}' dayside flux, we found a multiplicative factor of 1.05, which is close to 1 and thus showing consistency with Gaussian statistics.
                We computed the Lomb-Scargle periodogram \citep{Lomb1976, Scargle1982} of the series but could not identify any significant periodicity.
                
\begin{figure}
                \centering
                \includegraphics[width=\hsize,trim={0cm 0cm 0cm 0cm},clip]{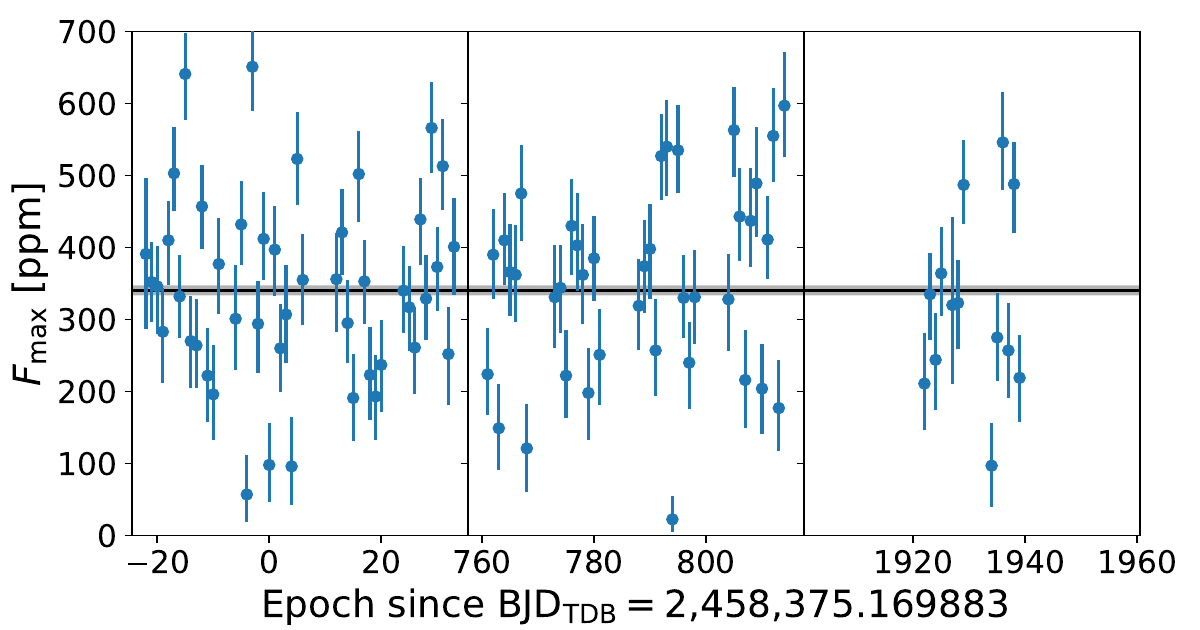}
                \caption{Dayside flux observed in the \tess{} passband over the 5 sectors: sectors~2 and~3 on the left,  sectors~29 and~30 in the center, and  sector~69 on the right. The horizontal black line and shaded areas mark the value $F_\text{max}^\text{\tess{}}={340.4}\pm{7.0}\,\text{ppm}$ from Table~\ref{tab:parameters}.}
                \label{fig:tess_odv}
                \end{figure}

                We split the \tess{} sectors into 110~short light curves with a duration of one planetary orbit. Given the \tess{} cadence, we had about 680~exposures for a full coverage of a planetary orbit, and we discarded 12 out of the 110 data sets that had fewer than 400~points (poor phase-curve coverage). For each of the 98~remaining light curves, we applied the same procedure as for the \cheops{} visits and obtained the $F_\text{max}^\text{TESS}$ values represented in Fig.~\ref{fig:tess_odv}. The scatter is larger than what is allowed from the uncertainties with 17~data points at $>3\,\sigma$ (all $\lesssim5\,\sigma$ but 1 at $9.5\,\sigma$). The errorbar multiplicative factor is this time of 2.21, further confirming the large inconsistency with Gaussian statistics. We investigated the outliers' light curves individually and found local variability likely of instrumental origin. No specific periodicity nor dominant timescale could be identified as a significant signal in the Lomb-Scargle periodogram. We note that this variability could also be of stellar origin and induced by supergranulation \citep[e.g. HAT-P-7\,b;][]{2022AJ....163..181L}.
                
\begin{figure}
                \centering
                \includegraphics[width=\hsize,trim={0cm 0cm 0cm 0cm},clip]{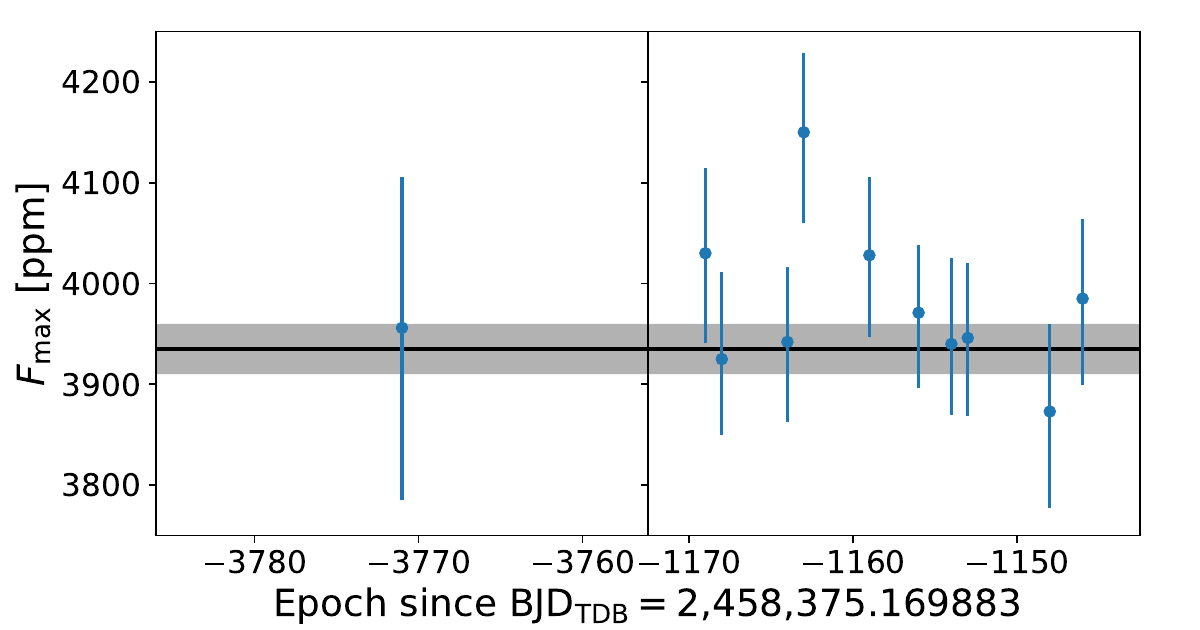}
                \caption{Dayside flux observed in the \spitzer{}/IRAC/Channel\,2 passband for programmes~50517 (left panel) and 11099 (right panel). The horizontal black line and shaded areas mark the value $F_\text{max}^\text{\spitzer{}\,2}={3935}\pm{25}\,\text{ppm}$ from Table~\ref{tab:parameters}.}
                \label{fig:spitzer_odv}
                \end{figure}

                We selected the \spitzer{} occultations of the IRAC Channel~2 and fitted them individually similar to what was done for \cheops{} and \tess{}. The dayside flux values we recover were all consistent with each other (see Fig.~\ref{fig:spitzer_odv}).
                The errorbar multiplicative factor is 0.88 for the \spitzer{} $F_\text{max}$ values, hence showing clear consistency with Normal distribution statistics.
                We compared our individual values to the ones published in \citep{2023AJ....165..104D} and found most of them being consistent ($<3\,\sigma$) with only 2 out of 11 showing a significant discrepancy ($4.1\,\sigma$ and $3.2\,\sigma$ for AOR keys 53516800 and 53515520, respectively).
                
                The analysis of the occultation depth variability of all three passbands did not reveal any significant signal on the short timescale. Thanks to the very long periods covered by our data sets (3~years for \cheops{}, 5~years for \tess{}, and 6.7~years for \spitzer{}), our results demonstrate long-term stability of the dayside of \planet{}.

\section{Atmospheric characterisation} \label{sec:atmos}
        
        \subsection{Planet thermal emission} \label{ssec:thermal_emission}
                
                We discussed in Section~\ref{ssec:param} the upper limits that we obtained on the minimum PC flux values, $F_\text{min}$, from both \cheops{} and \tess{} passbands. Given that we fixed the phase offset $\Delta\phi$ to 0, $F_\text{min}$ actually corresponds to the nightside flux. Assuming a black-body emission for the nightside of \planet{} and using synthetic stellar spectra from the PHOENIX library \citep{Husser2013_PHOENIX}, we could convert upper limits from flux to temperature. We selected the PHOENIX spectrum with the closest properties to our stellar parameters: $T_\text{eff}=6300\,K$, $\log g=4.5$ and $\left[\text{Fe}/\text{H}\right]=0.0$. We modelled the thermal emission of the planet following relationship:
                \begin{equation} \label{eq:nightside_flux}
                \frac{F_\text{night}}{F_\star}=\left(\frac{R_p}{R_\star}\right)^2\frac{\int_{\lambda=0}^{+\infty}\!\mathcal{S}\!\left(\lambda, T_\text{night}\right)\,\mathcal{T}_{\rm inst}\!\left(\lambda\right)\,{\lambda\ }d\lambda}{\int_{\lambda=0}^{+\infty}\!\mathcal{S}\!\left(\lambda, T_\star\right)\,\mathcal{T}_{\rm inst}\!\left(\lambda\right)\,{\lambda\ }d\lambda}
                ,\end{equation}
                where $R_p$ and $R_\star$ are the planetary and stellar radii, respectively, $\mathcal{S}\!\left(\lambda, T_\text{night}\right)$ and $\mathcal{S}\!\left(\lambda, T_\star\right)$ are the flux emission spectra of the planet and the star, respectively, and $\mathcal{T}_{\rm inst}\!\left(\lambda\right)$ is the instrumental passband. We downloaded the \cheops{} passband from the \cheops{} Archive Browser\footnote{\url{https://cheops.unige.ch/archive_browser/}}, and the \tess{} passband from the \tess{} Science Support Center\footnote{\url{https://heasarc.gsfc.nasa.gov/docs/tess/data/tess-response-function-v2.0.csv}}. From $F_\text{min}^\text{\cheops{}}<12.3\,\text{ppm}$ and $F_\text{min}^\text{\tess{}}<18.4\,\text{ppm}$, we could derive the nightside black-body temperatures, $T_\text{night}^\text{\cheops{}}<{2090}\,K$ and $T_\text{night}^\text{\tess{}}<{1970}\,K$, for the two instruments, leading to an overall limit of $T_\text{night}<{1970}\,K$.
                
\begin{figure*}
                        \centering
                        \sidecaption
                        \includegraphics[width=0.70\hsize]{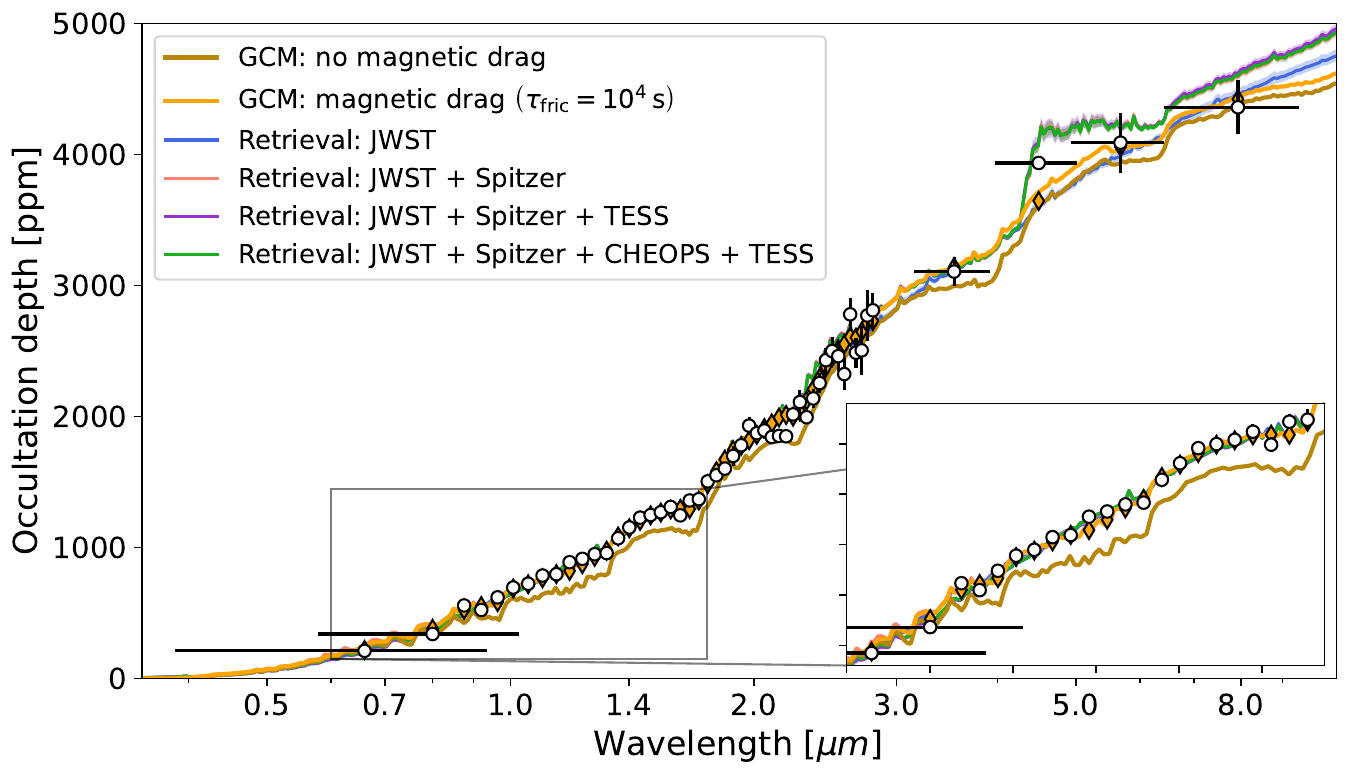}
                        \caption{Occultation depths as a function of wavelength. The black points with errorbars represent the measurement from this work (\cheops{}, \tess{} and \spitzer{}; 2 leftmost and 4 rightmost points) and from \cite{Coulombe2023} with \jwst{} (0.8--3\,\um). The GCM simulations {including TiO and VO} (Section~\ref{ssec:gcm}) with and without magnetic drag are shown in orange and light brown, respectively.
                        {The orange-filled diamonds mark the passband-integrated GCM values with magnetic drag ($\tau_\text{fric}=10^4\,\text{s}$).}
                        The retrieval runs (Section~\ref{ssec:retrieval}) are shown in {blue, pink, purple and green} (same colours as in Fig.~\ref{fig:bright_temp} and Table~\ref{table:retrievals}) depending on the data points included in the fit.
                        {An inset zoomed-in view of the \cheops{}-to-1.75\,\um{} range is shown in the lower right corner for convenience.}
                        }
                        \label{fig:docc_wl}
                \end{figure*}

                We performed a similar analysis for the dayside emission of \planet{} using Eq.~\ref{eq:nightside_flux} and replacing $F_\text{night}$ by $F_\text{day}=F_\text{max}$, and $T_\text{night}$ by the so-called brightness temperature $T_\text{b}$. We used $F_\text{max}$ from all passbands but \spitzer{}/IRAC Channels~3 and~4, as their wavelength coverage goes beyond the limit of 5.5\textmu m of the PHOENIX spectra. We obtained the following {brightness} temperatures: 
                $T_\text{b}^\text{CHEOPS}={3027\pm17}\,K$,
                $T_\text{b}^\text{TESS}={2956\pm10}\,K$,
                $T_\text{b}^\text{SPITZER\,1}={2957\pm58}\,K$, and
                $T_\text{b}^\text{SPITZER\,2}={3132\pm13}\,K$.
                The uncertainties are expected to be underestimated given the fact that we did not propagate the uncertainties of the stellar properties by using a single PHOENIX spectrum. Also, we might be slightly biasing our results by assuming a black-body emission for the planet. However, these results already show some interesting outcomes as the \tess{} and \spitzer{}/IRAC/Channel\,1 passband agree on a brightness temperature of {2960}\,K, whereas \cheops{} and \spitzer{}/IRAC/Channel\,2 would be capturing an excess of flux (converted into a higher temperature here). The excess in the \cheops{} passband might actually indicate the presence of reflected light as we probe visible wavelengths. As for \spitzer{}/IRAC/Channel\,2, the larger occultation depth could be explained by the presence of CO \citep{Brogi2023_W18b, Yan2023_W18b_W76b} that have emission lines at 4.5\textmu m.
                Using the framework of \cite{Cowan2011_AB_eps} and following the methodology detailed in Section~6.2 of \cite{Deline2022_WASP-189b}, we estimated the values of the Bond albedo $A_B$ and the heat redistribution efficiency $\varepsilon$ (more details in Appendix~\ref{sec:ab_eps}). As usual and expected for UHJs, we obtained very low values of both $A_B$ and $\varepsilon$ implying that the planetary atmosphere absorbs most of the incoming energy and redistributes it inefficiently. The narrowest constraints come from the \tess{} passbands with $A_B\leq{0.13}$ and $\varepsilon\leq{0.21}$. Interestingly, the maximum dayside temperature that this analytical approach allows is about 3060\,K for the extreme case where $A_B=\varepsilon=0$. This strengthens the hint for an excess of flux{, especially in the passband of} \spitzer{}/IRAC/Channel\,2.
                Given the several assumptions made for this analysis of the dayside emission, we consider the aforementioned results as indicative. In the following Sections~\ref{ssec:gcm} and~\ref{ssec:retrieval}, we analysed the dayside emission of \planet{} including the recently published JWST values \citep{Coulombe2023} to complete our 0.6-to-8.0\,\textmu m wavelength coverage of occultation depth measurements (see Fig.~\ref{fig:docc_wl}). We used two different approaches to characterise the planetary atmosphere: General Circulation Modelling (GCM) and atmospheric retrieval.

        \subsection{General circulation model} \label{ssec:gcm}
                
                We performed forward modelling of the atmosphere of \planet{} with the general circulation model (GCM) \texttt{ExoRad}, using the version with full radiative transfer \citep[\texttt{expeRT/MITgcm};][]{Carone2020,Schneider2022}\footnote{{Details of the GCM  are described in Section~\ref{sec: GCM_Detail}.}}. We generated synthetic emission spectra of the planetary dayside using the parameters listed in Table~\ref{tab:parameters} as inputs, and normalised the outcomes with the PHOENIX stellar spectra \citep{Husser2013_PHOENIX} up to 5.5\,\um{}, assuming $T_\text{eff}=6300\,K$, and a matching black-body SED at longer wavelengths. To calculate the radiative forcing in the model, we assumed solar metallicity and {equilibrium chemistry}. We included correlated-k tabulated opacities {combined to 11 spectral bins, corresponding to the S1 resolution as specified in \citet{Schneider2022} for the following species}: H$_2$O (from ExoMol\footnote{\url{https://www.exomol.com}}; \citealt{Tennyson2016_ExoMol, TennysonEtal2020jqsrtExomol2020}), Na \citep{Allard19_Na_K}, K \citep{Allard19_Na_K}, CO$_2$, CH$_4$, NH$_3$, CO, H$_2$S, HCN, SiO, PH$_3$ and FeH, as well as H$^-$ absorption suitable for an ionised atmosphere.
                
                We generated {several GCM simulations to derive dayside spectra. We first tested runs without TiO and VO opacities with strong magnetic drag $\tau_{\rm fric}$ and no drag. Here, we found that already the inclusion of magnetic drag alone yielded a vast improvement in the data fit. We then included} TiO and VO opacities \citep{McKemmish2016,McKemmish2019}. TiO and VO produce an upper atmosphere thermal inversion that impacts in particular the dayside emission in the optical as observed with CHEOPS and TESS. {We then tested various values for magnetic drag and including TiO/VO (see Table~\ref{tab:gcm_chi2}).} Magnetic drag is treated during the simulation by applying uniform friction to the {horizontal wind field $(u,v)$ via:
                \begin{align}
					&\frac{\rm du}{\rm dt}= -\frac{u}{\tau_\text{fric}},\\
					&\frac{\rm dv}{\rm dt}= -\frac{v}{\tau_\text{fric}},
				\end{align}
                where $u$ and $v$ are the zonal and meridional wind fields, respectively.}

                \begin{table}
                {
                \caption{Goodness-of-fit metrics for the GCM models.}
                \label{tab:gcm_chi2}
                \centering
                \resizebox{\columnwidth}{!}{
                    {
                    \begin{tabular}{cccrrrr}
                        \hline\hline
                        \multirow{2}{*}{TiO/VO} & Magnetic & \multirow{2}{*}{$\tau_\text{fric}$ [s]} & \multicolumn{2}{c}{All data} & \multicolumn{2}{c}{No Sp.Ch2} \\
                        & drag & & \multicolumn{1}{c}{$\chi^2$} & \multicolumn{1}{c}{$\chi_r^2$} & \multicolumn{1}{c}{$\chi^2$} & \multicolumn{1}{c}{$\chi_r^2$} \\
                        \hline
                        \xmark & \xmark & $+\infty$ & 2261.4 & 44.3 & 1889.4 & 37.8 \\
                        \xmark & \cmark & $10^4$    &  638.6 & 12.5 &  423.4 &  8.5 \\
                        \hline
                        \cmark & \xmark & $+\infty$ & 1448.1 & 28.4 & 1156.9 & 23.1 \\
                        \cmark & \cmark & $10^6$    &  498.7 &  9.8 &  322.9 &  6.5 \\
                        \cmark & \cmark & $10^5$    &  362.5 &  7.1 &  206.5 &  4.1 \\
                        \cmark & \cmark & $10^4$    &  265.2 &  5.2 &  130.4 &  2.6 \\
                        \hline\hline
                    \end{tabular}
                    }
                }
                \tablefoot{
                The inclusion or exclusion of TiO/VO and magnetic drag in the GCM models is indicated by the symbols \cmark and \xmark, respectively.
                The $\chi^2$ metric is computed as $\chi^2=\sum{\left(y_i-m_i\right)^2/\sigma_i^2}$, where $y_i$ is the passband-integrated GCM value, and $m_i$ and $\sigma_i$ are the occultation depth measurement and uncertainty, respectively.
                The reduced $\chi^2$ metric, $\chi_r^2$, is computed as $\chi_r^2=\chi^2/N$, where $N$ is the number of data points.
                The GCM models with TiO/VO (no drag and $\tau_\text{fric}=10^4$\,s) are shown in Fig.~\ref{fig:docc_wl}.
                The measured occultation depths used to compute the $\chi^2$ and $\chi_r^2$ values are reported in Table~\ref{tab:parameters}.
                The rightmost 2~columns entitled `No Sp.Ch2' report the metrics computed on all but the \spitzer{}/IRAC/Ch.\,2 data point to account of the GCMs not modelling the high CO/CO$_2$ emission peak.
                }
                }
                \end{table}
                
                 The inclusion of magnetic drag is motivated by the substantial degree of ionisation of the dayside atmosphere resulting in the presence of several ions such as Na$^+$, K$^+$, Ca$^+$, Fe$^+$, Al$^+$ and Ti$^+$ \citep{Helling2019_WASP-18b, Helling2021_HJ_clouds}. The partially ionised flow induces magnetic coupling that in turn affects the winds as already pointed out by \citet{Perna2010,Rodriguez-Barrera2015}. The necessity to account for the impact of magnetic fields in UHJs such as \planet{} has been further confirmed several times \citep{Wardenier2023,Beltz2022,Demangeon2024}. Recent JWST observations also support the impact of magnetic drag \citep[e.g. \planet{} in][]{Coulombe2023}. {Table~\ref{tab:gcm_chi2} clearly shows that a combination of drag and TiO/VO is needed and that the fit to the data improves with decreasing {$\tau_{\rm fric}=10^6,10^5,10^4$~s.} \citet{Coulombe2023} found with the GCM \texttt{SPARC/MITgcm} a lower value of  $\tau_{{\rm fric}}=10^3$~s. These authors tested, however, only two scenarios with {weak $\tau_{\rm fric}=10^6$~s and strong drag $\tau_{\rm fric}=10^3$~s,} respectively. They further did not perform a sensitivity study with the GCM with respect to drag and TiO, as we did in this paper. We thus note that the choice of {$\tau_{\rm fric}$} is currently still open for debate, in particular in the uniform drag assumption as implemented here \citep{Perna2010,Tan2019,Coulombe2023,Beltz2022}. We found with \texttt{expeRT/MITcgm} that {$\tau_{\rm fric}=10^4 $~s already effectively disrupts superrotation at the dayside and yields agreement with the JWST emission spectrum. We further note that $\tau_{\rm fric}=10^{4}$~s is currently the smallest drag time scale that was tested in \texttt{expeRT/MITcgm}. Smaller $\tau_{\rm fric}$} time scales as well as other implementations of magnetic field coupling are currently under investigation.} In any case, only full 3D GCMs can assess the impact of magnetic drag on the climate state that shapes horizontal heat transfer and thus the dayside emission. The resulting strong horizontal temperature gradient over the dayside, $\Delta T >1000$~K, also ensures that the TiO and VO abundances are not constant across the dayside.

Comparing the outcome with the measured occultation depths from \cheops{}, \tess{}, \jwst{} and \spitzer{} reveals {agreement when both TiO and VO and magnetic drag are used (see Fig.~\ref{fig:docc_wl} {and Table~\ref{tab:gcm_chi2}}). {Magnetic drag reduces circulation efficiency, resulting in a hotter dayside such that, even without TiO and VO, the GCM provides a relatively close match to the data, with a reduced $\chi^2$ ($\chi^2_r$) of 8.5\footnote{{It is important to note that GCMs are forward models that are {not fitted} to the data. Thus, given the very broad wavelength range covered in this analysis, we consider $\chi^2_r$ values below 10 as a reasonable match, and values below 5 to be in very good agreement.}}. The agreement with the data is nonetheless further improved when including TiO and VO, even with reduced drag. Our best GCM match to the data is obtained when both TiO/VO and a strong drag are present ($\chi^2_r = 2.6$). Therefore, both the magnetic drag and the presence of {a strong, local gas-phase absorber in the optical like} TiO and VO {known from cool star atmosphere modelling \citep[e.g.][]{Gustafsson2008,VanEck2017}} are needed to match the data from the optical to the IR wavelengths.}  Models without TiO and VO are not shown for clarity but underpredict the emitted planetary flux.}  We particularly note that the GCM output qualitatively matches with the atmospheric retrieval models (see Section~\ref{ssec:retrieval}), including the flattening of the spectrum between 1 and 2 micron due to H$_2$O dissociation. {In cooler planets, this wavelength region is dominated by water bands. We note, however, that while our GCM with TiO/VO and magnetic drag shows an upper atmosphere temperature inversion, it is not sufficiently hot for CO emission. Retrieval models show that temperatures higher than 3500~K are required to trigger CO emission (see Fig.~\ref{fig:bright_temp}). }

\subsection{Atmospheric retrieval}
\label{ssec:retrieval}

We performed atmospheric-retrieval analyses to characterise the thermal emission properties of {\planet}, hence, inferring its thermal and composition structure.  We employed the open-source {\pyratbay} package \citep{CubillosBlecic2021mnrasPyratBay} to perform the atmospheric modelling, spectral synthesis, and Bayesian posterior sampling.
Since the model only accounts for thermal emission, we primarily constrained the retrievals using the infrared {\jwst} and {\spitzer} occultations, while running additional fits with and without the optical {\cheops} and {\tess} occultations to assess how stellar reflected light affected the retrieval results.

The atmospheric model consists of a set of parameterized temperature and composition profiles as a function of pressure (81 layers between 100 and 10$^{-9}$ bar). We adopted the \citet{MadhusudhanSeager2009apjRetrieval} model to parameterise the temperature profile.
The atmosphere of the UHJ \planet{} is expected to exceed 2000\,K, with the hottest part of the dayside that could reach 3000\,K. Since disequilibrium-chemistry processes such as photochemistry and transport-induced quenching become less and less important with increasing effective temperature, at the extreme temperatures seen on \planet{}, the chemical reaction rates are fast enough to overcome the effect of disequilibrium chemistry \citep{KopparapuEtal2012apjWASP12bChemistry, Moses2014rsptaChemicalKinetics, MadhusudhanEtal2016ssrChemistryFormation, VenotEtal2018exaChemicalCompositionAriel}. If disequilibrium chemistry occurs at all, it would occur at high altitudes above the pressures probed by the observations presented in this work, and thus the modeled emission spectra of planets such as \planet{} would not be significantly impacted \citep{ShulyakEtal2020aaHotJupiterChemistry}\footnote{{More specifically, it takes less than $10^{-2}$\,s to reach thermo-chemical equilibrium for local gas phase temperatures exceeding 1000\,K \citep[][Sect. 6.3]{Rimmer2016}. }}. Thermochemical equilibrium is therefore a reasonable assumption to model the dayside atmospheric composition of \planet{} and its corresponding emission spectra. The chemical network consists of 45 neutral and charged species, accounting for the main H, He, C, O, N, Na, K, S, Si, Fe, Ti, and V-bearing species expected to form in the atmosphere.  We defined three retrieval parameters to sample the composition parameter space: the oxygen elemental abundance relative to solar values, [O/H]; the carbon abundance relative to solar values, [C/H]; and a catch-all parameter to scale all other metal abundances relative to solar values, [M/H].  Finally, we computed the height of the pressure layers assuming hydrostatic equilibrium. Table~\ref{table:retrievals} lists the retrieval parameters, priors, and posterior values.

{\renewcommand{\arraystretch}{1.3}
\begin{table*}
{
\caption{\planet{} retrieval analyses}
\label{table:retrievals}
\centering
{
\begin{tabular}{lccccc}
\hline\hline
\multirow{2}{*}{Parameter\tablefootmark{$\dagger$}} & \multirow{2}{*}{Prior\tablefootmark{$\ddagger$}} & \multicolumn{4}{c}{Posteriors\tablefootmark{$\S, \P$}} \\
& & \textcolor{royalblue}{JWST} & \textcolor{salmon}{JWST+Spitzer} & \textcolor{darkorchid}{JWST+Spitzer+TESS} & \textcolor{xkcdgreen}{JWST+Spitzer+CHEOPS+TESS} \\
\hline
$\log_{10}(p_1)$   & $(-9, 2)$    &  $-1.59^{+0.31}_{-0.37}$   &  $-3.06^{+0.34}_{-0.28}$    &  $-2.10^{+0.34}_{-0.66}$      &  $-2.97^{+0.40}_{-0.22}$  \\
$\log_{10}(p_2)$   & $(-9, 2)$    &  $1.30^{+0.45}_{-0.64}$    &  $1.11^{+0.37}_{-0.36}$     &  $1.01^{+0.64}_{-0.27}$       &  $1.62^{+0.21}_{-0.30}$   \\
$\log_{10}(p_3)$   & $(-9, 2)$    &  $0.73^{+0.81}_{-0.90}$    &  $0.79^{+0.64}_{-0.53}$     &  $0.91^{+0.56}_{-0.50}$       &  $1.23^{+0.44}_{-0.63}$   \\
$a_1$             & $(0.2, 2)$    &  $1.57^{+0.27}_{-0.34}$    &  $1.68^{+0.19}_{-0.24}$     &  $1.65^{+0.19}_{-0.21}$      &  $1.74^{+0.16}_{-0.29}$   \\
$a_2$             & $(0.2, 2)$    &  $0.304^{+0.070}_{-0.061}$ &  $0.285^{+0.029}_{-0.035}$  &  $0.240^{+0.073}_{-0.025}$    &  $0.312^{+0.020}_{-0.024}$ \\
$T_0$ (K)         & $(500, 4000)$ &  $2987.2^{+55.9}_{-58.2}$  &  $3842.2^{+69.3}_{-115.0}$  &  $3596.9^{+172.0}_{-136.3}$   &  $3826.1^{+86.0}_{-199.6}$ \\
{[C/H]}           & $(-2, 2.5)$   &  $-0.97^{+0.83}_{-0.65}$   &  $0.65^{+0.21}_{-0.17}$     &  $0.13^{+0.16}_{-0.15}$       &  $0.26^{+0.16}_{-0.13}$   \\
{[O/H]}           & $(-2, 2.5)$   &  $0.36^{+0.58}_{-0.37}$    &  $0.49^{+0.19}_{-0.16}$     &  $-0.06^{+0.15}_{-0.15}$      &  $0.08^{+0.15}_{-0.12}$   \\
{[M/H]}           & $(-2, 2.5)$   &  $0.82^{+0.53}_{-0.37}$    &  $0.10^{+0.16}_{-0.15}$     &  $-0.67^{+0.11}_{-0.12}$      &  $-0.48^{+0.10}_{-0.11}$  \\
\hline \hline
\end{tabular}
}
{
\tablefoot{
\tablefoottext{$\dagger$}{Pressure parameters are in bar units.}
\tablefoottext{$\ddagger$}{Uniform priors over the specified range.}
\tablefoottext{$\S$}{The reported retrieved values correspond to the marginal posterior distribution's median and boundaries of the 68\% central credible interval \citep{Andrae2010arxivErrorEstimation}.}
\tablefoottext{$\P$}{Each column corresponds to a retrieval considering different observational constraints (column headers colour-coded as in Fig.~\ref{fig:bright_temp}).}
}
}
}
\end{table*}
}

For the spectral synthesis, we considered line-sampled opacities from the dominant molecular species from the HITEMP \citep[\ch{CH4}, CO, and \ch{CO2};][]{RothmanEtal2010jqsrtHITEMP} and ExoMol databases \citep[\ch{H2O}, HCN, \ch{NH3}, TiO, VO, and \ch{C2H2};][]{Tennyson2016_ExoMol, TennysonEtal2020jqsrtExomol2020}. Prior to the retrievals, we tabulated the opacity line lists into a cross-section grid of temperatures, pressures, and wavenumbers (at a constant resolving power of $R=20\,000$). We pre-processed the larger line lists with the repack package to extract the dominant line transitions \citep{Cubillos2017apjRepack}.  Additionally, we considered opacities from the Na and K resonant lines \citep{BurrowsEtal2000apjBDspectra}, collision-induced absorption for \ch{H2}--\ch{H2} \citep{BorysowEtal2001jqsrtH2H2highT, Borysow2002jqsrtH2H2lowT, JorgensenEtal2000aaCIAH2He} and \ch{H2}--He \citep{BorysowEtal1988apjH2HeRT, BorysowFrommhold1989apjH2HeOvertones, BorysowEtal1989apjH2HeRVRT}, Rayleigh scattering for H, \ch{H2}, and He \citep{Kurucz1970saorsAtlas}, and \ch{H-} free-free and bound-free \citep{John1988aaHydrogenIonOpacity}.  We adopted a PHOENIX stellar model as the flux spectrum for {\host}.
We performed the Bayesian posterior sampling using the Nested Sampling algorithm \citep{Skilling2004_NS, Skilling2006_NS} as implemented in the \texttt{MultiNest} package \citep{Feroz2009_multinest, BuchnerEtal2014aaBayesianXrayAGN}, employing 1500 live points.

\begin{figure*}
    \centering
    \includegraphics[width=\linewidth,clip]{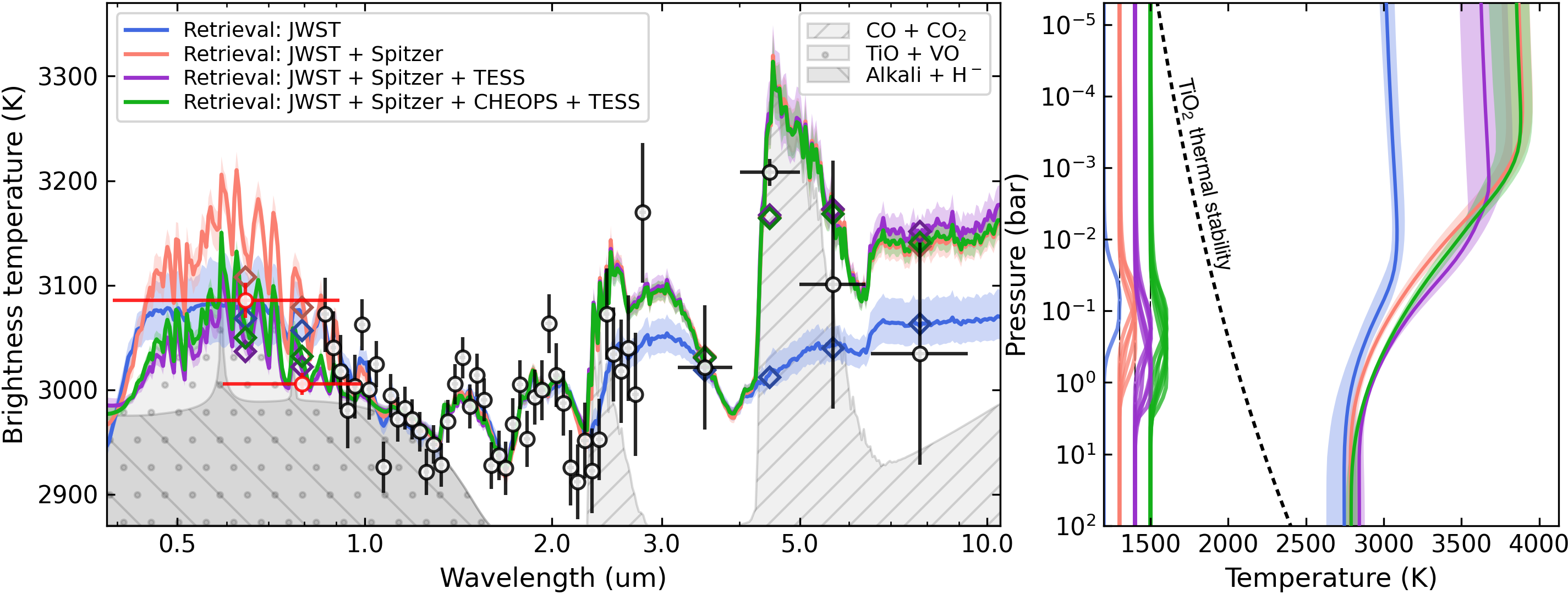}
    \caption{Brightness-temperature spectra derived from the atmospheric retrieval analyses. The circle markers with error bars show the observed occultation depths with uncertainties.
    {The two leftmost circles (red) correspond to \cheops{} and \tess{} passbands, while the four rightmost black points cover the \spitzer{}/IRAC channels.
    The {\jwst} spectroscopic observations are shown in black between 0.8 and 3.0\,\um{} and} have been binned down for better visualisation.
    The {coloured} solid curves with shaded areas denote the retrieved median spectra and span of the $1\sigma$ credible interval {for the four model fits (see legend)}.
    The diamond markers show the model spectra integrated over the photometric bands.
    {The grey shaded areas highlight the contributions of some relevant species to thermal emission: alkali/H$^-$ and TiO/VO in the blue end, and CO/CO$_2$ in the infrared.}
    {\it Right panel:} the median (solid curves) and $1\sigma$ credible interval (shaded area) of the temperature-profile posterior distributions (same colour coding as in the left panel).  The curves on the left edge show the normalized contribution functions that indicate the pressures mainly probed by the observations according to each retrieval analysis. {The dashed black line shows the thermal stability curve for \ch{TiO2}. Other thermal stability curves for species of interest, such as alkali metals (\ch{Na2S} or KCl), are situated at even lower temperatures around $\sim$1000\,K.}
    }
    \label{fig:bright_temp}
\end{figure*}

\begin{figure*}
    \centering
    \includegraphics[width=\linewidth,clip]{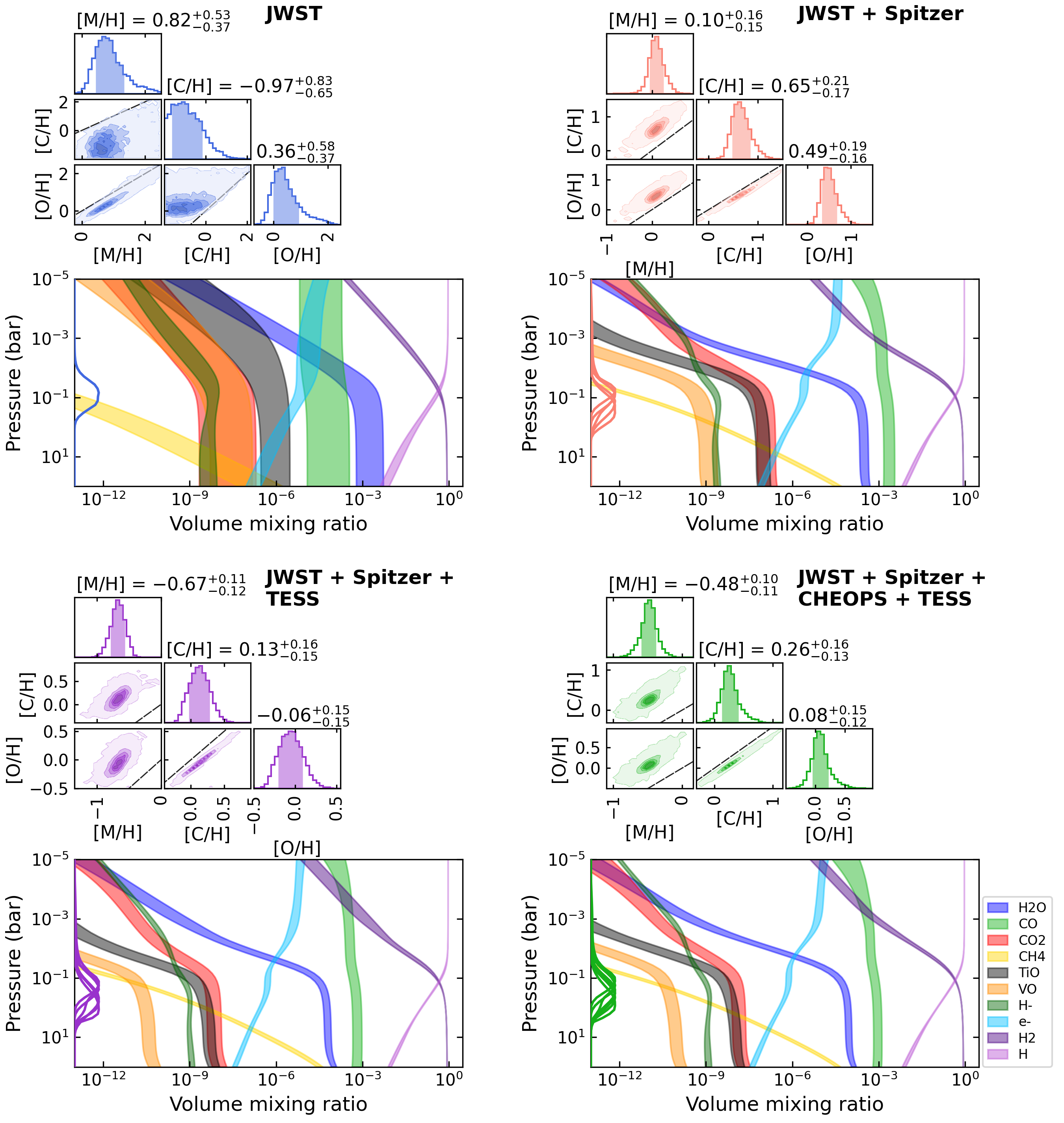}
    \caption{{Atmospheric retrieval compositions (each quadrant  shows one of the four retrieval runs, labelled in bold text).} {\it Top:} Pairwise and marginal posterior distributions for the abundance free parameters. The quoted values on top of the marginal histograms denote the median and $1\sigma$ credible interval for each parameter (span of the central 68\% percentiles, also denoted as the shaded areas). The black diagonal dashed lines denote the parameter space values where both parameter are scaled by the same amount (relative to solar abundances).
    {\it Bottom:} Volume mixing ratios (VMR) for each retrieval corresponding to the posterior distributions shown above. The shaded areas denote the $1\sigma$ credible-interval span in VMR for selected species (see labels at the right {of the bottom right quadrant}). The solid curves on the left edge of the VMR panels show the contribution functions that indicate the pressures probed by the observations.
    }
    \label{fig:retrieval_abundances}
\end{figure*}

To assess the impact of each observation in constraining the atmospheric properties, we performed {four} retrieval analyses considering different sets of observing constraints.  {First, we performed retrievals using only the {\jwst} occultations. Next, we added the {\spitzer} occultations to test the role of infrared absorbers. Finally, we included the TESS and CHEOPS occultations to test the impact of optical absorbers and reflected light.} As expected, the {\jwst} observations dominate the retrieval results, resulting in a consistent fit to the \ch{H2O} bands between 1.0 and 2.5~{\um}. However, we found that retrievals with and without the {\spitzer} constraints lead to widely different temperature and composition outcomes (see Figs.~\ref{fig:bright_temp} and~\ref{fig:retrieval_abundances}, and Table~\ref{table:retrievals}). The driving factor is the high signal-to-noise IRAC2 occultation depth at 4.5~{\um}.  The much larger brightness temperature at 4.5~{\um} than that of the {\jwst} occultations requires an atmosphere with significant CO/\ch{CO2} absorption {(as highlighted by the diagonally hatched area in left panel of Fig.~\ref{fig:bright_temp})}, which in turn leads to super-solar carbon and oxygen abundances of {3--5} times solar.
We also noted that {incorporating the {\spitzer} occultations to the retrieval constraints results in steeper temperature gradients.  This adjustment is expected given the stark brightness-temperature difference between the {\jwst} and the {\spitzer} 4.5-{\textmu m} observations} (Fig.~\ref{fig:bright_temp}, right panel).
In contrast, the retrieval of the {\jwst} data alone returns an atmosphere with solar-to-subsolar abundances, well in agreement with the previous analysis by \cite{Coulombe2023}.

Finally, we compared the retrieved thermal-emission spectra with the measurements in the {\cheops} and {\tess} passbands to look for the presence of reflected light coming from the planet. The left panel in Fig.~\ref{fig:bright_temp} displays the retrieval models integrated over all instrumental passbands as coloured diamonds.
Table~\ref{tab:ret_res} lists the thermal flux values retrieved for each one of the four retrieval models.
We detail below the comparison within the \cheops{} and \tess{} passbands in the cases for which the \spitzer{} data were included in the retrieval fit (pink, purple and green).
First, when only \jwst{} and \spitzer{} data points were included in the retrieval fit (pink model), we found that the measured occultation depth in the \tess{} passband was significantly below the retrieval value (at 4.1\,$\sigma$). Therefore, the inclusion of the \tess{} data point in the retrieval runs is necessary to properly model the data by mitigating the amount of thermal emission in the \tess{} passband.
With the \tess{} data point included, we compared two retrieval runs with and without fitting the \cheops{} data point (green and purple models, respectively).
When discarding \cheops{} from the fit, the retrieval model showed a smaller flux difference in the \tess{} passband ($-11.1 \pm 8.2$~ppm). If we convert the flux difference into a 3-$\sigma$ upper limit on the reflected light in the \tess{} passband, we get $F_\text{refl}^\text{\tess{}}<13.5$~ppm, which corresponds to a geometric albedo\footnote{The geometric albedo is defined as $A_g=\Delta F\, \left(R_p / a\right)^2$ for circular orbits, where $\Delta F$ is the normalised flux excess, $R_p$ is the planetary radius, and $a$ is the semi-major axis \citep{Sobolev1975,Charbonneau1999}.} of $A_g^\text{\tess{}}<0.017$.
The passband-integrated value of this model for \cheops{} gives a thermal flux of $190.7_{-3.5}^{+3.3}$\,ppm, which is smaller than the measured dayside flux by 2.6\,$\sigma$. Explaining this difference with reflected light means $F_\text{refl}^\text{\cheops{}} = 21.2 \pm 8.2$~ppm and a corresponding geometric albedo of $A_g^\text{\cheops{}}={0.027}\pm{0.011}$.
In our last run, we included all data points in the fit (green model) and analysed the effect on the limits of reflected light in both \tess{} and \cheops{} passbands. As listed in Table~\ref{tab:ret_res}, we obtained a larger flux difference for \tess{} and smaller for \cheops{}, at $-2.2\,\sigma$ and $+2.0\,\sigma$, respectively. These correspond to limits on geometric albedos of $A_g^\text{\tess{}}<0.009$ and $A_g^\text{\cheops{}}={0.020}\pm{0.010}$.

Our retrievals indicate that a significant fraction of the short-wavelength thermal emission originates from TiO/VO (Figure \ref{fig:bright_temp}, right panel).  Although the high planetary mass of \planet{} can potentially enhance the efficiency of cold-trapping -- hence depleting heavier metals from the atmosphere \citep{BeattyEtal2017ajKeppler13bColdTrapping} --, our retrieved thermal profiles lie well above the thermal stability curves for Ti-bearing and other refractory species.  This suggests that the dayside of \planet{} remains sufficiently hot to inhibit vertical cold-trapping \citep{SpiegelEtal2009apjTiO}, and thus allow for the presence of heavy metals like TiO, VO, and alkali metals in gas form in its atmosphere. While day-night cold-trapping remains a plausible scenario, additional phase-curve observations of the nightside emission are needed to evaluate its potential \citep{ParmentierEtal2013aa3DmixingColdTrap}. However, \citet{BeattyEtal2017ajKeppler13bColdTrapping} argued that observations of planets with daysides of 3000\,K and hotter like this planet appear to escape nightside trapping. Figure~\ref{fig:bright_temp} also elucidates why this would be the case. The dayside is too hot for TiO thermal stability even down to 100~bar. For such a high gravity and fast rotating planet, superrotation is expected to extend well down to 100~bar and deeper into the planet \citep{Carone2020}. Thus, an efficient horizontal heat and material transport is expected below 10~bar. Even if TiO would condense out (i.e. be thermally stable) on the nightside, it would evaporate again at depth and be transported towards the dayside, where it could again dominate the gas absorption in the optical wavelength range in agreement with the results in this work. This is also in line with the study of \cite{Helling2019_WASP-18b}, where they infer the absence of clouds on the dayside of \planet{} from cloud formation models.

{\renewcommand{\arraystretch}{1.3}
\begin{table*}
{
\caption{Retrieved thermal flux in the \cheops{} and \tess{} passbands.}
\label{tab:ret_res}
\centering
    {
    \begin{tabular}{cccccc}
        \hline\hline
        \multirow{2}{*}{Retrieval model} & \multirow{2}{*}{Passband} & Thermal flux & Flux diff. & \multirow{2}{*}{n$_\sigma$} & \multirow{2}{*}{$A_g$} \\
        & & [ppm] & [ppm] & & \\
        \hline
        \multirow{2}{*}{\textcolor{royalblue}{\jwst{}}} & \cheops{} & $204.1_{-10.5}^{+10.7}$ & $+7.8\pm13.1$ & $+0.6$ & $<0.061$ \\
        & \tess{} & $375.6_{-12.0}^{+12.3}$ & $-35.2\pm13.9$ & $-2.5$ & $<0.009$ \\
        \hline
        \multirow{2}{*}{\textcolor{salmon}{\jwst{}+\spitzer{}}} & \cheops{} & $222.1_{-7.8}^{+7.0}$ & $-10.2\pm10.9$ & $-0.9$ & $<0.029$ \\
        & \tess{} & $391.6_{-10.3}^{+9.5}$ & $-51.2\pm12.5$ & $-4.1$ & $\ll0$\tablefootmark{$\dagger$} \\
        \hline
        \multirow{2}{*}{\textcolor{darkorchid}{\jwst{}+\spitzer{}+\tess{}}} & \cheops{} & $190.7_{-3.5}^{+3.3}$ & $+21.2\pm8.2$ & $+2.6$ & $0.027\pm0.011$ \\
        & \tess{} & $351.5\pm4.3$ & $-11.1\pm8.2$ & $-1.4$ & $<0.017$ \\
        \hline
        \multirow{2}{*}{\textcolor{xkcdgreen}{\jwst{}+\spitzer{}+\cheops{}+\tess{}}} & \cheops{} & $196.1_{-3.4}^{+3.0}$ & $+15.8\pm8.1$ & $+2.0$ & $0.020\pm0.010$ \\
        & \tess{} & $358.2_{-4.3}^{+4.0}$ & $-17.8\pm8.2$ & $-2.2$ & $<0.009$ \\
        \hline\hline
    \end{tabular}
    }
\tablefoot{
The thermal flux is derived by integrating the retrieval model values in the corresponding passband (\cheops{} or \tess{}) following Eq.~\ref{eq:nightside_flux}.
The flux difference is computed with respect to the measured dayside fluxes listed in Table~\ref{tab:parameters}.
The n$_\sigma$ column reports the significance of the flux difference in number of $\sigma$.
The $A_g$ column shows the geometric albedo values corresponding to the flux difference assuming this difference is attributed to reflected light. When the significance on $A_g$ is $<2\,\sigma$, we report the 3-$\sigma$ upper limit. \\
\tablefoottext{$^\dagger$}{The 3-$\sigma$ upper limit is $-0.017$.}
}
}
\end{table*}
}

\begin{figure}
    \centering
    \includegraphics[width=\linewidth,clip]{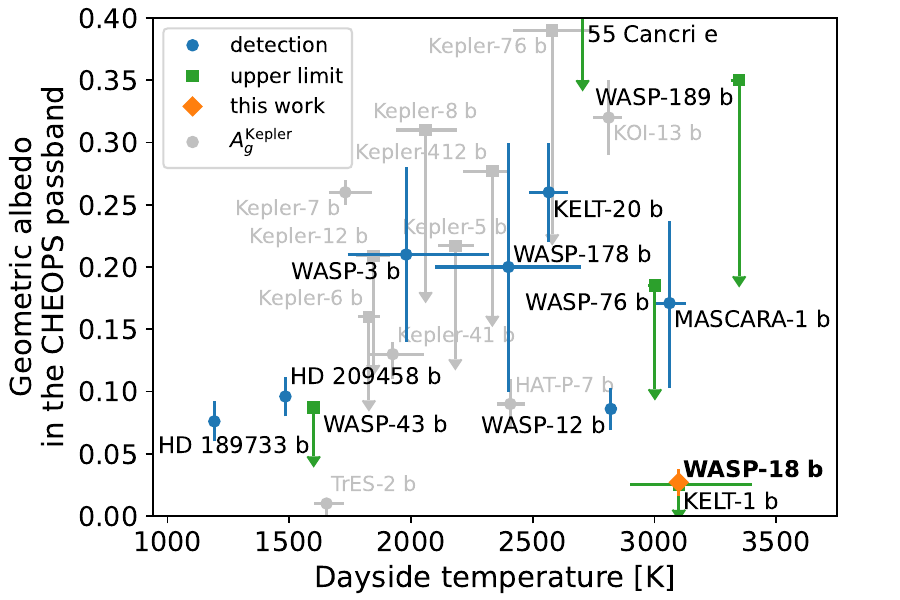}
    \caption{Geometric albedo of several hot and ultra-hot exoplanets as function of their estimated dayside temperature. The detection values (blue) or upper limits (green) are measured in the \cheops{} passband ($=A_g^\text{\cheops}$) and reported in the following works:
    \cite{2020A&A...643A..94L}, \cite{2022A&A...658A..75H}, \cite{Deline2022_WASP-189b}, \cite{Brandeker2022_HD209458b},
    \cite{2022A&A...668A..17S}, \cite{2022A&A...668A..93P}, \cite{2023A&A...669A..64D}, \cite{Krenn2023_HD189733b}, \cite{2024A&A...682A.102P}, \cite{2024A&A...683A...1S}, \cite{Demangeon2024}, \cite{2024A&A...685A..63A}{, and \cite{Scandariato2024_WASP-3b}.}
    {Measurements in the \kepler{} passband from \cite{Esteves2015_Ag} and \cite{Morris2024_Ag} are shown in grey.}
    }
    \label{fig:Ag_plot}
\end{figure}

We conclude that both the scenario including \tess{} only and the scenario including \cheops{}+\tess{} strongly constrain the amount of reflected light in the \tess{} passbands to extremely low values ($<0.017$). This is not unexpected as many atmospheres of UHJs have been reported to be very dark \citep[e.g.][]{Wong2021_TESS_PC}.
In the shorter wavelength \cheops{} band, we find tentative ($2.6\,\sigma$) evidence for reflected light, indicative of stronger atmospheric scattering than in the \tess{} passband. Even so, it's amplitude remains low, constrained to a geometric albedo below 0.059.
The overall reflectivity remains nonetheless very low as well for \cheops{}, in line with expectations for UHJs.
As the geometric albedo is wavelength-dependent, the difference in passbands between \cheops{} and \tess{} prevents us from performing a quantitative comparison between the two instruments. Fig.~\ref{fig:Ag_plot} shows how the reflected light detection of \planet{} computed in this work aligns well with previously published $A_g^\text{\cheops}$ values for other hot exoplanets
\citep{
2020A&A...643A..94L,
2022A&A...658A..75H,
Deline2022_WASP-189b,
Brandeker2022_HD209458b,
2022A&A...668A..17S,
2022A&A...668A..93P,
2023A&A...669A..64D,
Krenn2023_HD189733b,
2024A&A...682A.102P,
2024A&A...683A...1S,
Demangeon2024,
2024A&A...685A..63A,
Scandariato2024_WASP-3b}.
We also included measurements from the \kepler{} instrument \citep{Koch1998_Kepler, Borucki2010_Kepler} as its passband probes essentially the same wavelength range as \cheops{} \citep{Deline2020_CHEOPS}. Geometric albedos values in the \kepler{} passband have been reported in several publications \citep{Esteves2015_Ag, Angerhausen2015_Ag, Heng2021_Ag, Morris2024_Ag}. We selected values from the works of \cite{Esteves2015_Ag} and \cite{Morris2024_Ag} as they performed self-consistent analysis accounting for the contribution of thermal emission in the planetary dayside flux\footnote{\cite{Esteves2015_Ag} accounts for the thermal emission by assuming immediate re-radiation (i.e. zero heat redistribution or $\varepsilon=0$ in Eqs.~\ref{eq:tday_tnight} and \ref{eq:tday_tnight2}, which corresponds to a so-called greenhouse factor $f=\frac{2}{3}$) and Lambertian reflection of the planetary surface ($A_B=\frac{3}{2}\,A_g$). \cite{Morris2024_Ag} fits for thermal emission using spherical harmonics using values of the greenhouse factor $f \sim 0.7$ and deriving indirectly the Bond albedo $A_B$.}.

From our analysis, we derived a strong upper limit on the contribution of the reflected light of \planet{} in the \tess{} passband. Despite its significant overlap with the \cheops{} passband \citep{Deline2020_CHEOPS}, the latter has a much less stringent constraint with a hint of detection in \cheops{} at $2.6\,\sigma$. We explored the scattering properties of the atmosphere that could produce reflected light signals consistent with our computed geometric albedo values in both passbands.
Since the dayside of \planet{} is likely too hot for condensates to be formed, the only reflective component of the atmosphere would be Rayleigh scattering by molecular and atomic hydrogen as well as helium. We estimated the geometric albedo of the dayside based on the retrieval results shown on the left-hand side of Fig. \ref{fig:retrieval_abundances} and the corresponding median temperature profile depicted in Fig. \ref{fig:bright_temp}. Given the temperature profile and the stated element abundances, we calculated the chemical composition (assuming chemical equilibrium) and the {atmospheric absorption coefficients for the chemical species described above} and Rayleigh scattering {contributions by atomic and molecular hydrogen as well as helium}. For the pressure, we chose 0.1 bar, which is the region where the contribution function peak, as shown in Fig.~\ref{fig:bright_temp}. Based on the computed absorption and scattering coefficients we then obtained the wavelength-dependent geometric albedo $A_g$ following the analytic solution derived by \citet{Heng2021_Ag}. {The corresponding stellar and planet fluxes were then finally} integrated over the \cheops{} and \tess{} passbands to estimate the $A_g$-contribution of the corresponding measurements. 

\begin{figure}
    {
    \centering
    \includegraphics[width=\linewidth,clip]{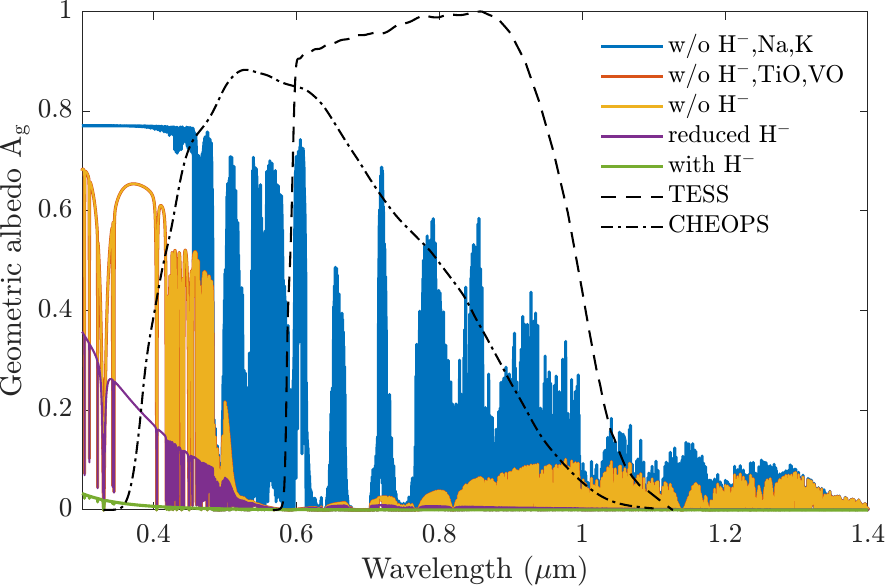}
    \caption{Estimated geometric albedo of \planet{} based on the retrieval results in comparison to the passbands of \cheops{} and \tess{}. {The green line shows the geometric albedo with an \ch{H-} abundance derived by assuming chemical equilibrium {(extremely low $A_g$ values, reaching 0.03 at short wavelengths)}. For the purple line, this abundance is artificially reduced by 98\% {(yielding $A_g$ values comparable to our results in the \cheops{} and \tess{} passbands)}, while for the yellow case the \ch{H-} continuum is entirely removed as an opacity source. {We further explored the effect of removing} TiO and VO (red) as well as {Na and K} (blue).} {The impact of the former is negligible and the red curve lies hidden behind the yellow model. The latter reveals however more significant scattering spectral features, especially in the \cheops{} and \tess{} passbands.} The strong increase of $A_g$ towards lower wavelengths is caused by the~$\lambda^{-4}$-dependence of Rayleigh scattering.}
    \label{fig:geometric_albedo}
    }
\end{figure}

The wavelength-dependent albedo and the passbands are shown in Fig. \ref{fig:geometric_albedo}. The atmospheric species that decisively affects the visible wavelength region for a hot planet such as \planet{} is the hydrogen anion~\ch{H-}. Its very strong continuum absorption tends to produce black-body like spectra with only very little contributions by scattering (see \citealt{Kitzmann2018}, for example). As the yellow curve in Fig.~\ref{fig:geometric_albedo} suggests, the geometric albedo of \planet{} would be quite low. The albedo still increases towards lower wavelengths due to the $\lambda^{-4}$-dependence of Rayleigh scattering, such that the contributions by scattering are higher at the blue end of the CHEOPS passband than in the one of TESS. However, with a passband-integrated geometric albedo for CHEOPS of~0.0013, the estimated value {remained} below the retrieved value of {$A_g^\text{\cheops{}}=0.027\pm0.011$.}

Since the most important species affecting the shortwave albedo is \ch{H-} {(see Fig.~\ref{fig:ag_vs_hminus})}, we performed two additional calculations that are depicted in Fig. \ref{fig:geometric_albedo}. In the first we artificially removed \ch{H-} (blue line). This leads to a strong increase in the geometric albedo, producing a CHEOPS $A_g$ value of~0.09, while the corresponding value for TESS would be~{0.017}. In the second additional calculation, the chemical-equilibrium abundance of \ch{H-} was reduced by 98\% (orange curve), which yields $A_g^\text{\cheops{}}={0.028}$ and $A_g^\text{\tess{}}=0.004$, more comparable to the retrieval results.

{Besides \ch{H-}, other short-wave absorbers, such as TiO and VO or the alkali metals Na and K can also impact the amount of scattered light and, thus, the geometric albedo. For the case of no \ch{H-} absorption, we also show the resulting geometric albedos when removing these species as opacity sources in Fig.~\ref{fig:ag_vs_hminus}. As the results suggest, the impact of TiO and VO on the geometric albedo is negligible for the atmosphere of \planet{}. The alkali metals, however, clearly have a stronger impact. When removing them, in addition to \ch{H-}, we obtain the analytical limit of 0.75 for the geometric albedo in case of Rayleigh scattering.

However, the impact of these species can only be noticed when \ch{H-} is not present. In Fig. \ref{fig:geometric_albedo_add} of Appendix \ref{sec:appendix_albedo} we show the impact of the additional shortwave absorbers on $A_g$ as a function of \ch{H-}. These results suggest that as long \ch{H-} is not removed entirely, the effect of all of these additional species on the geometric albedo is minimal.}

A \ch{H-} abundance lower than the one expected from chemical equilibrium can clearly be caused by non-equilibrium effects, including, for example, photochemistry. Furthermore, our estimated value of $A_g$ is based on a one-dimensional description. In the real three-dimensional atmosphere, the abundance of \ch{H-} should be expected to change considerably across the visible dayside since its abundance is very strongly temperature-dependent.

\section{Conclusion} \label{sec:conc}

        We performed a joint analysis of the phase-curve and occultation observations of the planet \planet{}, covering six passbands from the visible to the mid-infrared with the space-based instruments \cheops{}, \tess{}, and the four IRAC channels of \spitzer{}. We included several signals related to the planet in our modelling; namely, the transit, the occultation, the phase signal, the ellipsoidal variations, the Doppler boosting, and the light travel time. We derived new ephemerides with precisions of 1\,second and 1.4\,millisecond for the time of inferior conjunction and the orbital period, respectively. We measured a planetary radius of $R_p=1.1926\pm0.0077\,R_J$, reaching a precision of 0.65\% (equivalent to 550\,km).
        
        We interpreted a timing inconsistency with the time of superior conjunction from the \jwst{} observation \citep{Coulombe2023} as a consequence of the slight orbital eccentricity of the planet. Using {priors on the eccentricity and the argument of periastron from radial-velocity measurements \citep{Triaud2010_HJs, Nymeyer2011_WASP-18b, Csizmadia2019_WASP-18b}, we substantially improved the constraints on the argument of periastron with a value of $\omega=261.9^{+1.3}_{-1.4}\,\deg$ for an eccentricity of $e=0.00852\pm0.00091$.}
        Despite the relative youthfulness of the \host{} system, having such a short-period massive planet on an eccentric orbit is peculiar and might require further study to understand the formation and stability of the planetary orbit.

        We investigated the occultation depth values across the 15~years covered by our data sets and we were not able to detect either short-term nor long-term variability.

From the upper limits on the nightside flux in the \cheops{} and \tess{} passbands, we derived a maximum nightside temperature estimate of $T_\text{night}<2000\,K$.
Modelling the dayside emission with general circulation models (GCMs) showed that the presence of magnetic-driven friction and super-solar metallicity are necessary to explain our observations.
We also fit the occultation depths, including the data from \jwst{} \citep{Coulombe2023}, with an atmospheric retrieval model. We were able to show that new \spitzer{} data at 4.5\,\um{} reveal super-solar carbon and oxygen abundances, with significant \ch{CO}/\ch{CO2} absorption, and a steep inverted temperature gradient throughout the planetary atmosphere. Overall, we found that the retrievals provided a good fit to the observations. However, the difference in brightness temperature inferred from the CHEOPS and TESS occultation depths was not captured by any of the models. This could be indicative of missing physics in the retrievals, such as the omission of reflected stellar light or atmospheric variations across latitude and longitude. We note that while GCMs do help in investigating the 3D effects of the physics, they are not strict fits to the data. Ultimately, future phase-curve or dayside-mapping observations will be critical to better constrain the spatial variation of the physical properties of \planet{}, allowing us to fine-tune our GCMs and construct more sophisticated retrieval models.

        Finally, we explored the potential contribution of reflected light to the dayside brightness in the \cheops{} and \tess{} passbands. We strongly constrained the geometric albedo in \tess{} with a 3-$\sigma$ upper limit of $A_g^\text{\tess{}}<0.017$. In parallel, the \cheops{} passband showed a hint of flux excess at 2.6\,$\sigma$ that corresponds to a geometric albedo $A_g^\text{\cheops{}}=0.027\pm0.011$. This result suggests the presence of a very weak reflectivity of the planetary atmosphere.
        We modelled the scattering properties of \planet{} to match our derived $A_g$ values. We quantitatively reproduced this behaviour as the result of Rayleigh scattering in the presence of continuum absorption by H$^-$. The abundance of H$^-$ needs to deviate from chemical equilibrium to match the geometric albedo in the \cheops{} and \tess{} passbands, which can be caused by non-equilibrium effects (e.g. photochemistry) or significant temperature variations across the planetary dayside.

\addvspace{12pt}
\begin{spacing}{0.81}
{\noindent\tiny\itshape
Data availability.\\
}
{\tiny
Raw and detrended light curves are available at the CDS via anonymous ftp to \href{ftp://cdsarc.u-strasbg.fr}{cdsarc.u-strasbg.fr} (\href{ftp://130.79.128.5}{130.79.128.5}) or via \url{http://cdsweb.u-strasbg.fr/cgi-bin/qcat?J/A+A/}.
}
\end{spacing}

\begin{acknowledgements}\\
        This project has received funding from the European Research Council (ERC) under the European Union's Horizon 2020 research and innovation programme (project {\sc Four Aces}, grant agreement No. 724427).  It has also been carried out in the frame of the National Centre for Competence in Research ``PlanetS'' supported by the Swiss National Science Foundation (SNSF). A.De. acknowledges the financial support of the SNSF.\\
        P.E.C. is funded by the Austrian Science Fund (FWF) Erwin Schroedinger Fellowship, program J4595-N.\\
        LCa and CHe acknowledge financial support from the \"Osterreichische Akademie 1158 der Wissenschaften and from the European Union H2020-MSCA-ITN-2019 1159 under Grant Agreement no. 860470 (CHAMELEON). Calculations were performed using supercomputer resources provided by the Vienna Scientific Cluster (VSC).\\
        B.-O. D. acknowledges support from the Swiss State Secretariat for Education, Research and Innovation (SERI) under contract number MB22.00046.\\
        ML acknowledges support of the Swiss National Science Foundation under grant number PCEFP2\_194576.\\
        ABr was supported by the SNSA.\\
        MNG is the ESA CHEOPS Project Scientist and Mission Representative, and as such also responsible for the Guest Observers (GO) Programme. MNG does not relay proprietary information between the GO and Guaranteed Time Observation (GTO) Programmes, and does not decide on the definition and target selection of the GTO Programme.\\
        S.C.C.B. acknowledges support from FCT through FCT contracts nr. IF/01312/2014/CP1215/CT0004.\\
        GBr, VSi, LBo, VNa, IPa, GPi, RRa, and GSc acknowledge support from CHEOPS ASI-INAF agreement n. 2019-29-HH.0.\\
        The Belgian participation to CHEOPS has been supported by the Belgian Federal Science Policy Office (BELSPO) in the framework of the PRODEX Program, and by the University of Liège through an ARC grant for Concerted Research Actions financed by the Wallonia-Brussels Federation.\\
        S.G.S. acknowledge support from FCT through FCT contract nr. CEECIND/00826/2018 and POPH/FSE (EC).\\
        The Portuguese team thanks the Portuguese Space Agency for the provision of financial support in the framework of the PRODEX Programme of the European Space Agency (ESA) under contract number 4000142255.\\
        TWi acknowledges support from the UKSA and the University of Warwick.\\
        This project has received funding from the Swiss National Science Foundation for project 200021\_200726. It has also been carried out within the framework of the National Centre of Competence in Research PlanetS supported by the Swiss National Science Foundation under grant 51NF40\_205606. The authors acknowledge the financial support of the SNSF.\\
        YAl acknowledges support from the Swiss National Science Foundation (SNSF) under grant 200020\_192038.\\
        We acknowledge financial support from the Agencia Estatal de Investigación of the Ministerio de Ciencia e Innovación MCIN/AEI/10.13039/501100011033 and the ERDF “A way of making Europe” through projects PID2019-107061GB-C61, PID2019-107061GB-C66, PID2021-125627OB-C31, and PID2021-125627OB-C32, from the Centre of Excellence “Severo Ochoa” award to the Instituto de Astrofísica de Canarias (CEX2019-000920-S), from the Centre of Excellence “María de Maeztu” award to the Institut de Ciències de l’Espai (CEX2020-001058-M), and from the Generalitat de Catalunya/CERCA programme.\\
        DBa, EPa, and IRi acknowledge financial support from the Agencia Estatal de Investigación of the Ministerio de Ciencia e Innovación MCIN/AEI/10.13039/501100011033 and the ERDF “A way of making Europe” through projects PID2019-107061GB-C61, PID2019-107061GB-C66, PID2021-125627OB-C31, and PID2021-125627OB-C32, from the Centre of Excellence “Severo Ochoa'' award to the Instituto de Astrofísica de Canarias (CEX2019-000920-S), from the Centre of Excellence “María de Maeztu” award to the Institut de Ciències de l’Espai (CEX2020-001058-M), and from the Generalitat de Catalunya/CERCA programme.\\
        CBr and ASi acknowledge support from the Swiss Space Office through the ESA PRODEX program.\\
        ACC acknowledges support from STFC consolidated grant number ST/V000861/1, and UKSA grant number ST/X002217/1.\\
        ACMC acknowledges support from the FCT, Portugal, through the CFisUC projects UIDB/04564/2020 and UIDP/04564/2020, with DOI identifiers 10.54499/UIDB/04564/2020 and 10.54499/UIDP/04564/2020, respectively.\\
        This project was supported by the CNES.\\
        The Belgian participation to CHEOPS has been supported by the Belgian Federal Science Policy Office (BELSPO) in the framework of the PRODEX Program, and by the University of Liège through an ARC grant for Concerted Research Actions financed by the Wallonia-Brussels Federation.\\
        L.D. thanks the Belgian Federal Science Policy Office (BELSPO) for the provision of financial support in the framework of the PRODEX Programme of the European Space Agency (ESA) under contract number 4000142531.\\
        This work was supported by FCT - Funda\c{c}\~{a}o para a Ci\^{e}ncia e a Tecnologia through national funds and by FEDER through COMPETE2020 through the research grants UIDB/04434/2020, UIDP/04434/2020, 2022.06962.PTDC.\\
        O.D.S.D. is supported in the form of work contract (DL 57/2016/CP1364/CT0004) funded by national funds through FCT.\\
        MF and CMP gratefully acknowledge the support of the Swedish National Space Agency (DNR 65/19, 174/18).\\
        DG gratefully acknowledges financial support from the CRT foundation under Grant No. 2018.2323 “Gaseousor rocky? Unveiling the nature of small worlds”.\\
        M.G. is an F.R.S.-FNRS Senior Research Associate.\\
        KGI is the ESA CHEOPS Project Scientist and is responsible for the ESA CHEOPS Guest Observers Programme. She does not participate in, or contribute to, the definition of the Guaranteed Time Programme of the CHEOPS mission through which observations described in this paper have been taken, nor to any aspect of target selection for the programme.\\
        K.W.F.L. was supported by Deutsche Forschungsgemeinschaft grants RA714/14-1 within the DFG Schwerpunkt SPP 1992, Exploring the Diversity of Extrasolar Planets.\\
        This work was granted access to the HPC resources of MesoPSL financed by the Region Ile de France and the project Equip@Meso (reference ANR-10-EQPX-29-01) of the programme Investissements d'Avenir supervised by the Agence Nationale pour la Recherche.\\
        PM acknowledges support from STFC research grant number ST/R000638/1.\\
        This work was also partially supported by a grant from the Simons Foundation (PI Queloz, grant number 327127).\\
        NCSa acknowledges funding by the European Union (ERC, FIERCE, 101052347). Views and opinions expressed are however those of the author(s) only and do not necessarily reflect those of the European Union or the European Research Council. Neither the European Union nor the granting authority can be held responsible for them.\\
        GyMSz acknowledges the support of the Hungarian National Research, Development and Innovation Office (NKFIH) grant K-125015, a a PRODEX Experiment Agreement No. 4000137122, the Lend\"ulet LP2018-7/2021 grant of the Hungarian Academy of Science and the support of the city of Szombathely.\\
        V.V.G. is an F.R.S-FNRS Research Associate.\\
        JV acknowledges support from the Swiss National Science Foundation (SNSF) under grant PZ00P2\_208945.\\
        NAW acknowledges UKSA grant ST/R004838/1.\\
        \cheops{} is an ESA mission in partnership with Switzerland with important contributions to the payload and the ground segment from Austria, Belgium, France, Germany, Hungary, Italy, Portugal, Spain, Sweden, and the United Kingdom. The \cheops{} consortium would like to gratefully acknowledge the support received by all the agencies, offices, universities, and industries involved. Their flexibility and willingness to explore new approaches were essential to the success of this mission. \cheops{} data analysed in this article will be made available in the \cheops{} mission archive (\url{https://cheops.unige.ch/archive_browser/}).\\
        This paper includes data collected with the \tess{} mission, obtained from the MAST data archive at the Space Telescope Science Institute (STScI). Funding for the \tess{} mission is provided by the NASA Explorer Program. STScI is operated by the Association of Universities for Research in Astronomy, Inc., under NASA contract NAS 5–26555.\\
        This research has made use of the NASA/IPAC Infrared Science Archive, which is funded by the National Aeronautics and Space Administration and operated by the California Institute of Technology.\\
        This work has made use of data from the European Space Agency~(ESA) mission \gaia{} (\url{https://www.cosmos.esa.int/gaia}), processed by the \gaia{} Data Processing and Analysis Consortium~(DPAC, \url{https://www.cosmos.esa.int/web/gaia/dpac/consortium}). Funding for the DPAC has been provided by national institutions, in particular the institutions participating in the \gaia{} Multilateral Agreement.\\
        {We thank both referees for their insightful comments that helped improve the quality of this work.}
\end{acknowledgements}

\bibliographystyle{aa}
\bibliography{references}

\begin{appendix}

        {
        \section{\cheops{} observations}
            \begin{@twocolumnfalse}
                {\renewcommand{\arraystretch}{1.2}
                \begin{table*}[ht]
                \parbox{\textwidth}{
                        \caption{List of the \cheops{} observations.}
                        \label{tab:cheops_log}
                        \centering
                        \resizebox{\textwidth}{!}{
                        \begin{tabular}{ccccccc}
                                \hline\hline
                                \multirow{2}{*}{File key \tablefootmark{1}} & \multirow{2}{*}{UTC start \tablefootmark{2}} & \multirow{2}{*}{UTC end \tablefootmark{2}} & \multirow{2}{*}{Type} & \multirow{2}{*}{$N_\text{frames}$} & Efficiency \tablefootmark{3} & PSF location \\[.4ex]
                                & & & & & $\left[\%\right]$ & $\left(x, y\right)$ \\
                                \hline
                                \texttt{CH\_PR100016\_TG010701\_V0300} & 2020-09-05 02:23 & 2020-09-05 08:18 & Occultation & 277 & 65.0 & $\left(263, 842\right)$ \\
                                \texttt{CH\_PR100016\_TG010702\_V0300} & 2020-09-06 01:28 & 2020-09-06 07:24 & Occultation & 294 & 68.7 & $\left(263, 842\right)$ \\
                                \texttt{CH\_PR100016\_TG010703\_V0300} & 2020-09-22 01:37 & 2020-09-22 07:28 & Occultation & 292 & 69.4 & $\left(280, 828\right)$ \\
                                \texttt{CH\_PR100016\_TG010704\_V0300} & 2020-09-30 12:43 & 2020-09-30 19:01 & Occultation & 296 & 65.2 & $\left(280, 828\right)$ \\
                                \texttt{CH\_PR100016\_TG010705\_V0300} & 2020-10-01 11:15 & 2020-10-01 17:00 & Occultation & 292 & 70.7 & $\left(280, 828\right)$ \\
                                \texttt{CH\_PR100016\_TG010707\_V0300} & 2020-10-05 05:51 & 2020-10-05 11:16 & Occultation & 245 & 62.8 & $\left(280, 828\right)$ \\
                                \texttt{CH\_PR100016\_TG010708\_V0300} & 2020-10-08 00:47 & 2020-10-08 07:03 & Occultation & 315 & 69.8 & $\left(280, 828\right)$ \\
                                \texttt{CH\_PR100016\_TG010709\_V0300} & 2020-10-09 22:28 & 2020-10-10 04:00 & Occultation & 292 & 73.4 & $\left(280, 828\right)$ \\
                                \texttt{CH\_PR100016\_TG010710\_V0300} & 2020-10-16 12:29 & 2020-10-16 18:13 & Occultation & 307 & 74.3 & $\left(280, 828\right)$ \\
                                \texttt{CH\_PR100016\_TG010711\_V0300} & 2020-10-21 05:12 & 2020-10-21 11:05 & Occultation & 250 & 59.0 & $\left(280, 828\right)$ \\
                                \texttt{CH\_PR100016\_TG010712\_V0300} & 2020-10-24 01:30 & 2020-10-24 06:55 & Occultation & 279 & 71.5 & $\left(280, 828\right)$ \\
                                \texttt{CH\_PR100016\_TG010713\_V0300} & 2020-10-25 00:29 & 2020-10-25 06:13 & Occultation & 312 & 75.5 & $\left(280, 828\right)$ \\
                                \texttt{CH\_PR100016\_TG010714\_V0300} & 2020-10-26 21:27 & 2020-10-27 03:04 & Occultation & 291 & 72.0 & $\left(280, 828\right)$ \\
                                \texttt{CH\_PR100016\_TG010715\_V0300} & 2020-10-27 19:18 & 2020-10-28 01:19 & Occultation & 256 & 59.1 & $\left(280, 828\right)$ \\
                                \texttt{CH\_PR100012\_TG001201\_V0300} & 2021-08-18 23:30 & 2021-08-19 05:20 & Transit & 264 & 66.7 & $\left(291, 830\right)$ \\
                                \texttt{CH\_PR100016\_TG011801\_V0300} & 2021-09-17 14:30 & 2021-09-17 20:51 & Occultation & 273 & 59.9 & $\left(291, 830\right)$ \\
                                \texttt{CH\_PR100016\_TG011802\_V0300} & 2021-09-19 11:53 & 2021-09-19 17:41 & Occultation & 297 & 71.2 & $\left(291, 830\right)$ \\
                                \texttt{CH\_PR100012\_TG001202\_V0300} & 2021-09-19 23:34 & 2021-09-20 05:28 & Transit & 296 & 73.8 & $\left(291, 830\right)$ \\
                                \texttt{CH\_PR100012\_TG001203\_V0300} & 2021-09-26 13:37 & 2021-09-26 19:40 & Transit & 293 & 71.5 & $\left(291, 830\right)$ \\
                                \texttt{CH\_PR100016\_TG015101\_V0300} & 2021-10-03 13:07 & 2021-10-03 23:06 & Occultation & 464 & 64.6 & $\left(291, 830\right)$ \\
                                \texttt{CH\_PR100016\_TG015102\_V0300} & 2021-10-06 07:52 & 2021-10-06 18:09 & Occultation & 480 & 64.9 & $\left(291, 830\right)$ \\
                                \texttt{CH\_PR100016\_TG015103\_V0300} & 2021-10-14 19:14 & 2021-10-15 05:22 & Occultation & 484 & 66.4 & $\left(291, 830\right)$ \\
                                \texttt{CH\_PR100012\_TG001204\_V0300} & 2021-10-16 07:40 & 2021-10-16 14:10 & Transit & 276 & 62.6 & $\left(291, 830\right)$ \\
                                \texttt{CH\_PR100012\_TG001205\_V0300} & 2021-10-20 02:03 & 2021-10-20 10:36 & Transit & 374 & 64.5 & $\left(291, 830\right)$ \\
                                \texttt{CH\_PR100016\_TG015104\_V0300} & 2021-10-20 11:06 & 2021-10-20 21:05 & Occultation & 482 & 67.1 & $\left(291, 830\right)$ \\
                                \texttt{CH\_PR100016\_TG015105\_V0300} & 2021-10-21 09:36 & 2021-10-21 21:11 & Occultation & 546 & 65.5 & $\left(291, 830\right)$ \\
                                \texttt{CH\_PR100012\_TG001206\_V0300} & 2021-10-22 21:50 & 2021-10-23 04:20 & Transit & 303 & 68.7 & $\left(291, 830\right)$ \\
                                \texttt{CH\_PR100012\_TG001207\_V0300} & 2021-10-28 14:02 & 2021-10-28 21:58 & Transit & 349 & 64.9 & $\left(291, 830\right)$ \\
                                \texttt{CH\_PR100016\_TG015106\_V0300} & 2021-10-28 22:17 & 2021-10-29 08:26 & Occultation & 508 & 69.5 & $\left(291, 830\right)$ \\
                                \texttt{CH\_PR100012\_TG002701\_V0300} & 2022-09-04 05:42 & 2022-09-04 12:04 & Transit & 238 & 55.1 & $\left(291, 830\right)$ \\
                                \texttt{CH\_PR100012\_TG002702\_V0300} & 2022-09-29 16:21 & 2022-09-29 17:24 & --\tablefootmark{$\ast$} & 66 & 93.0 & $\left(291, 830\right)$ \\
                                \texttt{CH\_PR100012\_TG002703\_V0300} & 2022-10-02 12:04 & 2022-10-02 20:47 & Transit & 390 & 65.9 & $\left(291, 830\right)$ \\
                                \texttt{CH\_PR100012\_TG002704\_V0300} & 2022-10-12 21:04 & 2022-10-13 04:24 & Transit & 358 & 71.9 & $\left(291, 830\right)$ \\
                                \texttt{CH\_PR100016\_TG015701\_V0300} & 2022-10-13 04:53 & 2022-10-14 08:19 & Phase curve & 1297 & 65.7 & $\left(291, 830\right)$ \\
                                \texttt{CH\_PR100016\_TG015702\_V0300} & 2022-10-21 17:13 & 2022-10-22 03:27 & Occultation & 479 & 65.1 & $\left(291, 830\right)$ \\
                                \texttt{CH\_PR100012\_TG002705\_V0300} & 2022-10-23 04:47 & 2022-10-23 11:10 & Transit & 269 & 62.1 & $\left(291, 830\right)$ \\
                                \texttt{CH\_PR100016\_TG015703\_V0300} & 2022-10-26 09:51 & 2022-10-26 20:08 & Occultation & 483 & 65.3 & $\left(291, 830\right)$ \\
                                \texttt{CH\_PR100016\_TG015704\_V0300} & 2022-10-28 07:38 & 2022-10-28 20:01 & Occultation & 559 & 62.7 & $\left(291, 830\right)$ \\
                                \hline\hline
                        \end{tabular}
                        }
                        \tablefoot{
                        The time spent accumulating photons for each frame is referred to as the integration time or the exposure time, and is $t_\text{int}=53.0\,\text{s}$ for the transits (file key starting with \texttt{CH\_PR100012}) and $t_\text{int}=50.0\,\text{s}$ for the occultations and the phase curve (file key starting with \texttt{CH\_PR100016}). We note that the image read-out is performed in parallel of the next exposure leading to an effective data cadence equal to the integration time (duty cycle of 100\%).
                        \tablefoottext{1}{Each file key refers to a unique observation in the \cheops{} database.}
                        \tablefoottext{2}{UTC start and end are the starting and ending dates of the observation in UTC.}
                        \tablefoottext{3}{The efficiency represents the ratio between the observation time without interruptions (due to Earth occultation or SAA crossings) and the total observation duration.}
                        \tablefoottext{$\ast$}{Visit interrupted to perform a collision avoidance manoeuvre with the spacecraft.}
                        }
                        }
                \end{table*}
                }
                \clearpage
                \begin{figure*}[ht]
                    \parbox{\textwidth}{
                    \centering
                    \includegraphics[width=\hsize,trim={0cm 0cm 0cm 0cm},clip]{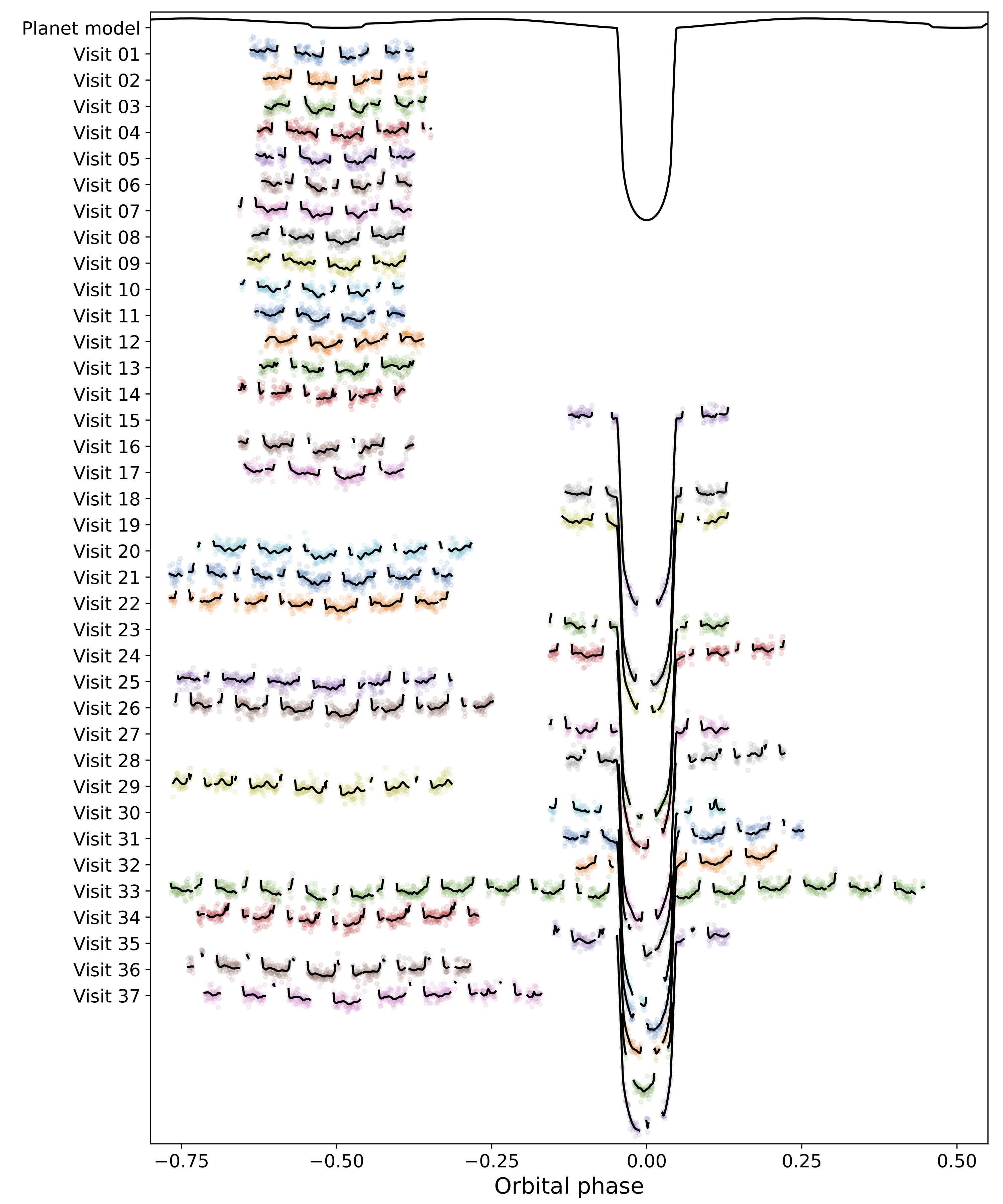}
                    \caption{\cheops{} raw light curves.
                    The 37~CHEOPS visits are displayed in coloured data points from top to bottom with their corresponding models shown as black lines. The planet model at the top (continuous black line) is the final model with only astrophysical signals (see Eq.~\ref{eq:final_model}).
                    }
                    \label{fig:cheops_raw_data}
                    }
                \end{figure*}
            \end{@twocolumnfalse}
            \clearpage
        }

        \section{\tess{} observations}
        {\renewcommand{\arraystretch}{1.1}
        \begin{table}[ht]
                        \caption{List of the times in $\text{BJD}_\text{TDB}$ used to discard \tess{} data points.}
                        \label{tab:tess_cut}
                        \centering
                        \begin{tabular}{cccc}
                                \hline\hline
                                Sector & Orbit & Start & End\\
                                \hline
                                02 & 12 & 2\,458\,368 & 2\,458\,369 \\
                                29 & 65 & 2\,459\,088 & 2\,459\,091 \\
                                29 & 66 & 2\,459\,111 & 2\,459\,114 \\
                                30 & 67-68 & 2\,459\,127 & 2\,459\,131 \\
                                69 & 145 & 2\,460\,182 & 2\,460\,184 \\
                                69 & 145 & 2\,460\,188 & 2\,460\,189 \\
                                69 & 145 & 2\,460\,192 & 2\,460\,194 \\
                                69 & 146 & 2\,460\,201 & 2\,460\,207 \\
                                \hline\hline
                        \end{tabular}
                \end{table}
                }
                                
                \begin{figure}[ht]
                        \centering
                        \includegraphics[width=.96\hsize,trim={0cm 0cm 0cm 0cm},clip]{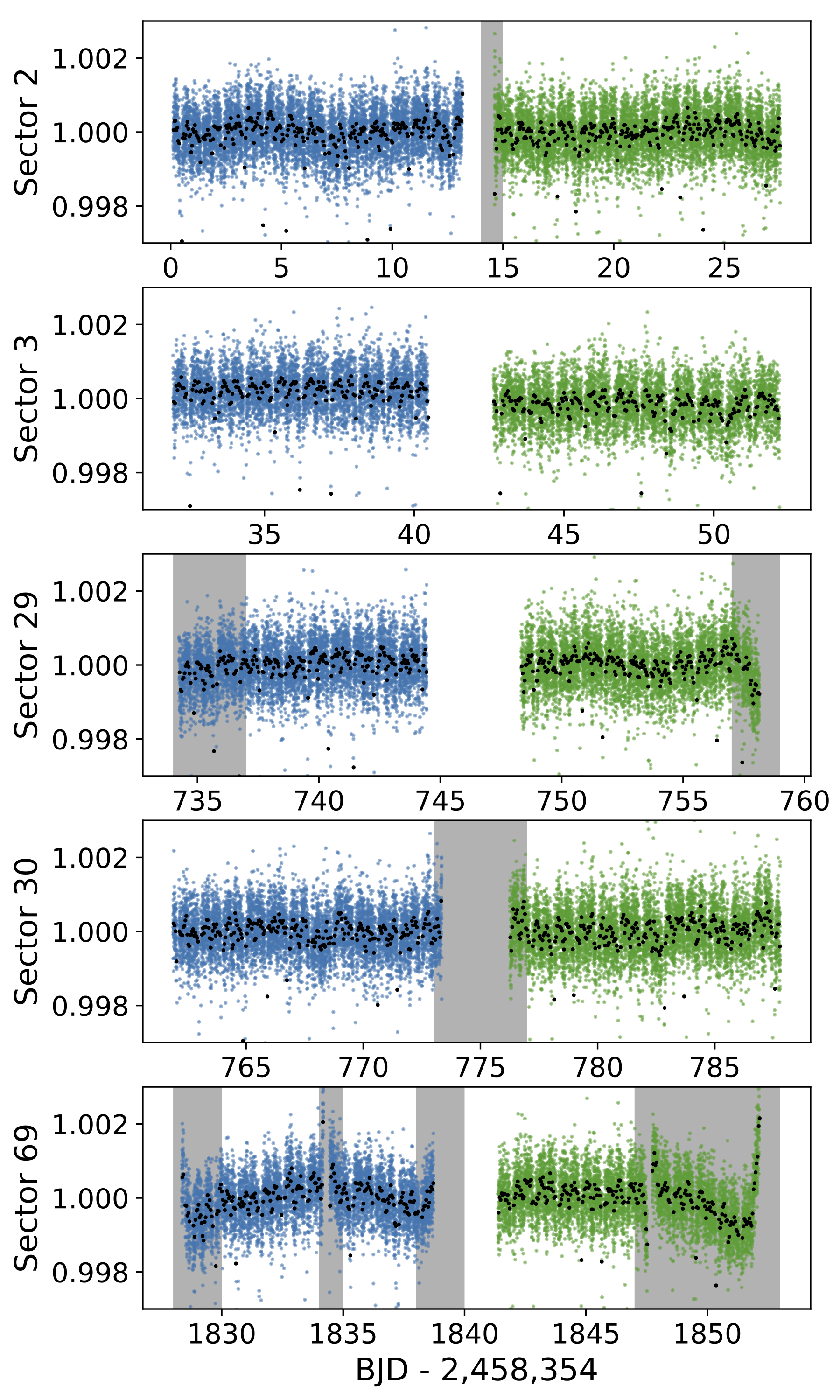}
                        \caption{\tess{} raw light curves extracted by the SPOC. The coloured points, blue and green, represent the photometric data of the two orbits of each sector. The black points are the data binned at a cadence of 30\,min. The shaded areas highlight the photometric ranges that were discarded due to the presence of remaining systematic trends. The times in $\text{BJD}_\text{TDB}$ of the different cuts are listed chronologically in Table~\ref{tab:tess_cut}. The light curve of sector 3 corresponds to the SAP flux (second panel) and the other sectors (2, 29, 30 and 69) are all PDCSAP fluxes.}
                        \label{fig:tess_raw_data}
                \end{figure}
            \vfill\break

        \section{\spitzer{} observations}

                {\renewcommand{\arraystretch}{1.3}
                \begin{table}[ht]
                    \caption{List of the \spitzer{} observations.}
                    \label{tab:spitzer_log}
                    \centering
                    \rotatebox{270}{
                    \begin{tabular}{ccccccc}
                        \hline\hline
                        Programme ID & PI & IRAC channels & Date & AOR keys & Type & Publications \\
                        \hline
                        \multirow{2}{*}{50517} & \multirow{2}{*}{J. Harrington} & 1, 3 & 2008-12-20 & 28775168 & Occultation & \multirow{2}{*}{\cite{Nymeyer2011_WASP-18b}} \\
                        & & 2, 4 & 2008-12-23 & 28775424 & Occultation & \\
                        \hline
                        \multirow{2}{*}{60185\tablefootmark{$\ast$}} & \multirow{2}{*}{P. F. L. Maxted} & 1 & 2010-01-24 & 38805504, 38805760 & Phase curve & \multirow{2}{*}{\cite{Maxted2013_WASP-18b}} \\
                        & & 2 & 2010-08-24 & 40269056, 40269312 & Phase curve & \\
                        \hline
                        \multirow{10}{*}{11099} & \multirow{10}{*}{L. R. Kreidberg} & 2 & 2015-09-08 & 53517824 & Occultation & \multirow{10}{*}{\cite{2023AJ....165..104D}} \\
                        & & 2 & 2015-09-09 & 53517568 & Occultation & \\
                        & & 2 & 2015-09-13 & 53517312 & Occultation & \\
                        & & 2 & 2015-09-14 & 53516800 & Occultation & \\
                        & & 2 & 2015-09-18 & 53516032 & Occultation & \\
                        & & 2 & 2015-09-20 & 53515776 & Occultation & \\
                        & & 2 & 2015-09-22 & 53515520 & Occultation & \\
                        & & 2 & 2015-09-23 & 53515008 & Occultation & \\
                        & & 2 & 2015-09-28 & 53517056 & Occultation & \\
                        & & 2 & 2015-09-30 & 53516288 & Occultation & \\
                        \hline\hline
                    \end{tabular}
                    }
                    \tablefoot{
                    \tablefoottext{$\ast$}{Data not included in the final analysis of this work due to the presence of strong red noise affecting significantly the quality of the data (more details in the text).}
                    }
                \end{table}
                }
                \clearpage

        {
        \section{Phase-curve model} \label{sec:pc_eq}
        
                We provide details about the expression of the phase-curve model used in this work and reported in Eq~\ref{eq:pc}.

                We start by defining the phase-curve signal of the planet as a function of the planetary phase angle $\alpha$:
                \begin{equation} \label{eq:pc_base}
                    F_\text{p} = a_\text{PC} \cos\!\left(\alpha\right) + b_\text{PC},
                \end{equation}
                where $a_\text{PC}$ and $b_\text{PC}$ are the phase-curve semi-amplitude and a constant offset, respectively. We defined those two parameters in such a way that the phase-curve signal $F_\text{p}$ reaches a maximum $F_\text{max}$ when $\alpha=0\deg$ and a minimum $F_\text{min}$ when $\alpha=180\deg$, which corresponds to rewriting the previous equation as follows:
                \begin{equation} \label{eq:pc_phase_angle}
                    F_\text{p} = \frac{1+\cos\!\left(\alpha\right)}{2}\left(F_\text{max}-F_\text{min}\right)+F_\text{min}.
                \end{equation}

                We then used the expression of the phase angle $\alpha$ as a function of the angular position of the planet $\theta$ and the orbital inclination $i$:
                \begin{equation} \label{eq:phase_angle}
                    \cos\!\left(\alpha\right) = -\sin\!\left(\theta\right)\,\sin\!\left(i\right).
                \end{equation}
                Here we have defined $\theta$ in such a way that the inferior conjunction occurs at $\theta=90\deg$. We note that when the planetary orbit is not viewed edge-on (i.e. $i\neq90\deg$), one has $0\deg<\alpha<180\deg$ and the phase-curve signal $F_\text{p}$ never reaches the values $F_\text{min}$ and $F_\text{max}$.

                The angular position of the planet can be written $\theta=\omega+\nu$, with $\omega$ and $\nu$ being the argument of periastron and the true anomaly, respectively. In order to account for potential phase shift between the superior conjunction ($\theta=270\deg$) and the phase-curve maximum value (e.g. due to a hotspot offset), we introduce the term $\Delta\phi$ in our expression of $\theta$:
                \begin{equation} \label{eq:theta_with_shift}
                    \theta = \omega + \nu - \Delta\phi.
                \end{equation}
                We chose to subtract the phase shift value, $\Delta\phi$, to ensure that a positive $\Delta\phi$ induces a positive time delay, which means that the phase-curve maximum will occur after the superior conjunction (e.g. westward hotspot offset).

                We obtain the phase-curve model reported in Eq.~\ref{eq:pc} by merging Eqs.~\ref{eq:theta_with_shift} and~\ref{eq:phase_angle} into Eq.~\ref{eq:pc_phase_angle}.
            \vfill\break
        }

        {
        \section{Ellipsoidal variations model} \label{sec:ev_eq}

                The ellipsoidal variations correspond to the photometric variability created by the rotation of a distorted star, which deformation is caused by the tidal forces of a close-by massive planet.

                In the work of \cite{Kopal1959_Binaries}, the light variation due to stellar distortion is derived in Section~{IV.2} of Chapter~{IV} as a series function of tesseral harmonics. Considering the tidal lag negligible, the tesseral harmonics reduce to Legendre polynomials and the light variation can be simplified (Eq.~{IV-2-37} of \citealt{Kopal1959_Binaries}) as shown below up to the third harmonic:
                \begin{equation}\begin{aligned} \label{eq:dl_kopal}
				    \delta\mathfrak{L} =\ 
                        &X_2^{(k)}\left(1+\frac{\beta_2}{4}\right)\left[\frac{v_1^{(2)}}{3}P_2\!\left(n_0\right)-w_1^{(2)}P_2\!\left(l_0\right)\right]\\
                        &- X_3^{(k)}\left(1+\frac{\beta_3}{10}\right)w_1^{(3)}P_3\!\left(l_0\right)\\
                        &+ \dots,
				\end{aligned}\end{equation}
                where:
                \begin{itemize}
				    \item $X_j^{(k)}$ are functions of the limb-darkening coefficients ($k=2$ for the linear law, $k=3$ for the quadratic law) detailed in Eqs.~IV-2-38 and IV-2-39 of \cite{Kopal1959_Binaries},
				    \item $\beta_j=\left[1+\eta_j\!\left(R_\star\right)\right]\,y$ from Eq.~IV-2-31 of \cite{Kopal1959_Binaries}, where $\eta_j\!\left(R_\star\right)\in\left[j-2; j+1\right]$ depends on the density distribution within the star (Eqs.~II-2-5 and II-2-6 of \citealt{Kopal1959_Binaries} with $a_1\equiv R_\star$), and $y$ is the gravity-darkening coefficient,
    				\item $P_n\!\left(x\right)$ is the $n^{th}$ Legendre polynomials, i.e. $P_2\!\left(x\right)=\frac{1}{2}\left(3x^2-1\right)$ and $P_3\!\left(x\right)=\frac{1}{2}\left(5x^3-3x\right)$,
    				\item $n_0=\cos\!\left(i\right)$ and $l_0=\cos\!\left(\psi\right)\sin\!\left(i\right)$, where $i$ is the orbital inclination (from the plane of the sky) and $\psi$ is the true anomaly reckoned from the moment of inferior conjunction\footnote{In \cite{Kopal1959_Binaries}, $\psi$ is defined as ``the true anomaly of the secondary component in the plane of the relative orbit, reckoned from the moment of superior conjunction (i.e., mid-primary minimum if the primary component is one of greater surface brightness)''. If one considers the planet as the secondary component and the star as the primary, then the mid-primary minimum actually corresponds to the inferior conjunction from the planetary perspective (i.e. the planet is in front of the star and $\omega+\nu=\pi/2$).}, i.e. $\psi=\omega+\nu-\pi/2$ and $l_0=\sin\!\left(\omega+\nu\right)\sin\!\left(i\right)$,
    				\item $v_1^{(2)}$ refers to the {rotational} distortion and will therefore be ignored (this term is time-independent),
    				\item $w_1^{(j)}$ refers to the {tidal} distortion, i.e. ellipsoidal variations, and $w_1^{(j)}=\Delta_j\frac{M_p}{M_\star}\left(\frac{R_\star}{a}\frac{1+e\cos\left(\nu\right)}{1-e^2}\right)^{j+1}$ from Eqs.~IV-2-33 and IV-2-56 of \cite{Kopal1959_Binaries} (with $a_1\equiv R_\star$, and $R\equiv a\frac{1-e^2}{1+e\cos\left(\nu\right)}$) where $\Delta_j=\frac{2j+1}{j+\eta_j\!\left(R_\star\right)}$ (Eq.~II-1-27 of \citealt{Kopal1959_Binaries}).
                \end{itemize}
                
                The two limits reported on $\eta_j\!\left(R_\star\right)$ depend on the density distribution within the star (Eqs.~II-2-4 to II-2-8 of \citealt{Kopal1959_Binaries}). The lower limit $j-2$ assumes a constant density distribution throughout the stellar interior. The upper limit $j+1$ represents the case where all the mass is concentrated at the center of the star. We follow \cite{Morris1985_EV} and thus \cite{Esteves2013_PC}, assuming the latter case to simplify the equations and obtain $\Delta_j=1$ and $\beta_j=\left(j+2\right)\,y$.

                When discarding the rotational distortion ($v_1^{(2)}=0$) and assuming a point-mass stellar density distribution ($\eta_j\!\left(R_\star\right)=j+1$), we can rewrite Eq.~\ref{eq:dl_kopal}, describing the light variation due to stellar distortion, as follows:
                \begin{equation}\begin{aligned} \label{eq:dl_kopal_ev}
				    \delta\mathfrak{L} =\ 
				    &X_2^{(k)}\left(1+y\right) \frac{M_p}{M_\star}\left[\frac{R_\star}{r\!\left(\theta\right)}\right]^{3} \frac{1-3\sin^2\!\left(\theta\right)\sin^2\!\left(i\right)}{2}\\
				    &X_3^{(k)}\left(1+\frac{y}{2}\right) \frac{M_p}{M_\star}\left[\frac{R_\star}{r\!\left(\theta\right)}\right]^{4} \frac{3\sin\!\left(\theta\right)\sin\!\left(i\right)-5\sin^3\!\left(\theta\right)\sin^3\!\left(i\right)}{2}\\
				    &+ \dots
                \end{aligned}\end{equation}
                Here, $r\!\left(\theta\right)=\frac{a\left(1-e^2\right)}{1+e\cos\left(\theta-\omega\right)}$ and $\theta=\omega+\nu=\psi+\pi/2$.

                We can further simplify the expression of the light variation in the case of a circular orbit ($e=0$) where one has $r\!\left(\theta\right)=a$. By discarding all but the time-dependent terms (i.e. the terms depending on~$\theta$), we can obtain from Eq.~\ref{eq:dl_kopal_ev} the following expression of $\delta\mathfrak{L}$:
				
				\begin{equation}\begin{aligned} \label{eq:dl_kopal_ev_circ}
					\delta\mathfrak{L} = \frac{M_p}{M_\star}\left(\frac{R_\star}{a}\right)^{\!3} \Bigg[
					&\frac{3}{4} X_2^{(k)} \left(1+y\right) \sin^2\!\left(i\right) \cos\!\left(2\theta\right)\\
					&+ \frac{3}{4} X_3^{(k)}\left(2+y\right) \frac{R_\star}{a} \left(1-\frac{5}{4}\sin^2\!\left(i\right)\right) \sin\!\left(i\right) \sin\!\left(\theta\right)\\
					&+ \frac{5}{16} X_3^{(k)}\left(2+y\right) \frac{R_\star}{a} \sin^3\!\left(i\right) \sin\!\left(3\theta\right)\\
					&+ \dots \Bigg]
				\end{aligned}\end{equation}
				
				This expression is the one reported in Eqs.~1-3 of \cite{Morris1985_EV} where the higher order terms (represented by the dots) have been neglected\footnote{The angle $\phi$ of \cite{Morris1985_EV} is equivalent of the angle $\psi$ of \cite{Kopal1959_Binaries}, which means $\phi=\psi=\theta-\pi/2$. We also have the term $k_1$ of Eq.~3 of \cite{Morris1985_EV} equal to zero because we do not consider perturbations from a third body in the system.}.

                We can finally write the expression of the light variation of Eq.~\ref{eq:dl_kopal_ev_circ} in a more compact way:
                \begin{equation}
                    \delta\mathfrak{L} = A_\text{EV}\,\left[\cos\!\left(2\theta\right) - A_1\sin\!\left(\theta\right) + A_3\sin\!\left(3\theta\right) + \dots\right],
                \end{equation}
                where
                \begin{align}
					&A_\text{EV} = \frac{3}{4} X_2^{(k)} \left(1+y\right) \frac{M_p}{M_\star}\left(\frac{R_\star}{a}\right)^{\!3} \sin^2\!\left(i\right),\\
                    &A_1 = \frac{1}{4} \frac{X_3^{(k)}}{X_2^{(k)}} \frac{2+y}{1+y} \frac{R_\star}{a} \frac{5\sin^2\!\left(i\right)-4}{\sin\!\left(i\right)},\\
                    &A_3 = \frac{5}{12} \frac{X_3^{(k)}}{X_2^{(k)}} \frac{2+y}{1+y} \frac{R_\star}{a} \sin\!\left(i\right).
				\end{align}
                Here, we retrieve the expressions of the Eqs.~\ref{eq:ev}, \ref{eq:ev_amp}, \ref{eq:ev_amp1} and \ref{eq:ev_amp3} of Section~\ref{sssec:star_model}.

                Based on the Eqs.~IV-2-15 and IV-2-39 of \cite{Kopal1959_Binaries}, we can write the functions of the limb-darkening coefficients $X_j^{(k)}$ for the quadratic limb-darkening law, i.e. for $k=3$. For this, we also need to take into account that quadratic LD law used in \cite{Kopal1959_Binaries} is different from the commonly used law reported in \cite{Manduca1977_LD} and implemented in the Python code \batman{} \citep{batman}. Indeed, the LD law from \cite{Manduca1977_LD} is $\mathcal{I}\!\left(\mu\right)/\mathcal{I}_0=1-u_1\left(1-\mu\right)-u_2\left(1-\mu\right)^2$, whereas the one from \cite{Kopal1959_Binaries} is $\mathcal{I}\!\left(\mu\right)/\mathcal{I}_0=1-\upsilon_1\left(1-\mu\right)-\upsilon_2\left(1-\mu^2\right)$. By applying the transformations $\upsilon_1=u_1+2\,u_2$ and $\upsilon_2=-u_2=$ in Eqs.~IV-2-15 and IV-2-39 of \cite{Kopal1959_Binaries}, we obtain:
                \begin{align}
					&X_2^{(3)}=\frac{2}{5}\,\frac{15+u_1+2\,u_2}{6-2\,u_1-u_2}\\
					&X_3^{(3)}=\frac{1}{7}\,\frac{35\,u_1+22\,u_2}{6-2\,u_1-u_2}
				\end{align}
                from which the expressions of $\alpha_\text{EV}$ and $\beta_\text{EV}$ of Section~\ref{sssec:star_model} are derived (see Eqs.~\ref{eq:ev_aev} and \ref{eq:ev_bev}).
                Note that when $u_2=\upsilon_2=0$, we get the expression for the linear limb-darkening law reported in \cite{Morris1985_EV} and \cite{Esteves2013_PC}.

            \vfill\break
        }

        \section{Bond albedo and heat redistribution efficiency} \label{sec:ab_eps}
        
                Here, we briefly explain an analytical method detailed in \cite{Deline2022_WASP-189b} to estimate the Bond albedo $A_B$ and the heat redistribution $\varepsilon$.
                
                We define the average effective temperature of a planet $\tilde{T}_p$ by:
                \begin{equation}
                        \tilde{T}_p^4 = \frac{1}{4\pi}\,\iint_\text{planet}T_p^4\,d\Omega\,,
                \end{equation}
                where $T_p$ is the local effective temperature in the planet atmosphere and $d\Omega$ describes a surface element of the atmosphere.
                
                Assuming the planet is emitting as a black body at thermal equilibrium (i.e. absorbed flux equals emitted flux), we can write the following relationship with the average effective temperature of the planet:
                \begin{equation}
                        \tilde{T}_p^4 = \frac{1-A_B}{4\,\sigma_\text{SB}} \left(\frac{R_\star}{a}\right)^2 \int_{\lambda=0}^{+\infty}\!\mathcal{S}\!\left(\lambda, T_\star\right) d\lambda\,,
                \end{equation}
                where $A_B$ is the Bond albedo, $\sigma_\text{SB}$ is the Stefan-Boltzmann constant, $R_\star$ is the stellar radius, $a$ is the semi-major axis, and $\mathcal{S}\!\left(\lambda, T_\star\right)$ is the flux emission spectrum of the star as a function of its temperature, $T_\star$, and the wavelength, $ \lambda$.
                
                The parametrisation of \cite{Cowan2011_AB_eps} allows one to estimate the dayside and nightside effective temperatures as functions of the heat redistribution efficiency $\varepsilon$ and $\tilde{T}_p$:
                \begin{align}
                        &T_\text{day} = \left(\frac{8-5\varepsilon}{3}\right)^\frac{1}{4}\tilde{T}_p\,, \label{eq:tday_tnight}\\
                        &T_\text{night} = \varepsilon^\frac{1}{4}\,\tilde{T}_p\,. \label{eq:tday_tnight2}
                \end{align}
                
                Therefore, from these equations, we can compute a relationship between $A_B$ and $\varepsilon$ for a given dayside or nightside temperature, and constrain the range of possible values for these two parameters. Usually, the measured and constrained value is the dayside temperature, $T_\text{day}$, from the occultation depth, and we derive the range of values $A_B=f\!\left(\varepsilon, T_\text{day}\right)$. In the framework of \cite{Cowan2011_AB_eps}, the maximum dayside temperature is reached when $A_B=\varepsilon=0,$ and one obtains
                \begin{equation}
                        T^4_\text{day, max} = \frac{2}{3\,\sigma_\text{SB}}\,\left(\frac{R_\star}{a}\right)^2\,\int_{\lambda=0}^{+\infty}\!\mathcal{S}\!\left(\lambda, T_\star\right) d\lambda\,.
                \end{equation}
            \vfill\break

        {
        \section{Detailed {\texttt{Exorad}} description}
        \label{sec: GCM_Detail}
        
        The {\texttt{Exorad}} GCM {used the \texttt{MITgcm} dynamical core to} solve the hydrostatic primitive equations \citep[e.g.][]{Showman2009} on a rotating sphere in an Arakawa C-type cubed-sphere grid that spans in the horizontal plane $128\times 64$ cells in longitude and latitude, respectively. In vertical direction it follows the formalism established in \citet{Carone2020}: in the vertical direction, 41 logarithmically spaced grid cells between $10^{-5}$~bar and 100~bar are combined with six linearly spaced grid cells between 100~bar and 700~bar, resulting in 47 vertical cells. 
        
        Following \citet{Showman2009}, a fourth-order Shapiro filter with dampening time scale $\tau_{shap}=25$~s is used to remove small grid scale noise after each time step from the horizontal velocity fields\footnote{We note that $\tau_{shap}$ corresponds to the dissipation time $\tau_{\nu}$ used in \citet{Heng2011} to compare horizontal dissipation in different dynamical cores.}. The GCM is stabilised against non-physical gravity wave reflection on top of the modelling domain by implementing a sponge layer between $10^{-4}$ and $10^{-5}$. The zonal horizontal velocity $u$ is damped by a Rayleigh friction term towards its longitudinally averaged mean $\bar{u}$ via:
        \begin{equation}
        \frac{\rm du}{\rm dt} = - k \left(u-\bar{u}\right)    
        ,\end{equation}
        where $t$ is the time and $k$ is the strength of the Rayleigh
        friction applied within the sponge layer, depending on pressure $p$ by:
        \begin{equation}
        k = k_{\rm top} \cdot \rm max \left[0,1- \left(\frac{p}{p_{\rm sponge}}\right)^2 \right]^{2}
        \end{equation}
        
        The control parameters $p_{\rm sponge}$ and $k_{\rm top}$ determine the position and strength of the applied Rayleigh friction in the sponge layer. In this paper, the default values $k_{\rm top} = 20$~days$^{-1}$ and $p_{\rm sponge} =10^{-4}$~bar are used.
        
        To stabilise the deep layers, basal drag is applied to the zonal $u$ and meridional wind velocity $v$ in pressure layers deeper than 400~bar via:
        \begin{align}
        &\frac{\rm du}{\rm dt} = -k_{\rm deep}\cdot u,\\
        &\frac{\rm dv}{\rm dt} = -k_{\rm deep}\cdot v,
        \end{align}
        where the control parameter $k_{\rm deep}$ is defined as 
        \begin{equation}
        k_{\rm deep} = k_{\rm bottom} \cdot \rm max \left[0, \frac{p-490\,\rm bar}{700\,\rm bar - 490\,\rm  bar} \right]
        \end{equation}
        with $k_{\rm bottom}=1$~day$^{-1}$.
        
        The model was run with a dynamical timestep $\Delta t=25$~s for 1000~days simulation time to ensure that the temperature structure does not evolve anymore in the modelling domain. Fluxes are recalculated every fourth dynamical timestep. 
        
        Performance tests for the sponge layer and basal drag can be found in \citet{Carone2020}. {Full radiative transfer was established with the \texttt{expeRT/MITgcm}-branch of the \texttt{ExoRad} framework.} A more detailed description of the radiative transfer implementation and performance tests can be found in \citet{Schneider2022}.
        }

        \clearpage
        \section{Additional geometric albedo calculations}
        \label{sec:appendix_albedo}

        \begin{figure}[h]
        { 
        {
          \centering
          \includegraphics[width=0.9\linewidth,clip]{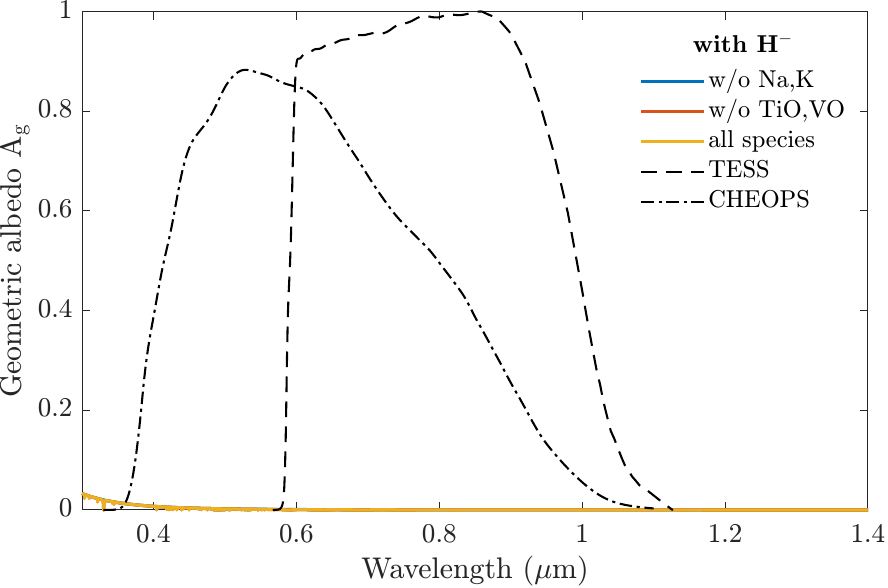}
          \includegraphics[width=0.9\linewidth,clip]{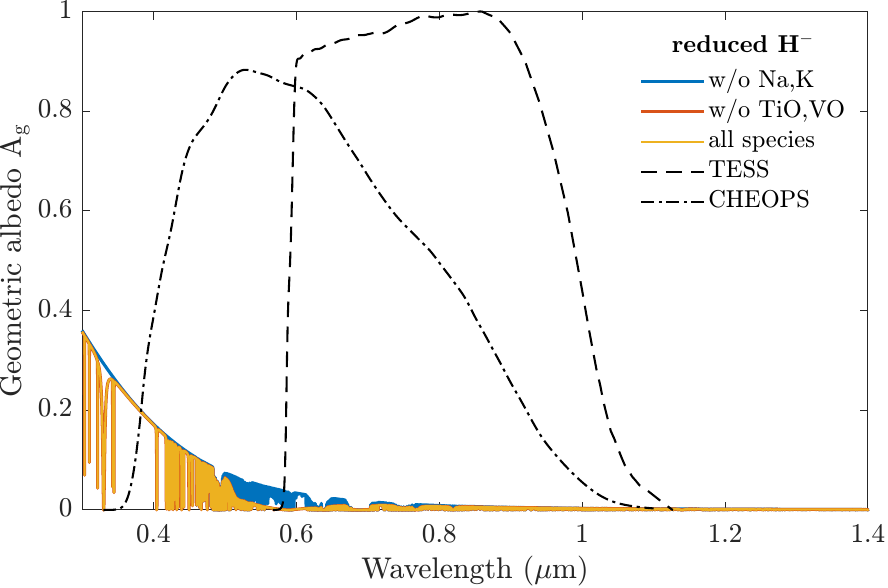}
          \includegraphics[width=0.9\linewidth,clip]{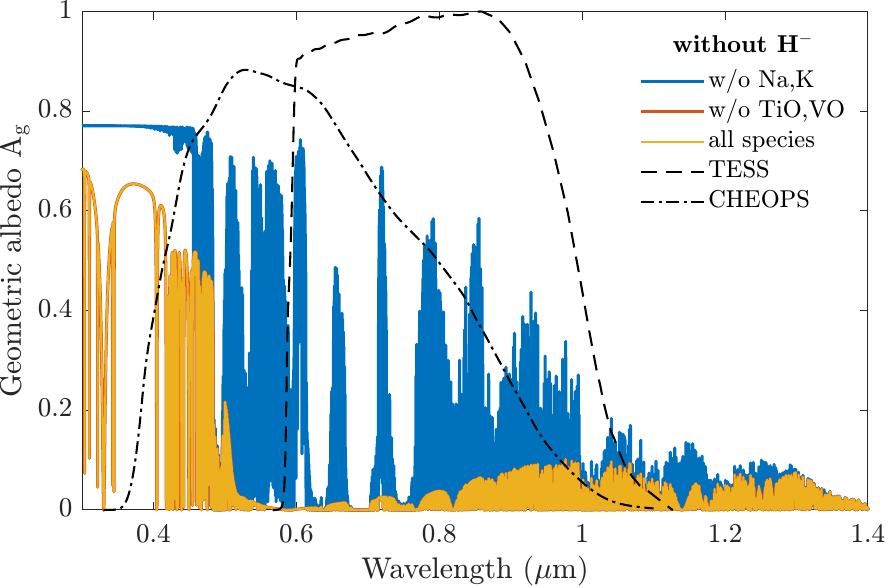}
          \caption{Estimated geometric albedo of \planet{} based on the retrieval results in comparison to the passbands of \cheops{} and \tess{}. The top panel shows the geometric albedo with the \ch{H-} abundance derived by assuming chemical equilibrium and for several species removed as opacity source {(note that all curves overlap as \ch{H-} dominates and other opacity sources are negligible)}. In the middle panel, \ch{H-} abundance is artificially reduced by 98\%, for which $A_g$ values are comparable to our results in the \cheops{} and \tess{} passbands (i.e. the red and yellow curves overlap, but the effect of removing Na and K starts to appear around 600\,nm).
          {In the lower panel,} the \ch{H-} continuum is entirely removed{, and the removal of alkali (Na and K) reveals even more strongly the scattering features, while the effect of TiO and VO remains negligible (overlap of red and yellow models)}. The strong increase of $A_g$ towards lower wavelengths is caused by the $\lambda^{-4}$-dependence of Rayleigh scattering. In the last case, when Na and K are removed as well, the geometric albedo converges towards its analytical limit of 0.75 for Rayleigh scattering.
          }
          \label{fig:geometric_albedo_add}
        }
        }
        \end{figure}
 
        \begin{figure}[ht]
        {
        \centering
        \includegraphics[width=\linewidth,clip]{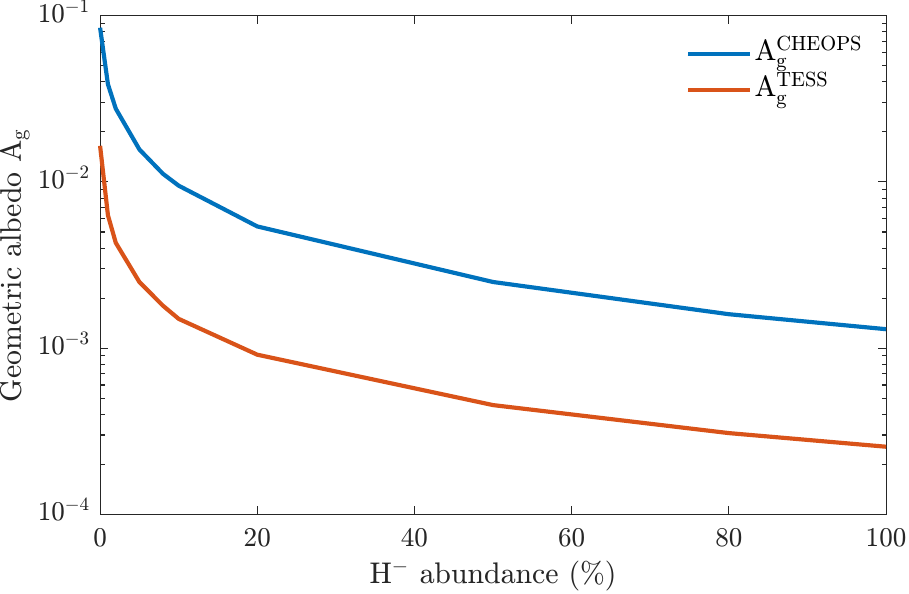}
        \caption{Estimated geometric albedo of \planet{} as a function of the abundance of \ch{H-}. The $A_g$ values are computed based on the retrieval results in comparison to the passbands of \cheops{} and \tess{}. The \ch{H-} abundance that best matches our measurements reported in Section~\ref{ssec:retrieval} lies around 2\% of the expected chemical equilibrium value (100\% on the x-axis), i.e. a reduction of 98\%.}
        \label{fig:ag_vs_hminus}
        }
        \end{figure}

        \clearpage
        \section{Joint-fit values of the systematic parameters}

                \begin{@twocolumnfalse}
                {\renewcommand{\arraystretch}{1.2}
                \begin{table*}[ht]
                    \parbox{\textwidth}{
                    \caption{Values of the systematic parameters for \cheops{}.}
                    \label{tab:cheops_syst}
                    \centering
                    {
                    \begin{tabular}{lccc}
                        \hline\hline
                        \multirow{2}{*}{Background} & $a_\text{bkg}$ [ppm] & $b_\text{bkg}$ [ppm] \\
                         & ${515}_{-{187}}^{+{186}}$ & ${695}\pm{96}$ \\
                        \hline\hline
                        \multirow{2}{*}{Roll angle} & $\sigma_\text{GP}$ [ppm] & $\rho_\text{GP}$ [$\deg$] \\
                         & ${107.3}_{-{5.8}}^{+{6.4}}$ & ${23.8}_{-{2.5}}^{+{2.8}}$ \\
                        \hline\hline
                        & $\sigma_w$ [ppm] & $f_0$ & $c_1$ [day$^{-1}$] \\
                        \hline
                        Visit 01 & ${120}_{-{30}}^{+{26}}$ & ${0.999843}\pm{0.000054}$ & -- \\
                        Visit 02 & ${126}_{-{27}}^{+{24}}$ & ${0.999802}\pm{0.000053}$ & -- \\
                        Visit 03 & ${86}_{-{50}}^{+{31}}$ & ${0.999789}_{-{0.000052}}^{+{0.000051}}$ & ${0.00067}_{-{0.00020}}^{+{0.00021}}$ \\
                        Visit 04 & ${126}_{-{27}}^{+{24}}$ & ${0.999791}\pm{0.000050}$ & -- \\
                        Visit 05 & ${138}_{-{25}}^{+{23}}$ & ${0.999806}\pm{0.000051}$ & -- \\
                        Visit 06 & ${20}_{-{58}}^{+{53}}$ & ${0.999786}_{-{0.000051}}^{+{0.000052}}$ & -- \\
                        Visit 07 & ${90}_{-{43}}^{+{29}}$ & ${0.999778}_{-{0.000049}}^{+{0.000050}}$ & -- \\
                        Visit 08 & ${122}_{-{27}}^{+{24}}$ & ${0.999775}\pm{0.000050}$ & -- \\
                        Visit 09 & ${128}_{-{25}}^{+{23}}$ & ${0.999806}_{-{0.000050}}^{+{0.000051}}$ & ${-0.00088}_{-{0.00021}}^{+{0.00022}}$ \\
                        Visit 10 & ${71}_{-{65}}^{+{39}}$ & ${0.999741}_{-{0.000051}}^{+{0.000050}}$ & -- \\
                        Visit 11 & ${26}_{-{57}}^{+{49}}$ & ${0.999764}\pm{0.000049}$ & -- \\
                        Visit 12 & ${114}_{-{28}}^{+{24}}$ & ${0.999808}\pm{0.000049}$ & -- \\
                        Visit 13 & ${110}_{-{31}}^{+{26}}$ & ${0.999780}_{-{0.000050}}^{+{0.000049}}$ & -- \\
                        Visit 14 & ${49}_{-{65}}^{+{46}}$ & ${0.999777}_{-{0.000051}}^{+{0.000050}}$ & -- \\
                        Visit 15 & ${119}_{-{29}}^{+{26}}$ & ${1.000084}\pm{0.000055}$ & -- \\
                        Visit 16 & ${22}_{-{56}}^{+{52}}$ & ${0.999740}_{-{0.000051}}^{+{0.000052}}$ & -- \\
                        Visit 17 & ${87}_{-{50}}^{+{31}}$ & ${0.999772}\pm{0.000051}$ & -- \\
                        Visit 18 & ${101}_{-{33}}^{+{26}}$ & ${1.000101}\pm{0.000051}$ & -- \\
                        Visit 19 & ${101}_{-{33}}^{+{26}}$ & ${1.000062}\pm{0.000050}$ & -- \\
                        Visit 20 & ${110}_{-{23}}^{+{20}}$ & ${0.999665}\pm{0.000049}$ & -- \\
                        Visit 21 & ${145}\pm{18}$ & ${0.999678}_{-{0.000049}}^{+{0.000050}}$ & -- \\
                        Visit 22 & ${78}_{-{40}}^{+{26}}$ & ${0.999679}_{-{0.000049}}^{+{0.000048}}$ & -- \\
                        Visit 23 & ${87}_{-{47}}^{+{30}}$ & ${1.000013}_{-{0.000049}}^{+{0.000050}}$ & -- \\
                        Visit 24 & ${75}_{-{50}}^{+{29}}$ & ${0.999891}\pm{0.000049}$ & -- \\
                        Visit 25 & ${73}_{-{46}}^{+{28}}$ & ${0.999661}_{-{0.000050}}^{+{0.000049}}$ & -- \\
                        Visit 26 & ${100}_{-{23}}^{+{20}}$ & ${0.999664}\pm{0.000049}$ & -- \\
                        Visit 27 & ${94}_{-{38}}^{+{27}}$ & ${1.000069}\pm{0.000049}$ & -- \\
                        Visit 28 & ${145}_{-{22}}^{+{21}}$ & ${0.999895}_{-{0.000050}}^{+{0.000051}}$ & -- \\
                        Visit 29 & ${176}\pm{17}$ & ${0.999659}\pm{0.000049}$ & -- \\
                        Visit 30 & ${80}_{-{62}}^{+{37}}$ & ${0.999937}\pm{0.000056}$ & -- \\
                        Visit 31 & ${72}_{-{49}}^{+{29}}$ & ${0.999928}\pm{0.000049}$ & ${0.00037}\pm{0.00012}$ \\
                        Visit 32 & ${35}_{-{56}}^{+{43}}$ & ${0.999888}\pm{0.000050}$ & ${0.00101}_{-{0.00014}}^{+{0.00015}}$ \\
                        Visit 33 & ${100}_{-{14}}^{+{13}}$ & ${0.999643}\pm{0.000048}$ & ${0.00012}\pm{0.00002}$ \\
                        Visit 34 & ${93}_{-{28}}^{+{23}}$ & ${0.999610}_{-{0.000049}}^{+{0.000050}}$ & -- \\
                        Visit 35 & ${64}_{-{65}}^{+{38}}$ & ${1.000089}_{-{0.000050}}^{+{0.000051}}$ & ${0.00117}_{-{0.00020}}^{+{0.00019}}$ \\
                        Visit 36 & ${47}_{-{57}}^{+{37}}$ & ${0.999669}_{-{0.000048}}^{+{0.000049}}$ & -- \\
                        Visit 37 & ${103}_{-{24}}^{+{20}}$ & ${0.999605}\pm{0.000050}$ & -- \\
                        \hline\hline
                    \end{tabular}
                    \tablefoot{
                    The background parameters $a_\text{bkg}$ and $b_\text{bkg}$ are the ones defined in Eq.~\ref{eq:bkg}.
                    The roll-angle GP hyperparameters $\sigma_\text{GP}$ and $\rho_\text{GP}$ are the standard deviation of the process and the correlation scale in roll-angle unit {($\deg$)}, respectively (see Section~\ref{ssec:systematics}).
                    The parameters $\sigma_w$, $f_0$ and $c_1$ are the additive noise jitter term, the flux normalisation factor and the flux linear slope, respectively. $\sigma_w$ is expressed in {parts-per-million (ppm)} and is added to all error bars quadratically. A negative value of $\sigma_w$ means that the error bars are shrunk. Each $f_0$ value is fitted after the corresponding data set has been normalised by its median value (i.e. all flux values are close to 1 and without units). The linear slope $c_1$ is expressed in normalised flux units per day.
                    }
                    }
                    }
                \end{table*}
                }
                \end{@twocolumnfalse}
                \clearpage
                        
                \begin{@twocolumnfalse}
                {\renewcommand{\arraystretch}{1.2}
                \begin{table*}[ht]
                    \parbox{\textwidth}{
                    \caption{Values of the systematic parameters for \tess{}.}
                    \label{tab:tess_syst}
                    \centering
                    {
                    \begin{tabular}{lcc}
                            \hline\hline
                            & $\sigma_w$ [ppm] & $f_0$ \\
                            \hline
                            Orbit 011 & ${113}_{-{23}}^{+{19}}$ & ${0.9996806}_{-{0.0000069}}^{+{0.0000070}}$ \\
                            Orbit 012 & ${89}_{-{40}}^{+{25}}$ & ${0.9996605}\pm{0.0000070}$ \\
                            Orbit 013 & ${39}_{-{63}}^{+{46}}$ & ${0.9998882}_{-{0.0000078}}^{+{0.0000079}}$ \\
                            Orbit 014 & ${150}_{-{18}}^{+{17}}$ & ${0.9994915}_{-{0.0000077}}^{+{0.0000076}}$ \\
                            Orbit 065 & ${203}\pm{16}$ & ${0.9997122}\pm{0.0000088}$ \\
                            Orbit 066 & ${261}\pm{12}$ & ${0.9996716}_{-{0.0000084}}^{+{0.0000085}}$ \\
                            Orbit 067 & ${210}_{-{13}}^{+{12}}$ & ${0.9996800}\pm{0.0000075}$ \\
                            Orbit 068 & ${267}\pm{11}$ & ${0.9996845}\pm{0.0000078}$ \\
                            Orbit 145 & ${198}_{-{17}}^{+{16}}$ & ${0.9996415}\pm{0.0000090}$ \\
                            Orbit 146 & ${150}_{-{26}}^{+{22}}$ & ${0.9997236}\pm{0.0000095}$ \\
                            \hline\hline
                    \end{tabular}
                    \tablefoot{The parameters $\sigma_w$ and $f_0$ are the additive noise jitter term and the flux normalisation factor, respectively. $\sigma_w$ is expressed in {parts-per-million (ppm)} and is added to all error bars quadratically. A negative value of $\sigma_w$ means that the error bars are shrunk. Each $f_0$ value is fitted after the corresponding data set has been normalised by its median value (i.e. all flux values are close to 1 and without units).}
                    }
                    }
                \end{table*}
                }
                \end{@twocolumnfalse}

                \begin{@twocolumnfalse}
                {\renewcommand{\arraystretch}{1.2}
                \begin{table*}[ht]
                    \parbox{\textwidth}{
                    \caption{Values of the systematic parameters for \spitzer{}.}
                    \label{tab:spitzer_syst}
                    \centering
                    {
                    \begin{tabular}{lccc}
                             \hline\hline
                            & $\sigma_w$ [ppm] & $f_0$ & $c_1$ [day$^{-1}$]\\
                            \hline
                            Visit 01 & ${-949}_{-{89}}^{+{102}}$ & ${0.997018}\pm{0.000076}$ & -- \\
                            Visit 02 & ${1690}_{-{160}}^{+{156}}$ & ${0.996087}_{-{0.000156}}^{+{0.000159}}$ & -- \\
                            Visit 03 & ${900}_{-{300}}^{+{220}}$ & ${0.996271}_{-{0.000076}}^{+{0.000078}}$ & ${-0.00356}\pm{0.00150}$ \\
                            Visit 04 & ${1351}_{-{158}}^{+{151}}$ & ${0.995746}_{-{0.000133}}^{+{0.000135}}$ & -- \\
                            Visit 05 & ${-321}_{-{27}}^{+{35}}$ & ${0.996462}\pm{0.000038}$ & -- \\
                            Visit 06 & ${216}_{-{49}}^{+{44}}$ & ${0.996468}_{-{0.000035}}^{+{0.000036}}$ & -- \\
                            Visit 07 & ${-176}_{-{44}}^{+{89}}$ & ${0.996450}\pm{0.000034}$ & ${-0.00142}\pm{0.00038}$ \\
                            Visit 08 & ${-514}_{-{15}}^{+{19}}$ & ${0.996469}\pm{0.000039}$ & ${-0.00118}\pm{0.00046}$ \\
                            Visit 09 & ${172}_{-{79}}^{+{55}}$ & ${0.996468}_{-{0.000035}}^{+{0.000036}}$ & ${-0.00097}\pm{0.00040}$ \\
                            Visit 10 & ${173}_{-{63}}^{+{49}}$ & ${0.996471}\pm{0.000034}$ & ${-0.00123}_{-{0.00038}}^{+{0.00037}}$ \\
                            Visit 11 & ${166}_{-{84}}^{+{57}}$ & ${0.996446}\pm{0.000035}$ & -- \\
                            Visit 12 & ${138}_{-{93}}^{+{58}}$ & ${0.996452}\pm{0.000033}$ & -- \\
                            Visit 13 & ${-425}_{-{10}}^{+{15}}$ & ${0.996427}_{-{0.000038}}^{+{0.000041}}$ & -- \\
                            Visit 14 & ${-447}_{-{15}}^{+{19}}$ & ${0.996439}_{-{0.000037}}^{+{0.000038}}$ & -- \\
                            \hline\hline
                    \end{tabular}
                    \tablefoot{The parameters $\sigma_w$, $f_0$ and $c_1$ are the additive noise jitter term, the flux normalisation factor and the flux linear slope, respectively. $\sigma_w$ is expressed in {parts-per-million (ppm)} and is added to all error bars quadratically. A negative value of $\sigma_w$ means that the error bars are shrunk. Each $f_0$ value is fitted after the corresponding data set has been normalised by its median value (i.e. all flux values are close to 1 and without units). The linear slope $c_1$ is expressed in normalised flux units per day.}
                    }
                    }
                \end{table*}
                }
                \end{@twocolumnfalse}
\clearpage
        
\end{appendix}

\end{document}